\documentclass[iop,revtex4,12pt,preprint]{aa}

\usepackage{graphics,graphicx}
\usepackage{helvet}
\usepackage{subfigure}
\usepackage[utf8]{inputenc}
\usepackage[T1]{fontenc}
\usepackage{soul} 
\usepackage[colorlinks,urlcolor=blue,citecolor=blue,linkcolor=blue,pdfusetitle]{hyperref}
\usepackage{amsmath, amssymb, gensymb}
\usepackage{morefloats}
\usepackage{float}
\usepackage{ulem}
\usepackage{url}
\usepackage{natbib}
\usepackage{microtype}
\usepackage{multirow}
\usepackage{morefloats}
\usepackage{rotating}
\usepackage{here}
\usepackage{xspace}
\usepackage{color}
\usepackage{lipsum}
\usepackage[a4paper]{geometry}
\usepackage{xspace}
\usepackage{tabularx}
\usepackage{supertabular}
\usepackage{caption}
\usepackage{float}
\usepackage{stmaryrd}

\def\Pf{\mathcal{P_{\rm{frac}}}}
\def\S{\mathcal{S}}
\def\StimesP{\S\times\Pf}

\def\eg {e.g.,\xspace} %e.g.,
\def\ie {i.e.,\xspace} %i.e.,
 \def\scaleGE{0.35} %for preprint 2 columns
 \def\scalepolaplots{0.31} %for preprint 2 columns

\begin{document}

\title{Physical conditions for dust grain alignment in Class 0 protostellar cores}
\subtitle{II. The role of the radiation field in models aligning/disrupting dust grains}

\author{V. J. M. Le Gouellec\inst{1,}\inst{2,}\inst{3} , A. J. Maury\inst{1,}\inst{4}, C. L. H. Hull \inst{5,}\inst{6,}\inst{8}, A. Verliat\inst{1}, P. Hennebelle\inst{1}, V. Valdivia\inst{7}}

\institute{
Laboratoire AIM, Paris-Saclay, CEA/IRFU/SAp - CNRS - Université Paris Diderot, 91191 Gif-sur-Yvette Cedex, France
\and
SOFIA Science Center, Universities Space Research Association, NASA Ames Research Center, Moffett Field, California 94035, USA
\email{valentin.j.legouellec@nasa.gov}
\and
European Southern Observatory, Alonso de C\'ordova 3107, Vitacura, Casilla 19001, Santiago , Chile
\and
Harvard-Smithsonian Center for Astrophysics, Cambridge, MA 02138, USA
\and
National Astronomical Observatory of Japan, NAOJ Chile, Alonso de Córdova 3788, Office 61B, 7630422, Vitacura, Santiago, Chile
\and
Joint ALMA Observatory, Alonso de Córdova 3107, Vitacura, Santiago, Chile
\and 
Department of Physics, Nagoya University, Furo-cho, Chikusa-ku, Nagoya, Aichi 464-8602, Japan
\and
NAOJ Fellow
}

\date{}

\abstract
{The polarized dust emission observed in Class 0 protostellar cores at high angular resolution with ALMA has raised several concerns about the grain alignment conditions in these regions.}
{We aim to study the role of the radiation field on the grain alignment mechanisms occurring in the interior ($\leq\,1000$ au) of Class 0 protostars.}
{
We produce synthetic observations of the polarized dust emission from a MHD model of protostellar formation, using the POLARIS dust radiative transfer tool, which includes dust alignment with Radiative Torques Alignment (RATs). We test how the polarized dust emission from the model core depends on the irradiation conditions in the protostellar envelope, by varying the radiation due to accretion luminosity propagating from the central protostellar embryo throughout the envelope.
The level of grain alignment efficiency obtained in the radiative transfer models is then compared to (sub-) millimeter ALMA dust polarization observations of Class 0 protostars.
}
{Our radiative transfer calculations have a central irradiation that reproduces the protostellar luminosities typically observed towards low- to intermediate-mass protostars, as well as super-paramagnetic grains, and grains $\geq\,10\,\mu$m, which are required to bring the dust grain alignment efficiencies of the synthetic observations up to observed levels.
We discuss together the characteristics timescales of the grain alignment physics, RAdiative Torque Disruption (RATD) of grains, and the typical time variability of accretion occurring in Class 0 protostellar cores. 
In our model, during an accretion burst or a steady-state phase of high luminosity from the protostellar embryo, RATD could have enough time to disrupt the largest grains in irradiated regions.
Finally, in high-luminosity conditions (with $L_{\star}\,\geq\,20\,$L$_\odot$, in our model) we find that the alignment of grains with respect to the anisotropic component of the radiation field ($k$-RAT) could drive inefficient alignment for grains $\gtrsim\,10\mu$m. However, given the high grain alignment efficiency observed in protostellar envelopes, large grains are most likely aligned with the magnetic field and thus potentially subject to rotational disruption, depending on their tensile strength.
}
{Our radiative transfer calculations show that irradiation plays an important role in the mechanisms that dictate the size range of aligned grains in Class 0 protostars. Regions of the envelope that are preferentially irradiated harbor strong polarized dust emission but can be affected by the rotational disruption of dust grains, thus controlling the population of the largest aligned grains. Episodes of high luminosity could affect grain alignment and trigger grain disruption mechanisms.
}


\titlerunning{Constraining the radiation field via dust polarization}
\authorrunning{Le Gouellec et al.}

\maketitle

\section{Introduction}

Class 0 protostars are the youngest objects in the protostellar phase leading to the formation of solar-type stars. They are embedded protostars, in which most of the mass still is under the form of a dense cold envelope surrounding the central protostellar embryo in the making \citep{Andre1993,Andre1994}. Because of their embedded nature, their spectral energy distribution (SED) peaks at sub-millimeter wavelengths, as most of the accretion luminosity produced in the center gets reprocessed by the dense envelope. The accretion of the envelope material is also accompanied by the ejection of material via (bi)polar jet/outflow system. Both of the radiative energy liberated at the accretion shock on the protostellar embryo and the mechanical energy transported by the jet and outflow, contribute to open and shape outflow cavities.

The luminosity of a protostar is the sum of the luminosity of the central protostellar photosphere and the accretion luminosity, which corresponds to the fraction of energy carried by the infalling matter that is radiated away at the accretion shock. Generally, constraints on protostellar accretion come from the total luminosity of protostars, given that the dusty envelope absorbs the radiative energy coming from the center and re-emits it at (far)-IR and sub-millimeter wavelengths. Several surveys \citep{Kenyon1990b,Kenyon1994,Evans2009,Dunham2010a,Dunham2013} supported the existence of a luminosity problem because the measured total luminosities are on average too low compared to what can be predicted by protostellar accretion models \citep{Myers1998,Young2005}, that consider a typical protostellar lifetime \citep{Dunham2015,Kristensen2018}. While \citet{Offner2011b,Myers2011,Myers2014} constructed models that reasonably predict the observed luminosities without including outburst activity, there are several evidences suggesting that protostellar luminosities can be variable on timescales of months/years \citep{Fischer2019,LeeYH2021,Park2021,Zakri222,Fischer2022}, which is also supported by chemical studies of protostellar envelopes \citep{Visser2012,Jorgensen2013,Visser2015,Anderl2016,Frimann2016}, and variability of emission line outflow tracers as a function of the distance from protostars \citep{Arce2001,Plunkett2015}. Such variability could point to episodic accretion \citep{Baraffe2009,Vorobyov2010}, episodic ejection (see, \eg recent models from \citealt{Commercon2022}), viscous and magnetic instabilities at the boundary between the stellar magnetosphere and the accretion disk \citep{Kulkarni2008,DAngelo2012,Takasao2019}, binary interactions \citep{Bonnell1992}, or instabilities of the structure of the protostellar embryo itself \citep{Vaytet2018}.

The accretion shock converts kinetic energy of the infalling circumstellar gas into a ultraviolet radiation field which is rapidly reprocessed by the high density gas and grains surrounding the protostellar embryo. Hence, the protostellar accretion rate and conditions responsible for reprocessing the accretion radiation in the first inner au are key to set the temperature and alignment efficiency of dust grains, which emission is in turn widely used to probe both the mass of material in protostellar environments but also the magnetic field topology through their polarized emission. Indeed, aspherical grains can align their minor axis parallel the ambient magnetic field lines via the phenomenon of “Radiative Torques Alignment” (RATs; \citealt{Draine1996,Draine1997,LazarianHoang2007,Andersson2015}). 
Paramagnetic dust grains get internally aligned via, \eg the nuclear relaxation, the inelastic relaxation, and the Barnett internal relaxation processes \citep{Purcell1979,Lazarian1999a,Lazarian1999c,LazarianHoang2007,Hoang2009,Hoang2022b}, and interact with the external magnetic field such that the grain angular momentum performs a Larmor precession around the ambient magnetic field \citep{Dolginov1976}. Then, the radiative torques, that result from the different scattering efficiencies of an impinging photon beam decomposed in a right- and left-hand circular polarized light, are driving the alignment between the grain angular momentum and the magnetic field. However, when embedded in a strong radiation field, grain precession can preferably happen around the anisotropic component of the radiation field rather than the magnetic field orientation ($k$-RATs; \citealt{LazarianHoang2007,Tazaki2017}).

The efficiency of the RAT grain alignment mechanism is a balance between the radiative torques efficiency (that are spinning dust grains up) and the collisional de-alignment of dust grains due to gas pressure (the rotational damping is also affected by the IR emission of dust grains, but it negligible in protostellar envelopes; \citealt{Draine1998,Hoang2020b}). In practice, this balance determines for which grains the local conditions enable their alignment with magnetic field lines via RATs.
A dust grain is considered aligned if its alignment is sufficiently stable, \ie if the RAT-induced grain's angular momentum $J_{\rm{max}}(\psi)$ (where $\psi$ is the angle between the radiation field vector and the magnetic field) is sufficiently large compared to the grain thermal angular velocity $J_{\rm{th}}$, \ie for $J_{\rm{max}}(\psi)/J_{\rm{th}}\approx3$ \citep{Hoang2008}. This ratio of angular momentum defines the condition for grains to be aligned, which translates into a critical grain size $a_{\rm{align}}$. Dust grains larger than $a_{\rm{align}}$ can be considered aligned, and this condition is thus set by the environment parameters, \ie radiation field, gas density, and temperature. 
Given the large mean wavelength of the photons propagating in the envelope of protostars from the central object (\ie the mean wavelength of the radiation field that has been reddened by the inner envelope material only; of the order of $\sim$ 1-10 $\mu$m, see \citealt{Hoang2020b}), it was found that dust grains, in order to be aligned, must be larger than the typical size of ISM dust grains, \ie larger than $\sim$ 10 $\mu$m \citep{LeGouellec2019a,Valdivia2019,Hull2020a}. Homogeneous and high level of grain alignment efficiency was found in the diffuse ISM \citep{Planck2018XII,Reissl2020}. Looking at the inner regions of protostellar cores, \citet{LeGouellec2020} found that the grain alignment efficiency in the envelope of Class 0 protostellar cores is on average relatively high, at the order of perfectly aligned grains (see also \citealt{Kuffmeier2020}). To reproduce the polarization fractions observed in the ISM and in dense environments, it was proposed that the efficiency of RATs can be highly increased for grains containing super-paramagnetic inclusions, such as iron inclusions \citep{Yang2021,ChauGiang2022}. 
Indeed, while the magnetic relaxation of ordinary paramagnetic grains has been found to inefficiently align thermally rotating grains with the magnetic field (mechanism introduced by \citealt{DavisGreenstein1951}, consisting in dissipation of the grain rotational energy due to the rotating magnetization with respect to the grain main axis), superparamagnetic relaxation can enhance the alignment degree induced by RATs
\citep{Lazarian2008,Hoang2016,Hoang2016b}. Namely, the super-paramagneticity of dust grains increases the fraction of grains eligible to RAT-induced supra-thermal rotation, up to which dust grains can be considered perfectly aligned.

ALMA observations revealed polarized dust emission in several Class 0 circumstellar envelopes at $\sim$ 50 $-$ 1000 au scale \citep{Hull2017b,Cox2018,Maury2018,Ko2019,Kwon2019,LeGouellec2019a,Sadavoy2018a,Sadavoy2018b,Sadavoy2019,Takahashi2019,Hull2020a}, and noted the prominence of dust polarization features following outflow cavity walls.
\citet{LeGouellec2020} exhibited regions of low polarization with quite organized magnetic fields, noting that the grain alignment efficiency does not seem to be totally homogeneous throughout the inner envelope, and appear to depend on the local physical conditions, like the irradiation field (see \citealt{Pillai2020}). In this work, we propose to focus on the specific role of the radiation field on the grain alignment in protostellar envelopes. This in turn allows us to discuss the properties of the aligned dust grains. In \citet{LeGouellec2023} (Paper I in the following), we compared ALMA dust polarization and radiation-sensitive molecular line observations for a handful of Class 0 protostellar objects, exhibiting a variety of morphologies for the radiation field morphologies and polarized intensity enhancement. In Class 0 protostars where the spatial distribution of the circumstellar material is rather isotropic, and only one clear outflow emanates from the central object, we find that observations suggest the local irradiation is potentially correlated to the regions of enhanced polarized dust emission.

This paper is structured as follows. In Section \ref{sec:p3_RT}, we present the MHD model of a protostellar core, that we synthetically observe with the polarization radiative transfer code POLARIS \citep{Reissl2016}, which incorporates the RAT grain alignment mechanism. 
We analyze the synthetic dust polarization maps and compare the grain alignment efficiency resulting from different levels of irradiation with those found in \citet{LeGouellec2020} in Section \ref{sec:RT_results}.
Finally, in Section \ref{sec:p3_disc}, we discuss the results of the radiative transfer calculations in light of the radiation field characteristics, dust evolution, and accretion variability in Class 0 protostars.
We draw our conclusions in Section \ref{sec:p3_ccl}.

\section{Model of a collapsing protostellar core}
\label{sec:p3_RT}

In this section, we present the radiative transfer calculations we have performed onto a radiation non-ideal magneto-hydrodynamic (MHD) simulation. This simulation, representing a prototypical star forming object, is then used a series of radiative transfer calculations, where a range of several parameters are explored, namely the protostar luminosity $L_\star$, the maximum dust grain size $a_{\textrm{max}}$, and the fraction of grains aligned in suprathermal rotation $f_{\rm{high}-J}$.

\subsection{RAMSES MHD simulation}

\begin{figure}[!tbh]
\centering
%\hspace{-0.7cm}
\subfigure{\includegraphics[scale=0.31,clip,trim= 5cm 1cm 3cm 2.5cm]{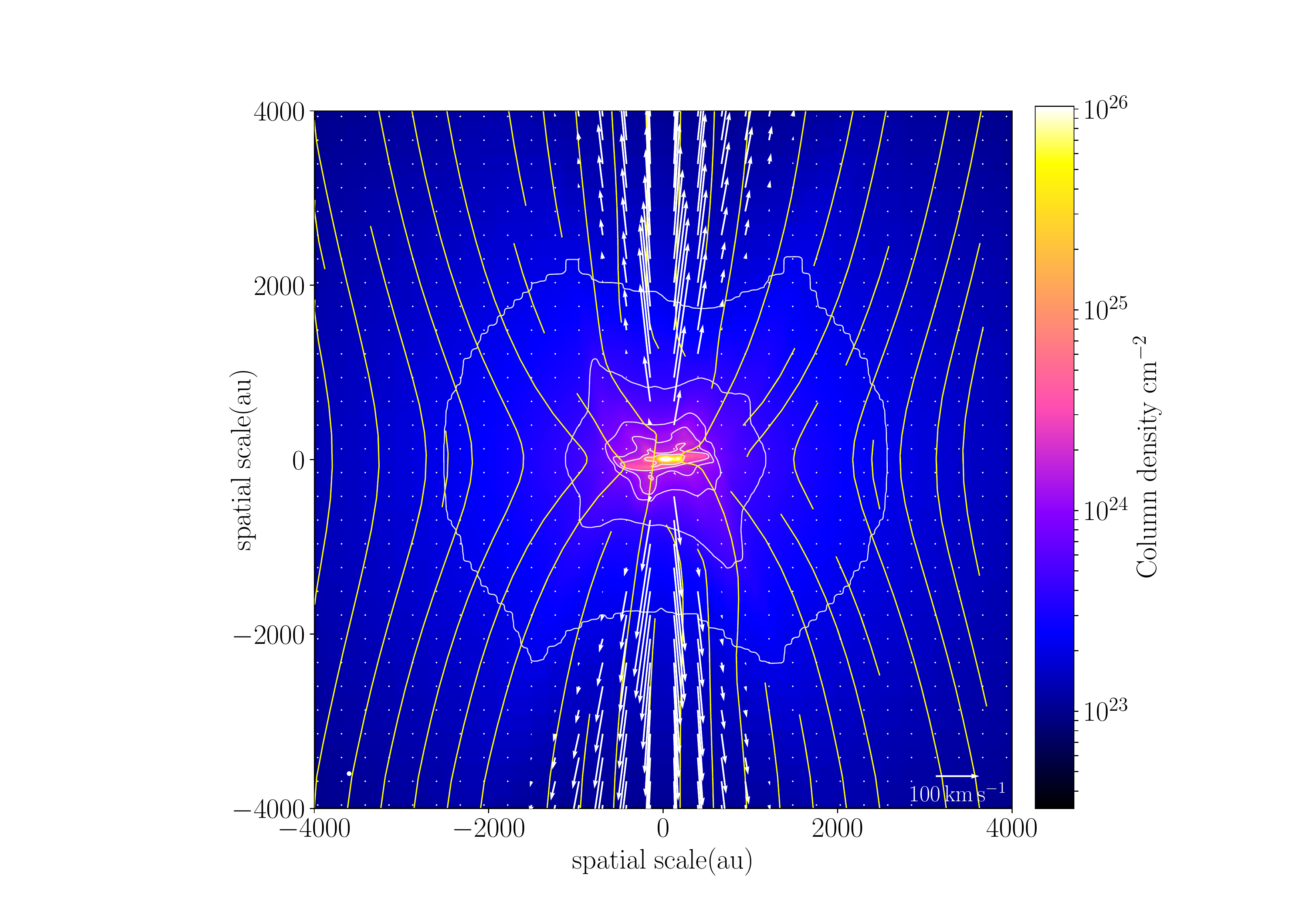}}
\subfigure{\includegraphics[scale=0.31,clip,trim= 5cm 1cm 3cm 2.5cm]{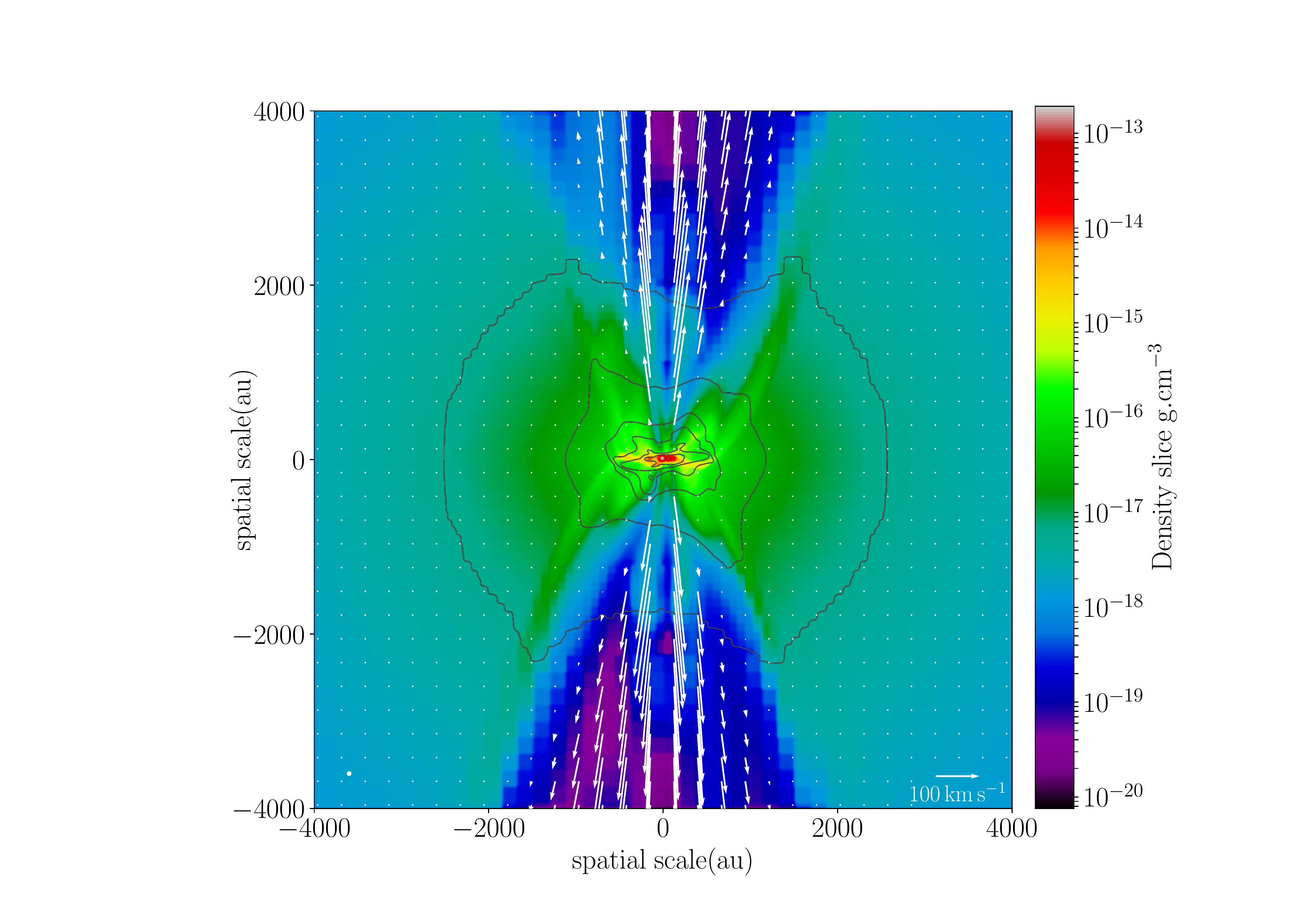}}
\caption[Snapshot of our MHD simulation]{\small Snapshot of our MHD simulation. \textit{Top panel}: The system is seen edge-on. The color scale and white contours represent the gas column density. The yellow streamlines trace the orientation of the density weighted average magnetic field along the line-of-sight. The white arrows represent the velocity field in a slice in the simulation centered on the sink particle. \textit{Bottom panel}: The contours and white arrows are the same as in the top panel. The colorscale represents a slice of gas mass density, centered on the sink particle. In both panels, the little white circle is the full width half maximum of a circular gaussian kernel we use to slightly smooth the highest resolution of the AMR grid, for visibility.}
\label{fig:simu}
\end{figure}

We use a detailed model of the evolution of a protostellar object that includes many of the relevant physics and allows us to accurately test how the central luminosity gets reprocessed in realistic envelope conditions. Further details and consideration are presented in Appendix \ref{sec:app_MHD_RT_details}. We use the RAMSES code \citep{Teyssier2002,Fromang2006,Commercon2011} with the implementation of sink particles \citep{Krumholz2004,BleulerTeyssier2014}, ambipolar diffusion \citep{Masson2012}, and Adaptive Mesh Refinement (AMR) to simulate the gravitational collapse of a protostellar core, with radiation non-ideal magneto-hydrodynamic calculations. The reference scenario we adopt here follows the collapse of a magnetized (the mass-to-flux ratio $\mu$ is 5), intermediate-mass starless core of 30\,M$_\odot$ initial mass, without initial turbulence, with an initial density profile of $\rho\,\propto$\,1/(1+$r^2$). We use the hybrid method developed by \citet{Risse2020}, which consists of a hybrid radiative transfer method using the gray M1 closure relation (that implements the zeroth
and first moments of the equation of radiative transfer; \citealt{Levermore1984,Rosdahl2013,Rosdahl2015}) for the radiation emanating from the protostellar embryo, and the gray flux-limited diffusion approach \citep{Levermore1981,Commercon2011b,Commercon2014} for photons emitted elsewhere in the simulation. After it has formed, the central protostellar embryo is represented by a sink particle. A jet is implemented by hand at the creation of the sink particle (the method used is the one of \citealt{Verliat2022}). The resolution of the highest level of the AMR grid is $\sim$5 au. 

\begin{table*}[!tbph]
\centering
\small
\caption[Radiative transfer calculations details]{Radiative transfer calculations details}
\label{t.RT_details}
\setlength{\tabcolsep}{0.25em} %\vspace*{0.1in}
\begin{tabular}{p{0.13\linewidth}c|c|c|c|c|cc}
\hline \hline &&&&&& \tabularnewline
 & Set I &  Set II &  Set III &  Set VI &  Set V &  Set VI \tabularnewline[0.1cm]
\hline  &&&&&& \tabularnewline
\small{$f_{\rm{high}-J}$} & 1.0 & 0.0, 0.25, 0.5, 0.75, 1.0 & 1.0 & 0.25 & 1.0 & 0.25 \tabularnewline[0.2cm]
\hline &&&&&& \tabularnewline
\small{Luminosity ($L_{\odot}$)} & 20 & 20 & 1, 5, 20, 50, 100 & 20 & 1, 5, 20, 50, 100 &  1, 5, 20, 50, 100 \tabularnewline[0.2cm]
\hline &&&&&& \tabularnewline
\small{$a_{\textrm{max}}$ ($\mu$m)} & 0.5, 2, 10, 30, 50 & 10 & 10 & 0.5, 2, 10, 30, 50 & 2 & 10 \tabularnewline[0.2cm]
\hline 
\end{tabular}
% \vspace{-0.15in}
\caption*{\small The values of the parameters we vary in the POLARIS radiative transfer calculations, \ie fraction of aligned grains at the high-$J$ attractor point $f_{\rm{high}-J}$, luminosity of the central object, and maximum dust grain size $a_{\textrm{max}}$ ($\mu$m). Each set gathers five radiative transfer runs where only one parameter varies. Each radiative transfer run produces synthetic observations at 0.87, 1.3 and 3 mm, and assumes that the RAT mechanism is responsible for aligning the dust grains with respect to the magnetic field.
}
\end{table*}

We present in Fig. \ref{fig:simu} the snapshot of this simulation that we use from now on. The simulation is 38.52 kyr old, and the central sink, which mimics the central protostellar embryo, is 14.8 kyr old, and has a mass of 1.2 M$_\odot$. As no initial turbulence is implemented, the collapse proceeds isotropically and the core exhibits a symmetric structure, which favors the accumulation of material toward the equatorial mid-plane. The jet rapidly clears out outflow cavities, which also causes the accumulation of material toward the outflow cavity walls, visible in the column density map and slice of mass density in Fig. \ref{fig:simu}, top and bottom panel, respectively. The line-of-sight we use from now on is an edge-on projection of the core.

\subsection{Radiative transfer of the dust emission with POLARIS}

\subsubsection{Radiative transfer parameters}

We perform radiative transfer calculations on this simulation using the POLARIS code \citep{Reissl2016}, which calculates the local dust temperature and dust grain alignment efficiency of oblong dust grains with respect to the magnetic field orientation following the RAT theory developed in \citet{LazarianHoang2007,Hoang2014b}. The POLARIS code computes first the propagation of photons alongside the dust temperature via a Monte Carlo analysis. Then, using the density, radiation field, temperature, and dust grain properties, POLARIS computes the RAT-induced grain angular momentum and compares it to the  thermal angular velocity to derive the $a_{\textrm{align}}$ parameter in each cell of the grid. Finally POLARIS solves the radiative transfer for the four Stokes parameters, at a given wavelength and line-of-sight. The only source of heating we include is the one emanating from the central source. We assume the impact of the external heating from ISRF or close by young stars can be neglected in those embedded objects, because of the small spatial scales targeted by these synthetic observations. The luminosity from the protostar photosphere and the accretion shock is modeled, for simplicity reasons, as a blackbody source located at the sink particle, whose temperature we vary in our different sets of calculations, and is derived as follows.

The parameters of the equivalent protostellar embryo are derived from the models developed by \citet{Kuiper2013} (see also \citealt{Hosokawa2009}), that provide the radius and luminosity of the central protostellar object ($R_\star$ = 1.23 R$_\odot$ and $L_\star$ = 0.58 L$_\odot$ for our MHD model snapshot). Then the effective accretion luminosity needs to be derived. This is, however, a very degenerate quantity for several reasons: a fraction of the accretion energy can be radiated away through mechanical processes such as jets, the accretion is known to be episodic and can undergo bursting phases of accretion, and the accretion rate onto the sink may be different from the actual accretion rate of material onto the central protostellar embryo (because the accretion can be regulated by a disk, or happen directly form the inner envelope; see \citealt{LeeYN2021}). Therefore, the determination of the exact value of the luminosity escaping from the sink particle is degenerate. In order to account for the variable accretion activity, we explore a range of luminosities, from 1 to 100 L$_\odot$, by varying the temperature of the central blackbody $T_\star$, and choosing for $R_\star$, the equivalent radius derived from the models cited above (where $T_\star$ and $R_\star$ are the parameters implemented in POLARIS that set the characteristics of the central blackbody, \ie the source of heating). In addition, we also note that in our simulation the accretion rates onto the sink are large, with average values of $\dot{M}_{\rm{acc}}\,\sim\,10^{-5}-10^{-4}\,\rm{M}_\odot\,\rm{yr}^{-1}$ (the accretion rate onto the sink in our simulation can vary by $\sim$1$-$2 orders of magnitude), compared with the common bolometric luminosities of low- and intermediate-mass Class 0 star-forming objects (\ie $\dot{M}_{\rm{acc}}\,\sim\,10^{-7}-10^{-5}\,\rm{M}_\odot\,\rm{yr}^{-1}$; \citealt{Evans2009}). 
This may be due to the simulation being fairly massive, and/or the spatial resolution that do not allow us to resolve the disk below 5 au, which is where the accretion mechanisms are occurring.
In order to reproduce different states of accretion, \ie from steady or low accretion activity, to episodes of high accretion activity, we thus adopt different central luminosities (values of $L_\star$ from 1 to 100 L$_\odot$) for the same MHD model snapshot that represent the fraction of accretion kinetic energy that is radiated away in the accretion luminosity.

% scale in preprint 2 columns 0.31
\begin{figure*}[!tbh]
\centering
\hspace{-0.1cm}
\subfigure{
\includegraphics[scale=\scalepolaplots,clip,trim= 5cm 1cm 3cm 2.5cm]{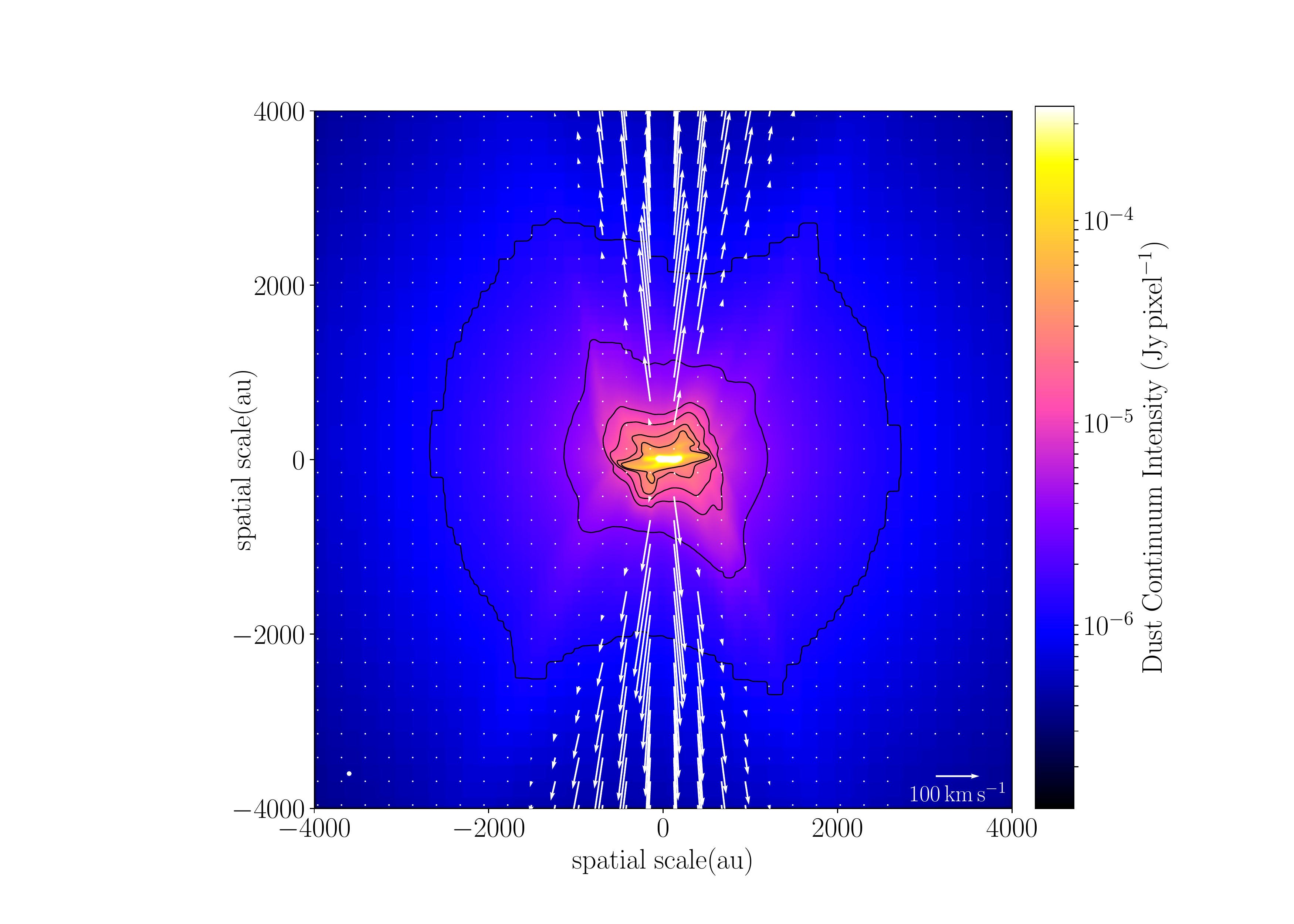}}
\subfigure{
\includegraphics[scale=\scalepolaplots,clip,trim= 5cm 1cm 3cm 2.5cm]{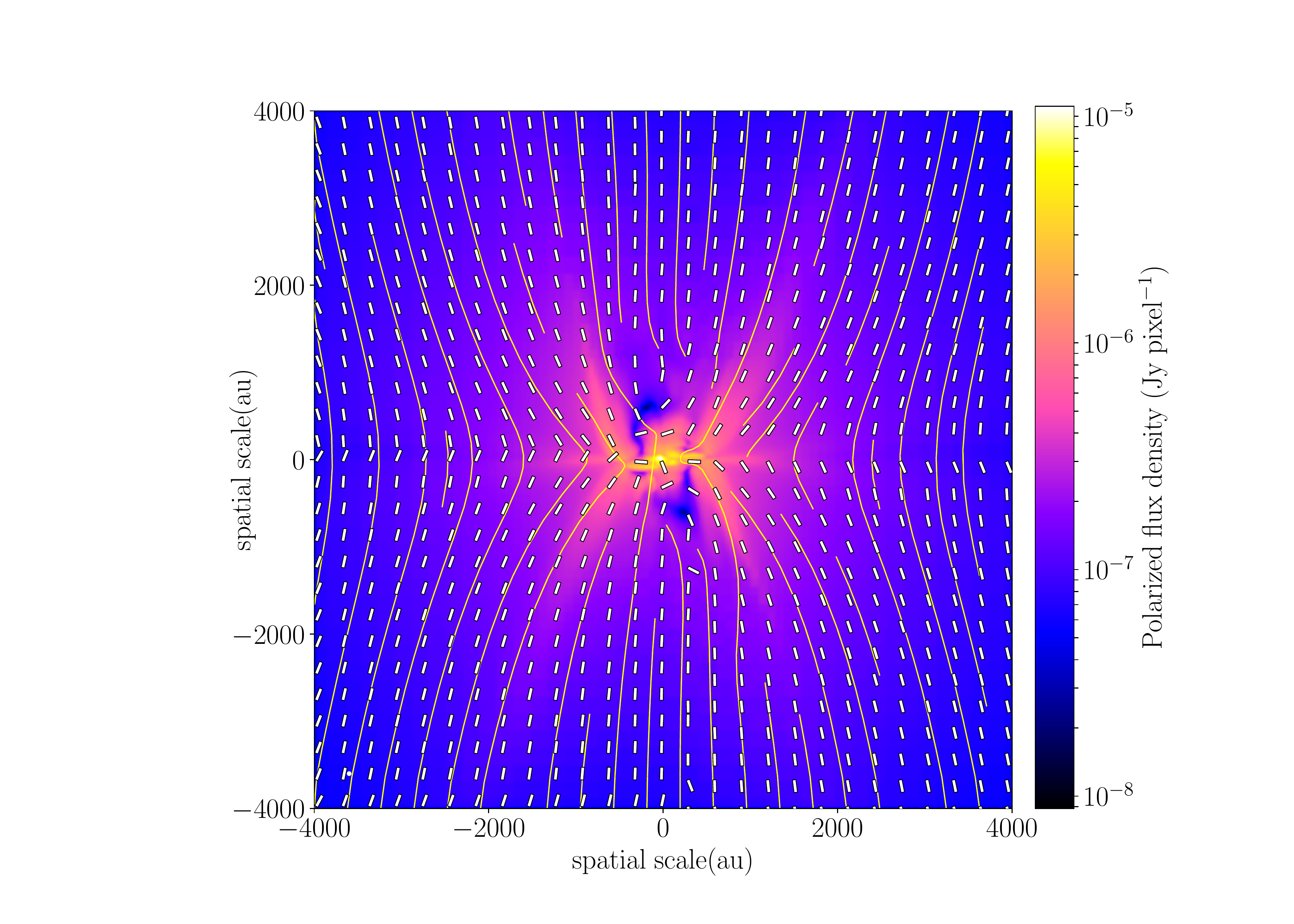}}
\subfigure{
\includegraphics[scale=\scalepolaplots,clip,trim= 5cm 1cm 3cm 2.5cm]{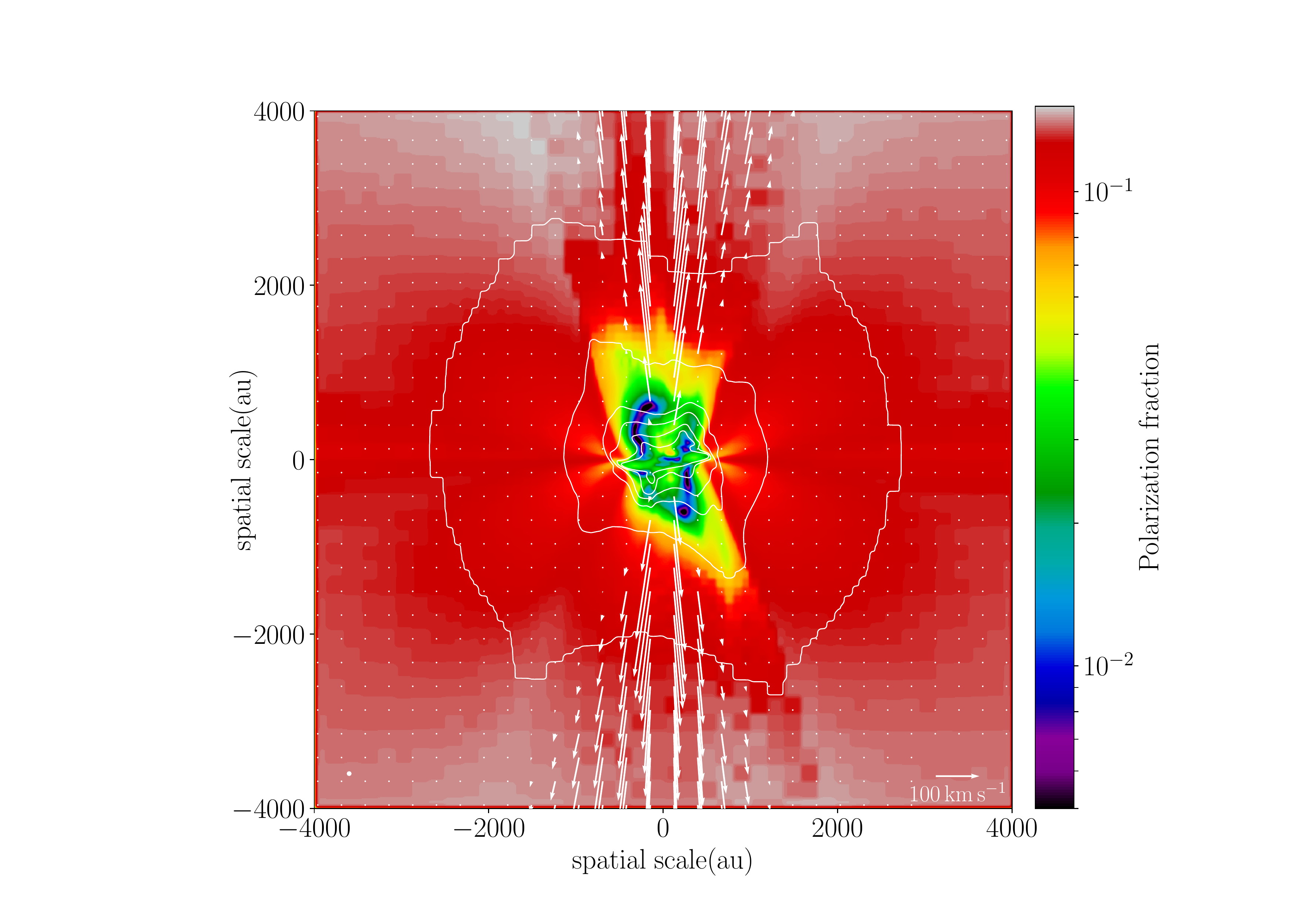}}
\subfigure{
\includegraphics[scale=\scalepolaplots,clip,trim= 5cm 1cm 3cm 2.5cm]{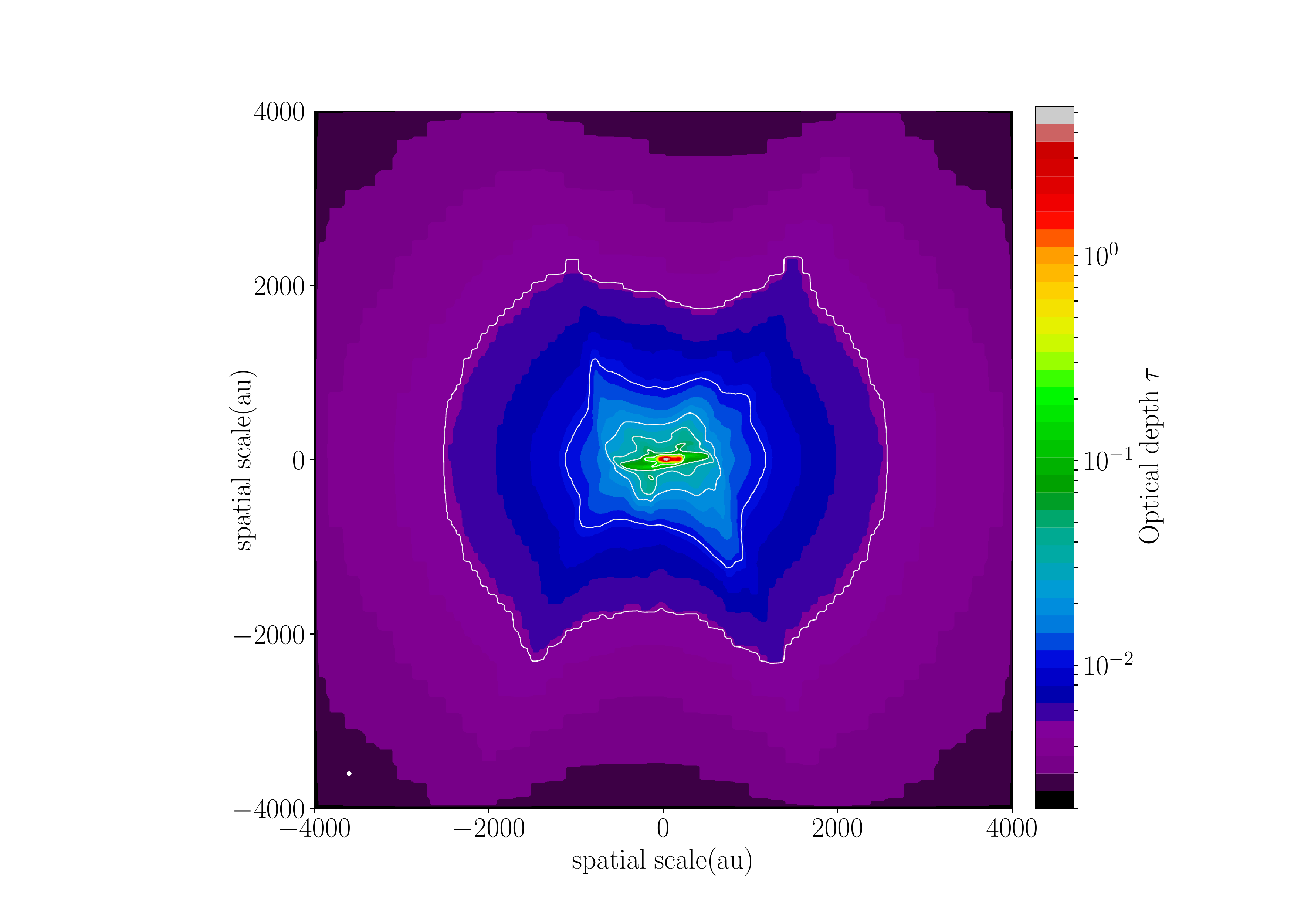}}
\caption[Dust polarization radiative transfer results of the reference case]{\small Dust polarization radiative transfer results of the fiducial case ($a_\textrm{max}\,=\,10\,\mu$m, $f_{\rm{high}-J}\,=\,1$, and $L_\star\,=\,20\rm{L}_\odot$). The plots are from one run of POLARIS, where our simulation has been synthetically observed in dust polarization at 0.87 mm. \textit{Top left panel}: The color scale and black contours represent the dust continuum total intensity (Stokes $I$). The white arrows represent the velocity field in a slice in the simulation centered on the sink particle. \textit{Top right panel}:  The color scale represents the polarized intensity ($\sqrt{Q^2+U^2}$). The yellow streamlines trace the orientation of the density-weighted average magnetic field along the line-of-sight. The line segments represent the polarization position angle orientations. \textit{Bottom left panel}: The color scale represents the fractional polarization $\Pf$. \textit{Bottom right panel}: The color scale is the optical depth computed during the ray-tracing of the radiative transfer. In the bottom panels, the white contours trace the total intensity.}
 \label{fig:polaris_canonical}
\end{figure*}

\begin{figure*}[!tbh]
\centering
%\hspace{-0.7cm}
\subfigure{\includegraphics[scale=0.31,clip,trim= 5cm 1cm 3cm 2.5cm]{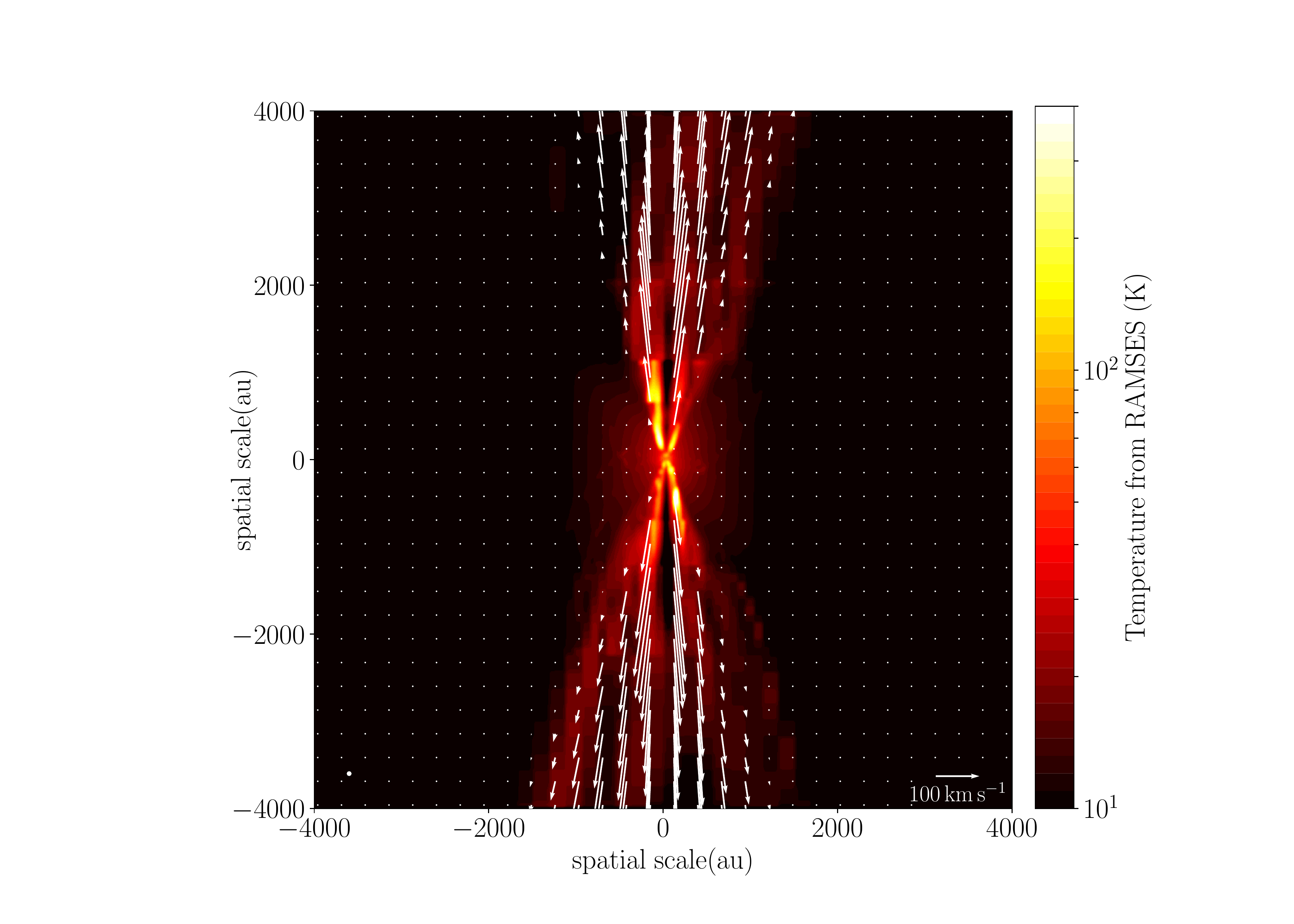}}
\subfigure{\includegraphics[scale=0.31,clip,trim= 5cm 1cm 3cm 2.5cm]{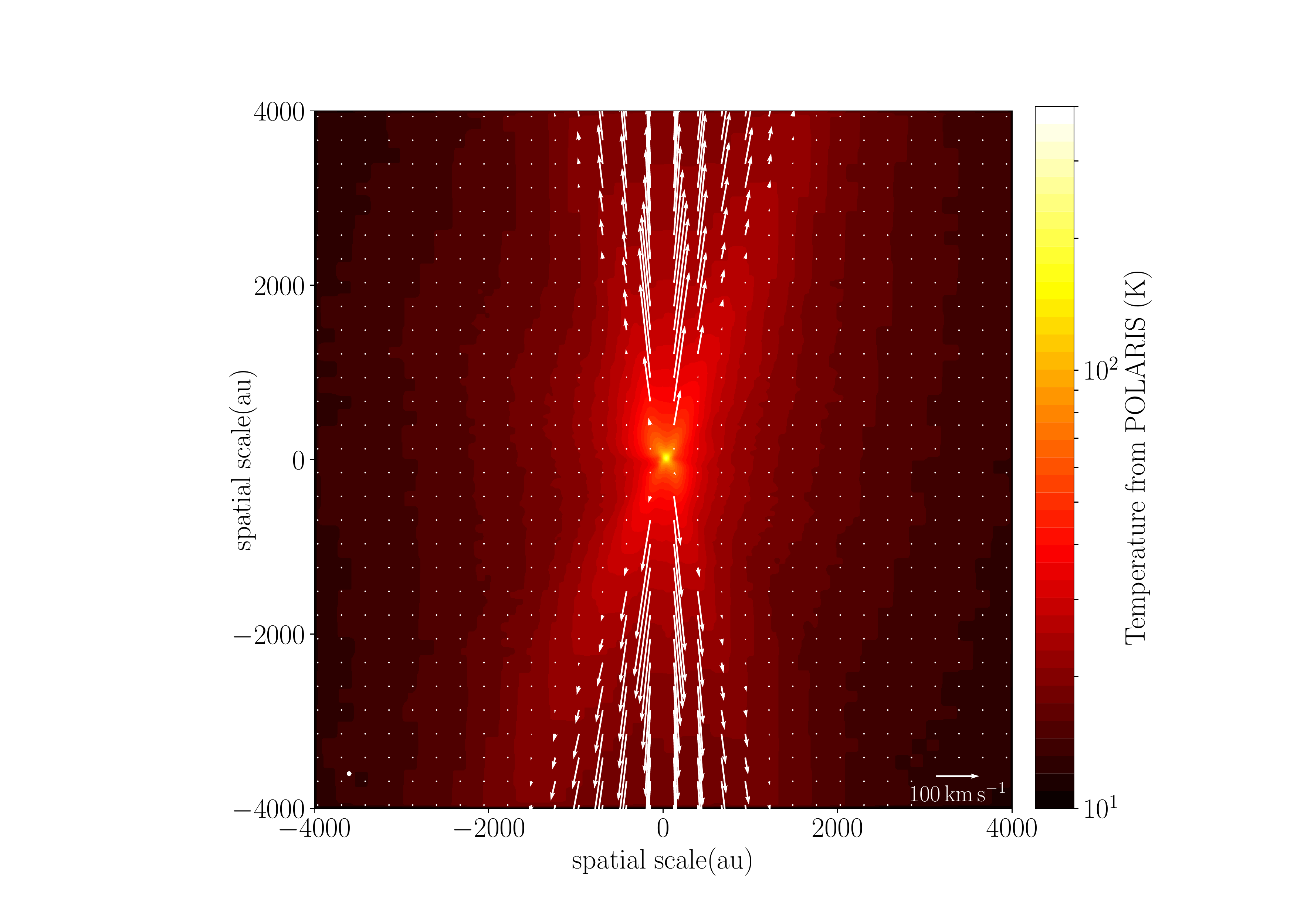}}
\caption[Temperature maps from RAMSES and POLARIS]{\small Temperature maps from RAMSES and POLARIS. The left panel presents the temperature map of the RAMSES simulation cube before the POLARIS radiative transfer, from the slice centered on the sink particle (
in the simulation the only central source of heating is the luminosity of the central protostellar object with $L_\star$ = 0.58 L$_\odot$). The right panel shows the temperature resulting from the radiative transfer performed by POLARIS for the reference case ($a_\textrm{max}\,=\,10\,\mu$m, $f_{\rm{high}-J}\,=\,1$, and $L_\star\,=\,20\rm{L}_\odot$).
The point is not to compare the temperature values across both maps, but to highlight that the heating mechanisms are different---in the RAMSES simulation the dynamics of the gas induce a significant heating toward a thin layer surrounding the high velocity jet.
}
 \label{fig:polaris_canonical_temp}
\end{figure*}

We use a population of dust grains with typical ISM composition, with 62.5\% astronomical silicates and 37.5\% graphite grains (MRN dust \citealt{Mathis1977}; optical properties from \citet{Weingartner2001}). This composition governs the ultimate maximum number of aligned grains, as silicates can be aligned with the magnetic field much more easily than graphite or carbonaceous grains (\citealt{Andersson2015} and references therein, see however a recent work by \citealt{Lazarian2020}). The dust grains are assumed oblate with an aspect ratio of 0.5 \citep{Hildebrand_Dragovan1995} and they follow a standard MRN-like distribution \citep{Mathis1977} with cutoff sizes of $a_\textrm{min}\,=\,2\,\textrm{nm}$ and various values of $a_\textrm{max}$. As the maximum grain size seems to be a key parameter that controls the resulting amount of polarized flux obtained in synthetic observations \citep{Valdivia2019}, we vary this parameter $a_\textrm{max}$ from 0.2 to $50\,\mu\textrm{m}$ (see Table \ref{t.RT_details}). We choose these values in light of recent works that have hinted at the presence of grains larger than the typical $\sim$\,0.5\,$\mu$m ISM dust grain maximum size in Class 0/I envelopes \citep[e.g.,][]{Miotello2014,Valdivia2019,LeGouellec2019a,Galametz2019,AgurtoGangas2019,Hull2020a,Nakatani2020,Ohashi2021}. Indeed, the reprocessed radiation field impinging on the dust grains in the protostellar envelope needs to comprise photons whose wavelength are comparable to the size of dust grains in order to efficiently align the grains via RATs. Given the density of the inner envelope structures, only low-energy submillimeter photons can propagate, which in turn corresponds to grains with sizes >1 $\mu$m. We assume a gas-to-dust ratio of 100. 

The radiative transfer computations carried out by the POLARIS code include several approximations to describe the effects of grains' paramagneticity on the grain dynamics. Among the aligned grains via RATs (\ie grains with $a\,\geq\,a_{\textrm{align}}$), a given fraction of grains aligned at supra-thermal rotation is specified, \ie the fraction of aligned grains at the high-$J$ attractor point, $f_{\rm{high}-J}$ \citep{LazarianHoang2007}.
Grains are considered imperfectly internally aligned at low-$J$, and perfectly internally aligned at high-$J$. For the $1\,-\,f_{\textrm{high-J}}$ fraction of low-$J$ aligned grains, the internal alignment is inefficient because of internal thermal fluctuations in the grains \citep{Lazarian1997}. To model the imperfect internal alignment of these grains, a Boltzmann distribution is used to describe the precession of grain's major axis of inertia with its angular momentum. This version of POLARIS does not include the effect of wrong internal alignment of low-$J$ aligned grains, which means we might overestimate the alignment efficiency of low-$J$ aligned grains in our study. In addition, we consider that the high-$J$ aligned grains are perfectly internally aligned. This is true for certain conditions or gas density, irradiation, and level of iron inclusions locked in the dust. Indeed, the amount of iron inclusions in dust increases significantly the grain magnetic susceptibility, which increases the rate of Barnett relaxation. Following Section 4 of \citet{Hoang2022b}, we see that for the typical conditions of our model, \ie $n_{\textrm{H}}\,\sim\,10^{6}-10^{8}$ cm$^{-3}$ and $u_{\rm{rad}}$/$u_{\rm{ISRF}}\,\sim\,10^{2}-10^{5}$ for the envelope $\lesssim\,2000$ au (see Section \ref{sec:RT_results}), grains whose size $a\,\lesssim\,10-50\,\mu$m should be efficiently internally aligned, depending on the level of iron inclusions locked in the dust. We do not consider maximum grain sizes $a_{\textrm{max}}\,\geq\,50\,\mu$m. 
Grains are considered perfectly externally aligned in both the low-$J$ and high-$J$ cases in our radiative transfer models, which also represents an approximation, especially for the low-$J$ aligned grains whose magnetic relaxation is less efficient.
In our use of POLARIS, this makes the $f_{\textrm{high-J}}$ parameter the parameter that describes the magnetic response of dust grains, because super-paramagnetic grains (\ie grains with iron inclusions; \citealt{Jones1967}) are mainly driven to high-$J$ state due to the effect of radiative torques and increased efficiency of magnetic relaxation. In a given model, $f_{\textrm{high-J}}$ is fixed for all grains. However, see \citet{ChauGiang2022} who varied $f_{\textrm{high-J}}$ using a step function based on the local magnetic relaxation conditions.
Table \ref{t.RT_details} presents all the sets of radiative transfer runs that we have performed, varying $a_\textrm{max}$, $f_{\rm{high}-J}$, and the luminosity, $L_\star$. Because the focus of the paper is the role of the radiation field on the dust polarization in protostars, we present the impact of $a_\textrm{max}$, and $f_{\rm{high}-J}$, on the resulting dust polarization in Appendices \ref{sec:app_RT_amax} and \ref{sec:app_RT_highJ}, respectively.

\subsubsection{Synthetic observations}

We choose a line of sight where the model is seen edge-on to produce with POLARIS the output dust emission maps from the model, at 0.87 mm, 1.3mm and 3mm, at a distance of 400\,pc, in maps 8000\,au in size with pixel sizes of 8\,au. We present our fiducial case of the dust polarization maps calculated by one POLARIS radiative transfer run in Fig. \ref{fig:polaris_canonical}. This fiducial case corresponds to the radiative transfer with $a_\textrm{max}\,=\,10\,\mu$m, $f_{\rm{high}-J}\,=\,1$, and $L_\star\,=\,20\textrm{L}_\odot$, performed at 0.87 mm. We note that the cavity walls exhibit enhanced emission in the dust continuum emission map, which is an important feature generally observed with ALMA toward star forming objects \citep{Maury2018,Kwon2019,LeGouellec2019a,Hull2020a}. This is likely due to the implementation of the jet, that has pushed aside material that accumulated in a shell-like structure, at the junction between the collapsing envelope and the outflowing gas. This enhancement in the dust continuum map is less visible when analyzing the column density map (see Fig. \ref{fig:simu}). The radiation field in the direction of the outflow heats the dust in the cavity walls, enhancing the total (Stokes I) thermal dust emission. Our radiative transfer results also reproduced the enhanced polarized dust emission toward the outflow cavity walls, another common signature observed in ALMA observations \citep{Hull2017b,Maury2018,Ko2019,Kwon2019,LeGouellec2019a,Hull2020a}. The line of sights toward the base of the cavities (up to $\sim\,700$ au from the center), and the inner mid-planes (up to $\sim\,400$ au), exhibit low fractional polarization ($\sim\,1$\%), which is explained by the highly disorganized magnetic field and high gas volume density in these two regions, respectively. In contrast, the outflow cavity walls, and outer mid-planes, exhibit enhanced signal in polarized dust emission, up to $\sim\,15$\%. Analyzing the optical depth map, we can assume the emission is optically thin in the envelope, except for the inner $\sim\,$200 au, that exhibit optical depth larger than unity.
% , which correspond to the dense inner circumstellar disk.

The enhancement that we obtain toward the cavity walls in total intensity and polarized dust emission is explained by the propagation of the photons computed by POLARIS. Figure \ref{fig:polaris_canonical_temp}, that presents the 2D-slices of temperature before and after the radiative transfer calculation for the fiducial case (left and right, respectively), shows how the material in outflow cavities is preferentially heated, compared with the equatorial mid-planes. In the radiative transfer results, the temperature reaches $\sim$100K in the central 100 au, and at 1000 au from the protostar, reached $\sim$40K in the direction of the cavities but only $\sim$ 20 K in the direction of equatorial mid-planes. 
The differences in the morphology of the temperature maps between before and after the radiative transfer are large toward the outflow cavities. 
During the non-ideal MHD simulation, the central source of heating only corresponds to the luminosity of the central protostellar object ($L_\star$ = 0.58 L$_\odot$ in the simulation; as the point is to discuss only the different heating mechanisms, the fiducial model with $L_\star$ = 20 L$_\odot$ is used for the radiative transfer results shown in Fig. \ref{fig:polaris_canonical_temp}).
However, RAMSES takes into account the dynamics of the gas to compute the local temperature. 
Given the velocity in the corresponding cells, the values of the ion-neutral gas friction parameter gets significantly high, and causes the encountered high gas temperature in a thin layer along the cavity walls. This physics is not accounted for in POLARIS, which only derive the local heating given the propagation of photons, and density/opacity of cells encountered in the photons propagation pathways. 
In other words, the dynamics of the outflow can represent a source of heating, and in turn a modification in the local physical conditions for grain alignment, that we do not have access to in our radiative transfer calculations (see Section \ref{sec:shocks}).

\begin{figure}[!tbh]
\centering
%\hspace{-0.7cm}
\subfigure{\includegraphics[scale=0.31,clip,trim= 5cm 1cm 3cm 2.5cm]{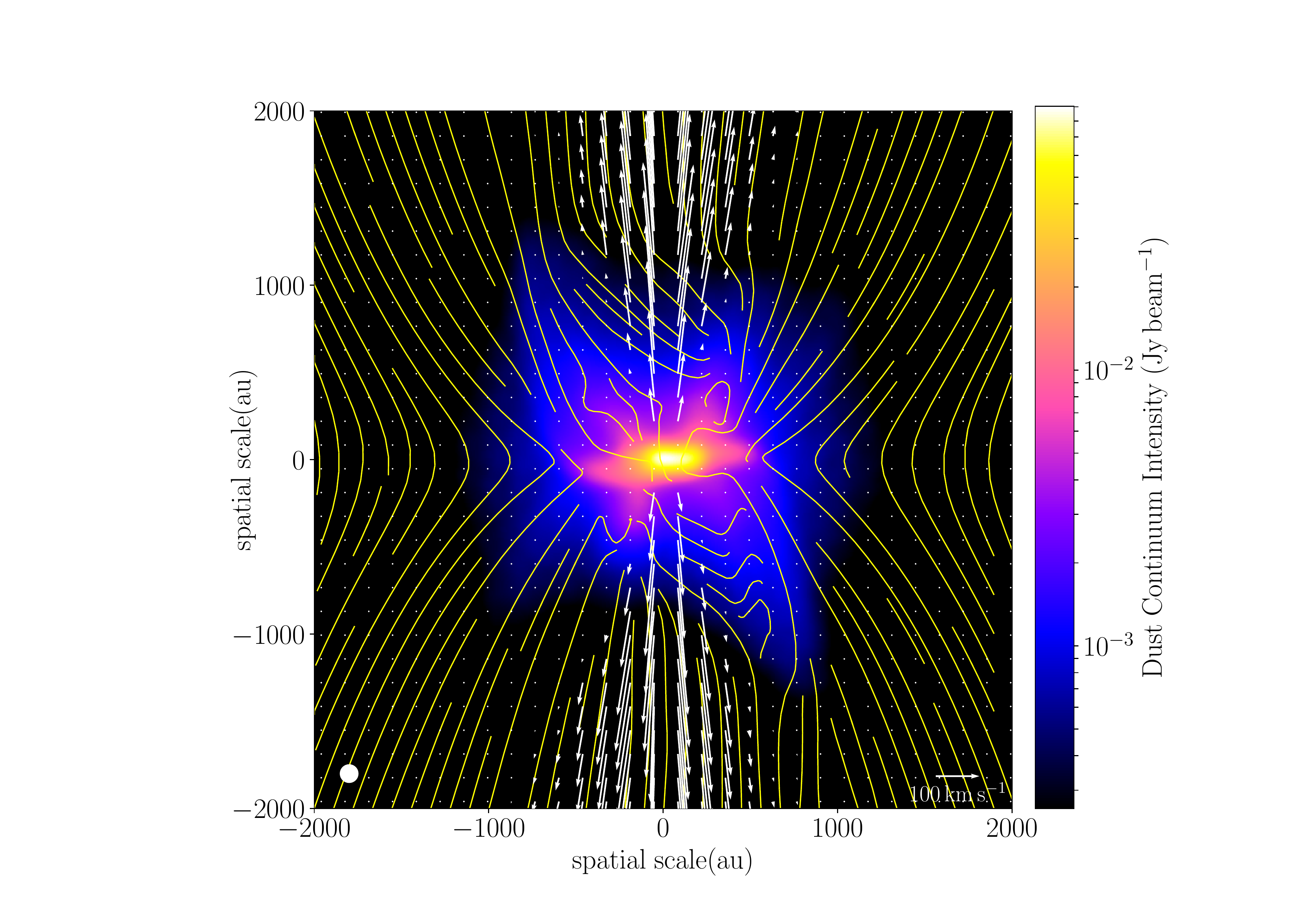}}
\subfigure{\includegraphics[scale=0.31,clip,trim= 5cm 1cm 3cm 2.5cm]{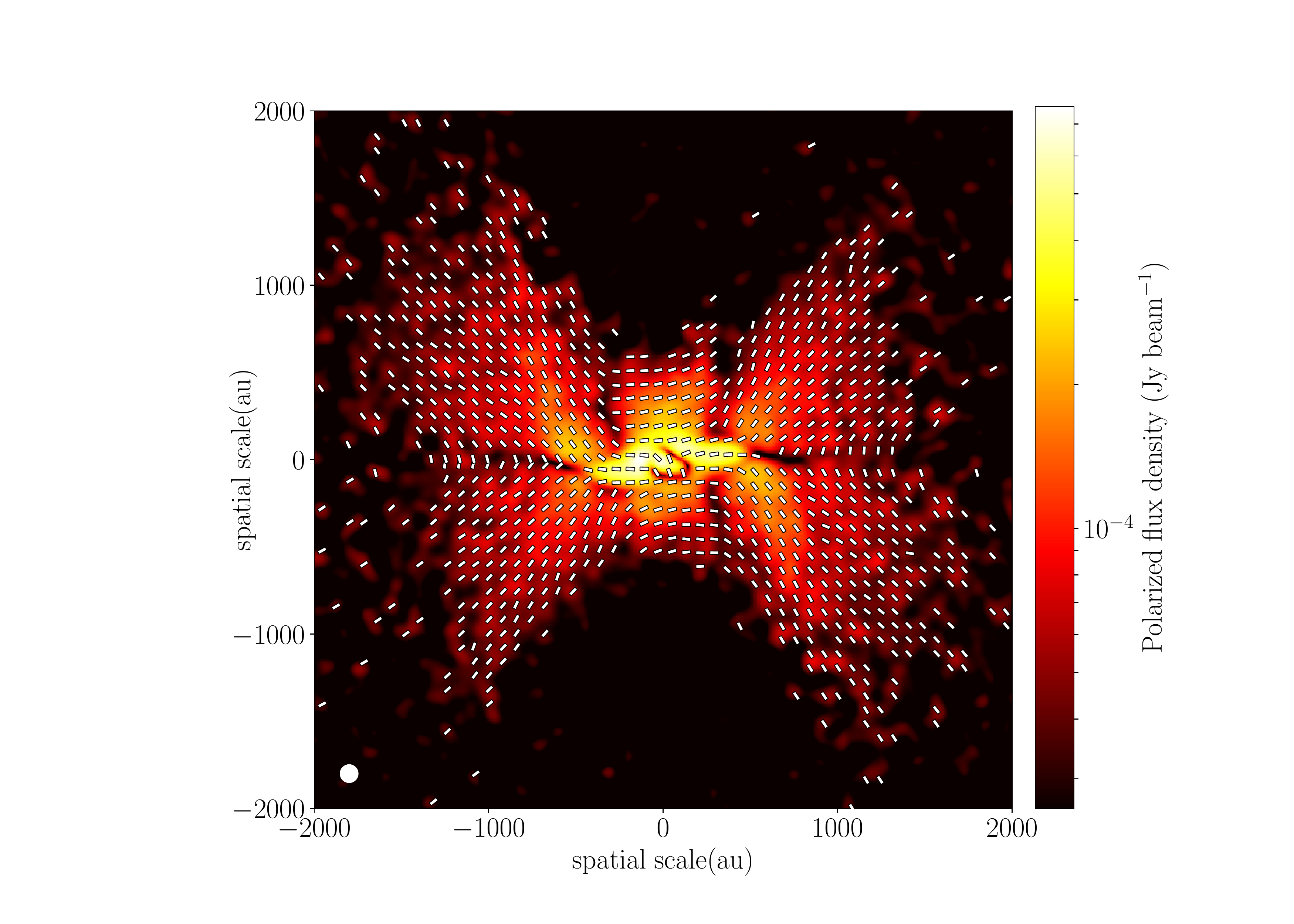}}
\caption[Dust polarization radiative transfer results of the reference case after spatial filtering]{\small Dust polarization radiative transfer results performed at $\lambda\,=\,0.87\,\mu$m with $f_{\rm{high}-J}=1.0$, $L_\star$=100 L$_{\odot}$, and $a_{\textrm{max}}=10\;\mu$m, after spatial filtering with the CASA simulator, mimicking ALMA interferometric observations. \textit{Top panel:} Synthetically observed dust continuum total intensity (Stokes $I$, color scale) is plotted with the velocity field of the central slice in the MHD simulation (white arrows), and the density weighted magnetic field lines (yellow lines). Stokes $I$ is shown when $I$>3\,$\sigma_I$, where $\sigma_I\,=\,0.10$\,mJy beam$^{-1}$. \textit{Bottom panel:} The color scale is the total linearly polarized intensity, which is shown where $P$\,>\,3\,$\sigma_P$, where $\sigma_P\,=\,10\,\mu$Jy beam$^{-1}$. The line segments represent the magnetic field orientations inferred from the dust polarization map. They are plotted where $P$\,>\,5\,$\sigma_P$. The white circle at the bottom left corner, is the resolution of the ALMA synthetic observations, and is 100\,au in size.}
\label{fig:polaris_canonical_filtered}
\end{figure}

We use the CASA (version 5.8) simulator to interferometrically filter the synthetically observed maps, mimicking ALMA observations. For each simulation, we combine synthetic observations from ALMA configurations C-2, C-4, and C-6, with an exposure time of three hours per antenna configuration. After an appropriate slight smoothing, the resulting synthesized beams (resolution elements) of these filtered maps have an effective size of 100\,au, when observed at 0.87 mm (and 140\,au, when observed at 1.3mm). We present in Fig. \ref{fig:polaris_canonical_filtered} an example of the resulting polarized dust emission maps after the ALMA synthetic observations. By doing so, the large scale emission (\ie the emission corresponding to low spatial frequency signal) is removed and a standard sky noise model is applied on the data. We recover a significant part of the total intensity signal inside the central $\sim\,$2000 au, while the detection of the polarized intensity suffers more from the resulting sensitivity (as noticed in \citealt{Reissl2017,Kuffmeier2020,LeGouellec2020}, the intrinsic grain alignment efficiency is generally lower in radiative transfer models compared to observations; see Section \ref{sec:align_eff}). The main features in the polarized intensity map are the polarized outflow cavity walls and equatorial mid-plane. We note also that $\sim$\,100-200\,au above and below from the equatorial plane, we recover a horizontal apparent magnetic field, suggesting a significant toroidal magnetic field component (also seen in Fig. \ref{fig:polaris_canonical} and \ref{fig:simu}). As shown in  \citep{LeGouellec2020}, interferometric filtering  has a strong impact on the polarization, as it artificially increases the polarization fraction values. Indeed, Stokes $I$ and Stokes $Q$ \& $U$ have different power spectrum (when analyzed as a function of spatial scales), due to the fact that Stokes $Q$ \& $U$ are sensitive to the total intensity as well as the grain alignment conditions and magnetic field morphology. Therefore, to compare these synthetic observations with observations and discuss the grain alignment efficiency, this instrumental filtering is necessary (see Section \ref{sec:align_eff}).

%we present the results from the sets III, V, and VI of radiative transfer runs, \ie the resulting polarized intensity, irradiation field, and temperature maps, before filtering. The results sets I, II, and IV, are presented

\section{Analysis of the synthetic polarized dust emission maps}
\label{sec:RT_results}

\subsection{Dust polarization maps}
\label{sec:dust_pol_maps}

\begin{figure*}[!th]
\centering
% \vspace{-0.8cm}
\includegraphics[scale=0.59,clip,trim= 3cm 3.3cm 2.2cm 0.7cm]{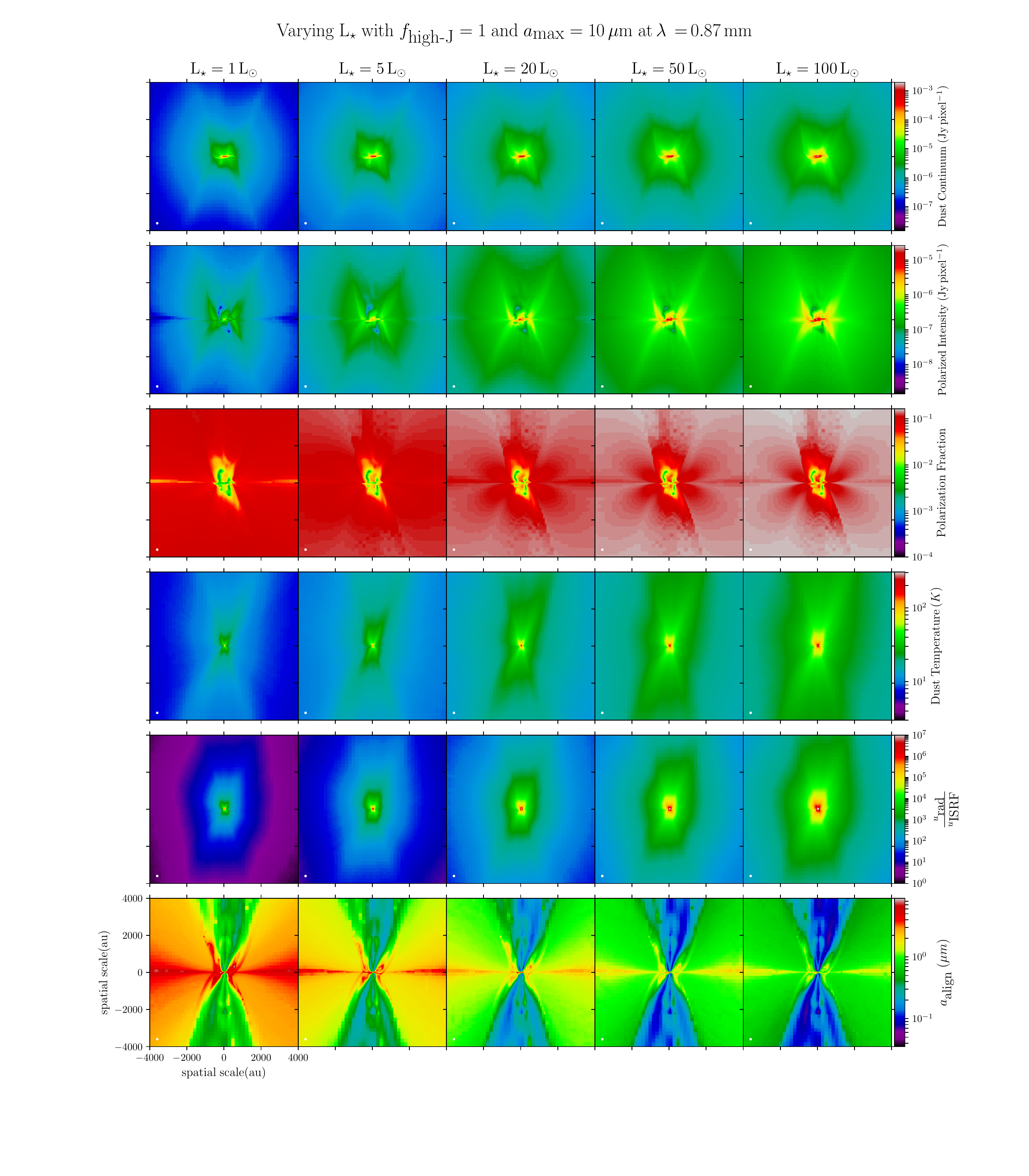}
\vspace{-0.3cm}
\caption{\small Effects of the central luminosity $L_\star$ on the radiative transfer results at 0.87 mm, with the fixed parameters of $a_{\rm{max}}\,=\,10\,\mu$m, and $f_{\rm{high}-J}\,=\,1$. This corresponds to the set III of Table \ref{t.RT_details}. Each column is one radiative transfer run of POLARIS, with $L_\star$\,=\, 1, 5, 20, 50, and 100\,L$_\odot$. Each row is a quantity provided the radiative transfer, from the first to the sixth row: total intensity Stokes $I$, polarized intensity $P$, polarization fraction $\Pf$, 2D temperature slice obtained at the center, 2D radiation field slice obtained at the center u$_{\rm{rad}}$/u$_{\rm{ISRF}}$, and the 2D slice of the $a_{\rm{align}}$ parameter obtained at the center.
}
\vspace{-0.2cm}
\label{fig:RT_maps_vrad}
\end{figure*}

We present in Fig. \ref{fig:RT_maps_vrad} the set III of radiative transfer runs performed at 0.87 mm (while we vary $L_\star$, $f_{\rm{high}-J}$ = 1 and $a_{\rm{max}}$ = 10 $\mu$m; see Table \ref{t.RT_details}). For each value of $L_\star$ (\ie 1.0, 5.0, 20, 50 , and 100 L$_{\odot}$), we show the total intensity, polarized intensity, polarization fraction, dust temperature, radiation field, and minimum size of aligned dust grains ($a_{\rm{align}}$). For the radiation field, we plot $u_{\rm{rad}}$/$u_{\rm{ISRF}}$, where $u_{\rm{rad}}$=\,$\int_{0}^{\infty} u_\lambda\,d\lambda$, with $u_\lambda$ is the spectral energy density at the wavelength $\lambda$, and $u_{\rm{ISRF}}\,=\,8.64\times10^{-13}$\,erg\,cm$^{-3}$ is the interstellar radiation field from \citet{Mathis1983}.

In the temperature and radiation fields maps, Figure \ref{fig:RT_maps_vrad} shows the progressive heating of the inner envelope as $L_\star$ increases. In the most luminous case we consider ($L_{\star}\,=\,100\,$L$\odot$), the temperature reaches $\sim$\,70 K and the radiation field reaches $\sim\,10^4\,u_{\rm{ISRF}}$, at 1000\,au in the direction of the outflow cavities. In the lowest luminous case ($L_{\star}\,=\,1\,$L$\odot$), the polarized intensity does not reach observable values, while in the higher luminous cases ($L_{\star}\,\ge\,20\,$L$\odot$), the polarization fraction in the outflow cavity walls can reach values of $\sim\,15\,\%$, with substantial amount of polarized flux, allowing detection after filtering the emission from large scales, as expected when observing the model with an interferometer such as ALMA (see Section \ref{sec:align_eff}). As the luminosity $L_\star$ increases, the efficiency of the radiative torques increases as well, and more and more grains get aligned, \ie $a_{\rm{align}}$ decreases with increasing irradiation field strength. The equatorial mid-planes remain poorly irradiated compared to the outflow cavities and cavity walls, and exhibit the highest values of $a_{\rm{align}}$ in the core. Indeed, given the high dust column density in the mid-planes, the radiation field is reprocessed very quickly toward lower energy photons. This is not totally hindering the production of polarized flux in the mid-planes: while the signal is significantly lower than in the cavity walls, the magnetic field lines appear very organized in the mid-planes, which allow the emission to remain significantly linearly polarized.

\subsection{Dust grain alignment efficiency}
\label{sec:align_eff}

\begin{figure*}[!tbh]
\centering
% \vspace{-0.6cm}
\subfigure{\includegraphics[scale=\scaleGE,clip,trim= 1.2cm 0.5cm 2cm 1.5cm]{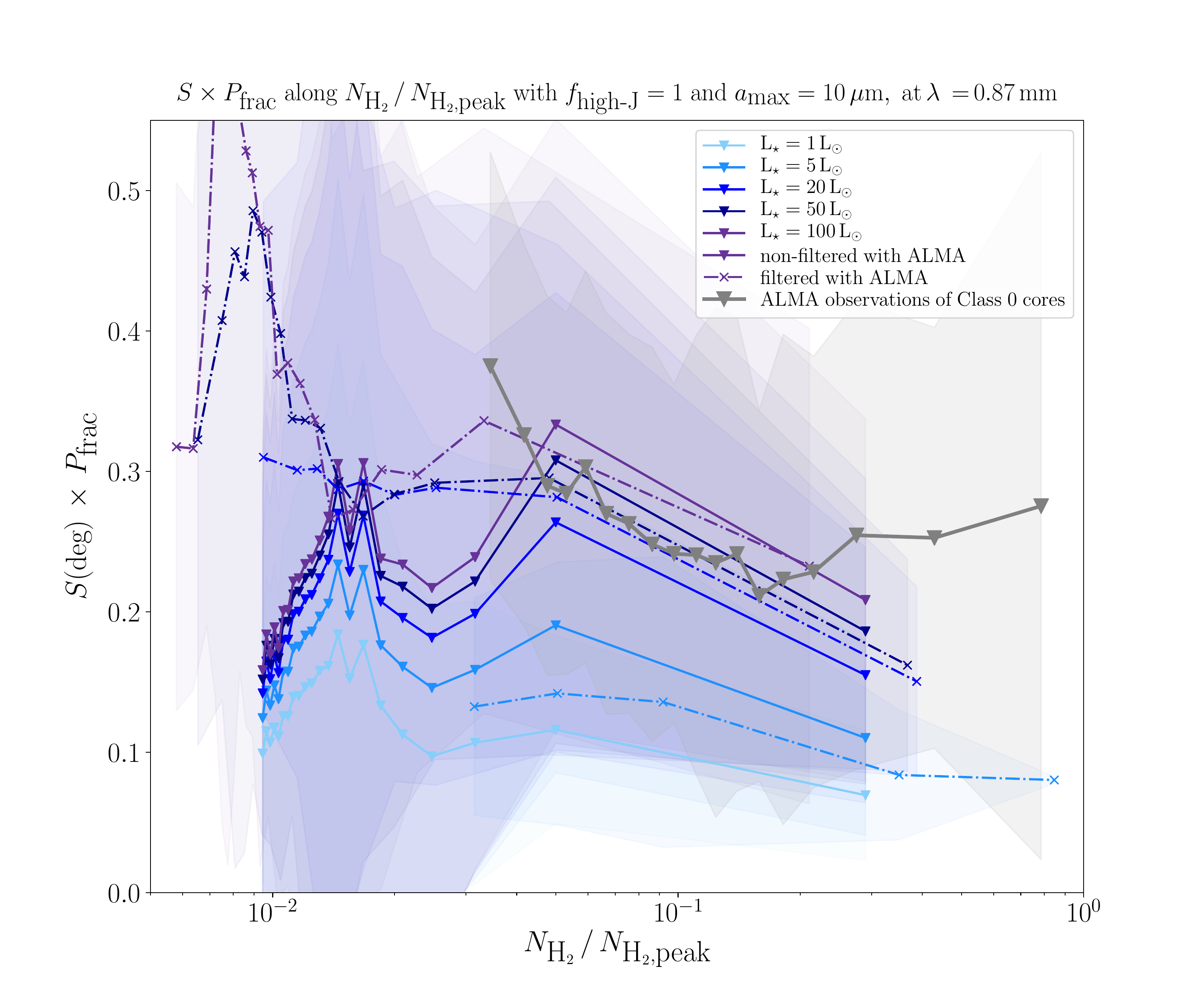}}
\subfigure{\includegraphics[scale=\scaleGE,clip,trim= 3.58cm 0.5cm 2cm 1.5cm]{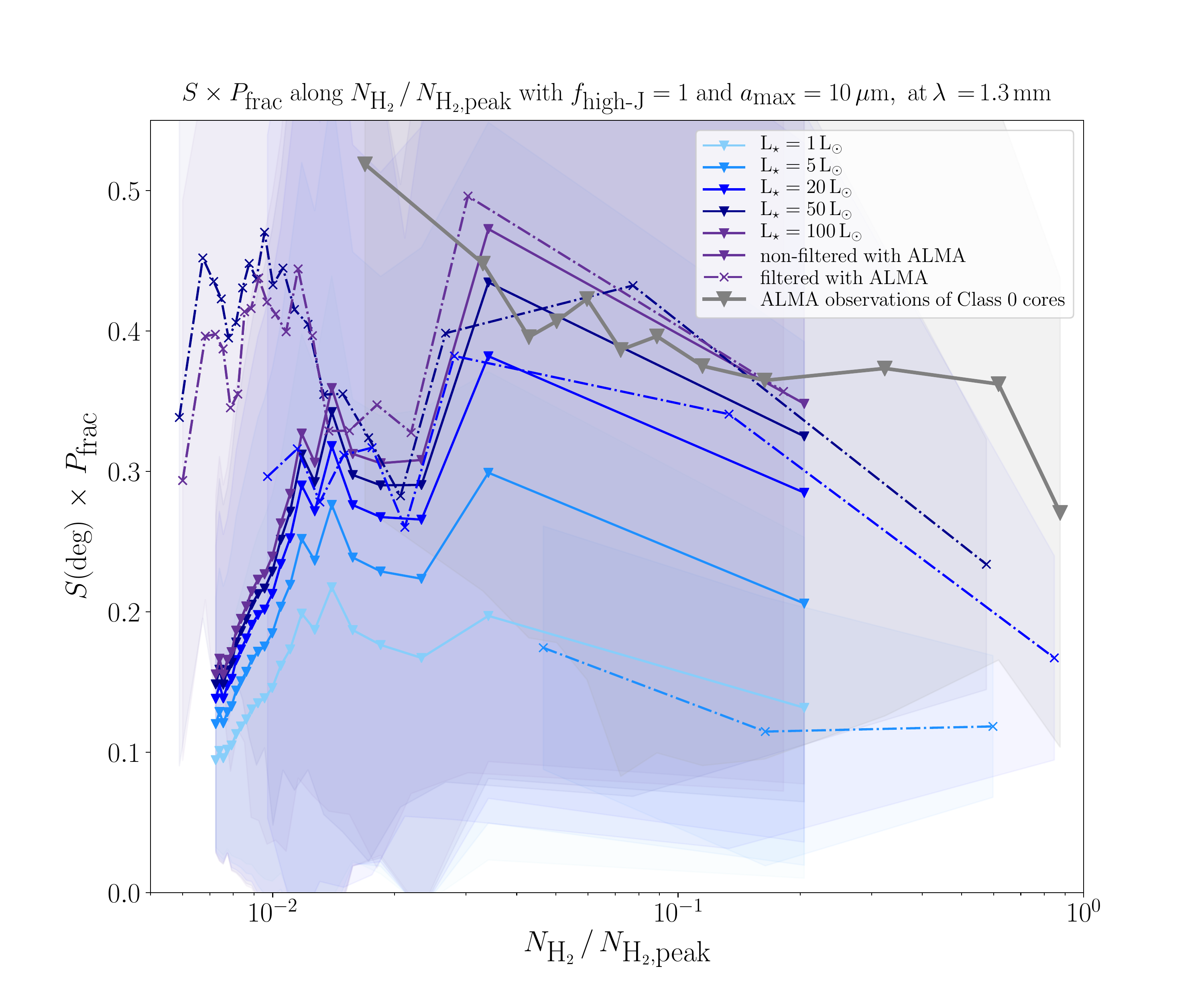}}
\caption{\footnotesize Comparison of the evolution of $\StimesP$ as a function of normalized column density $N_{{\textrm{H}_2}}/N_{{\textrm{H}_2,\textrm{peak}}}$, between ALMA observations, and our models, before and after spatial filtering. The left (right) panel corresponds to the results of the radiative transfer performed at 0.87 mm (1.3mm). The distributions of $\StimesP$ values from our models are in different tints of blue, while the distributions of $\StimesP$ values from the ALMA observations presented in \citet{LeGouellec2020} are in grey (where we split the 0.87 and 1.3 mm observations). The solid lines are the mean of the $\StimesP$ values in a given bin of normalized column density, and the corresponding shaded area represents the standard deviation. These two plots correspond to the set III of models (see Table \ref{t.RT_details}), \ie we fix $f_{\rm{high}-J}$ = 1 and $a_{\rm{max}}$ = 10 $\mu$m, and we vary $L_\star$ in the range 1.0, 5.0, 20, 50, 100 L $_\odot$. The dashed colored lines correspond to the filtered models, the solid lines are from the non-filtered models.
}
\label{fig:RT_maps_GE1}
\end{figure*}

\begin{figure*}[!tbh]
\centering
% \vspace{-0.4cm}
\subfigure{\includegraphics[scale=\scaleGE,clip,trim= 1.2cm 0.5cm 2cm 1.5cm]{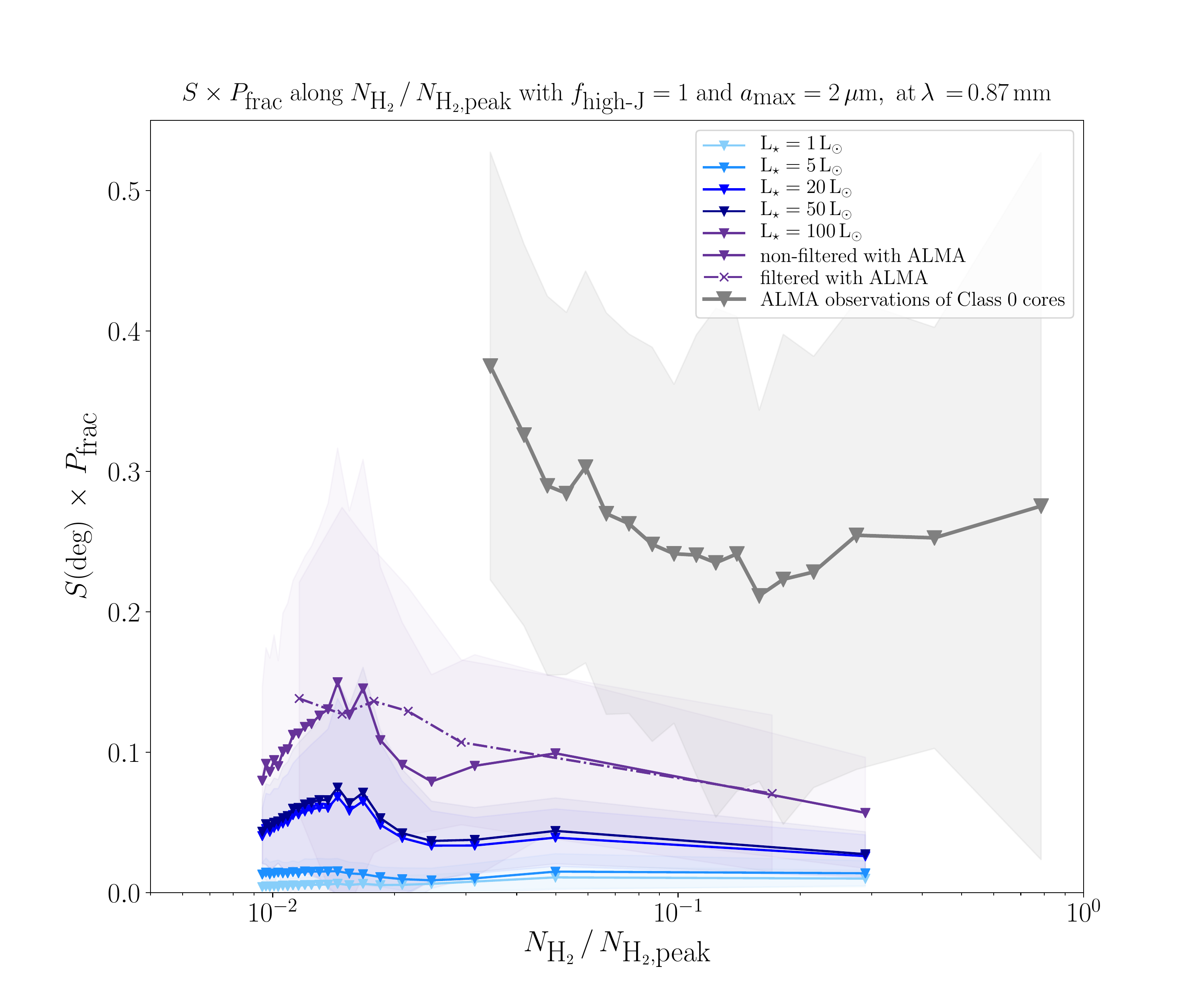}}
\subfigure{\includegraphics[scale=\scaleGE,clip,trim= 3.58cm 0.5cm 2cm 1.5cm]{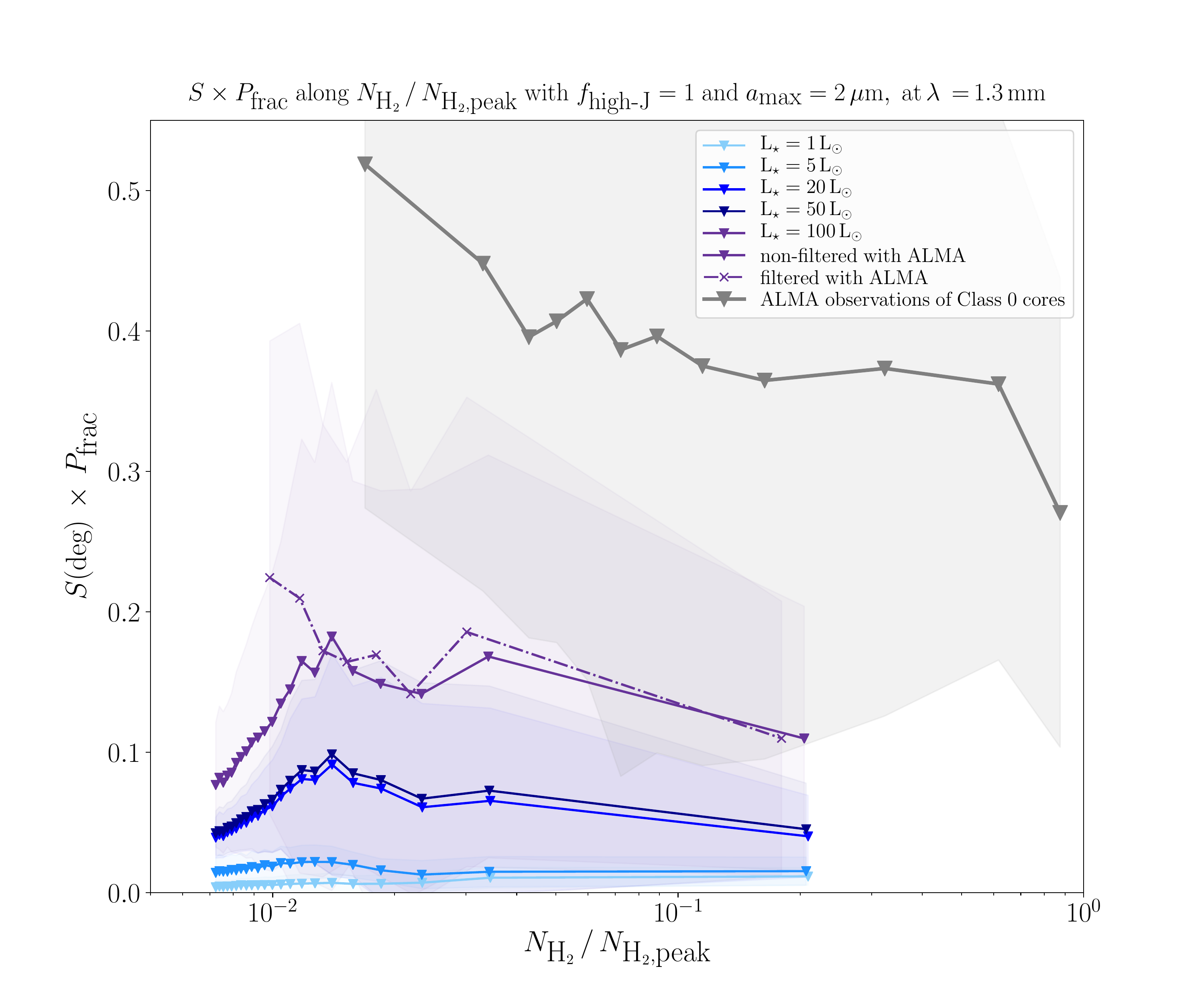}}
\subfigure{\includegraphics[scale=\scaleGE,clip,trim= 1.2cm 0.5cm 2cm 1.5cm]{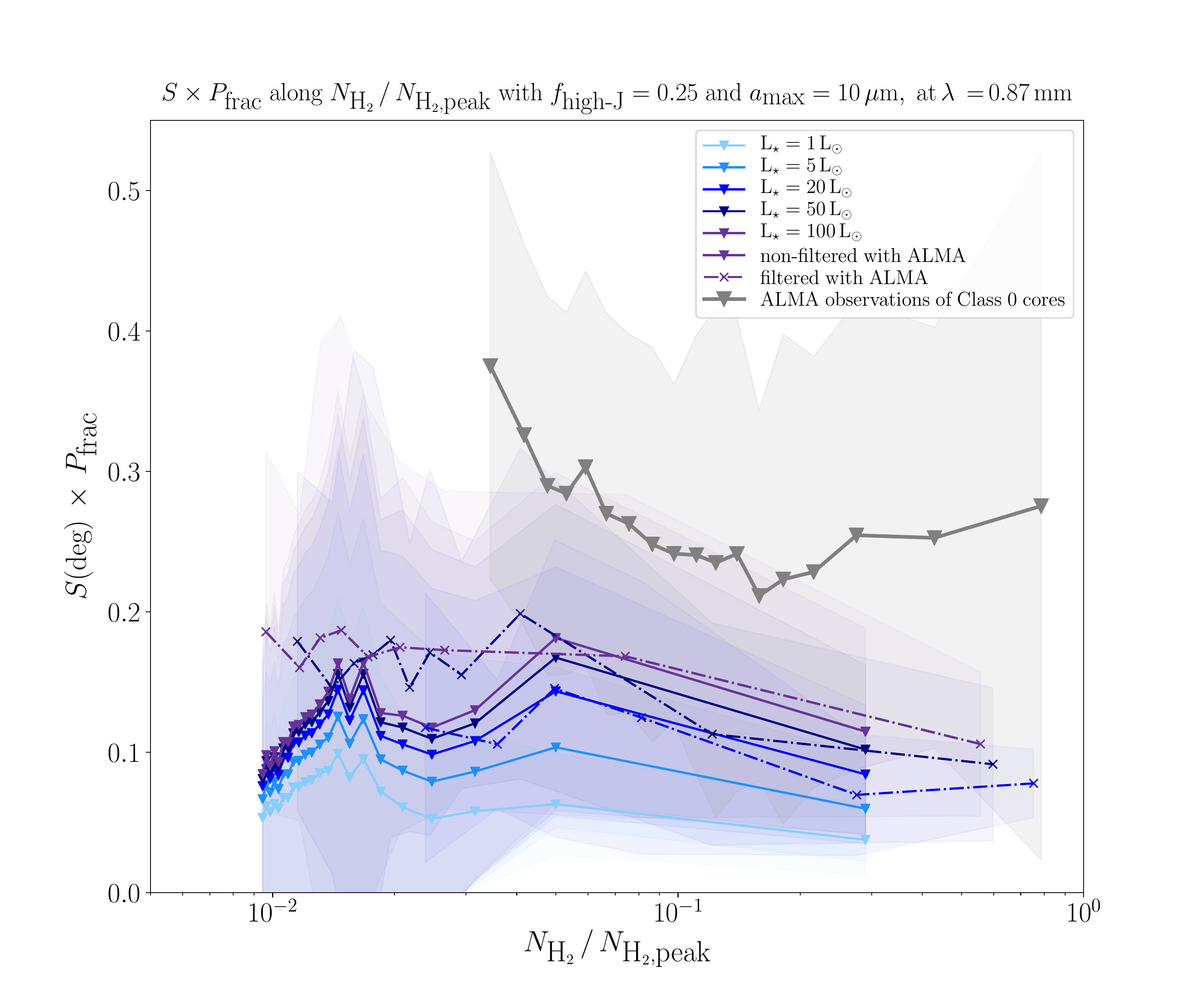}}
\subfigure{\includegraphics[scale=\scaleGE,clip,trim= 3.58cm 0.5cm 2cm 1.5cm]{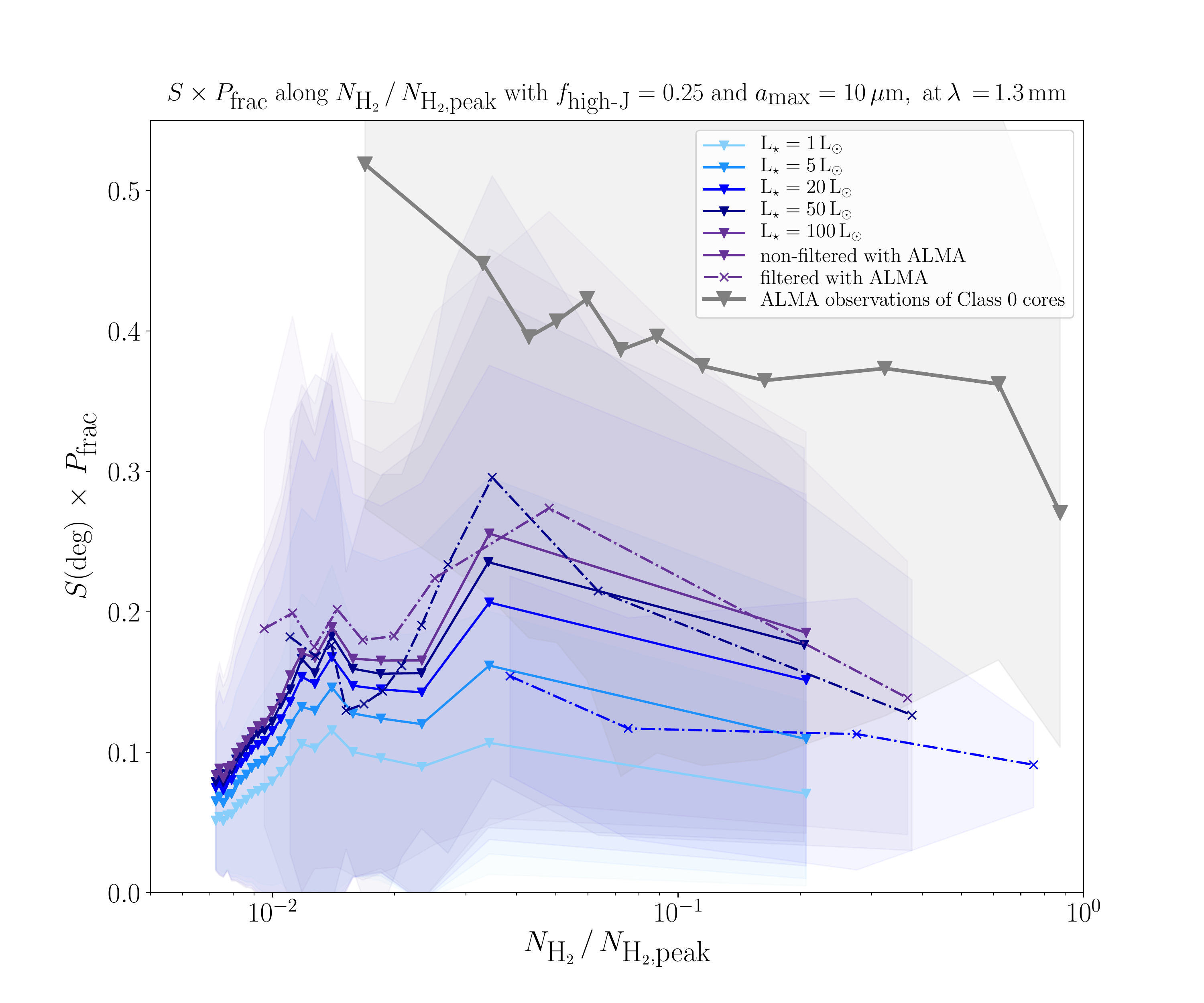}}
\caption{\footnotesize Comparison of the evolution of $\StimesP$ as a function of normalized column density $N_{{\textrm{H}_2}}/N_{{\textrm{H}_2,\textrm{peak}}}$, between ALMA observations, and our models, before and after spatial filtering. Same as Figure \ref{fig:RT_maps_GE1} for the sets IV, V (see Table \ref{t.RT_details}), where we use smaller values for $a_\textrm{max}$ (2 $\mu$m, \textit{top row}) and for $f_{\rm high-J}$ (0.25, \textit{bottom row}).
}
\label{fig:RT_maps_GE2}
\end{figure*}

To derive the average grain alignment efficiency in our models, we use the method presented in \citet{LeGouellec2020}, which consists in using the product $\StimesP$ as an estimation of the dust grain alignment efficiency, where $\S$ is the dispersion of polarization angles in the plane of the sky (computed over the 8 neighboring pixels at each location, with maps that have been regridded to 4 pixels per beam area) and $\Pf$ is the polarization fraction (sensitive to the grain alignment efficiency, and disorganization of the magnetic field in the line of sight).
To do so, we derive the distributions of $\StimesP$ values as a function of column density in our models, before and after spatial filtering with the CASA simulator, and compare them with the $\StimesP$ distributions obtained in \citet{LeGouellec2020}. 

There are many different cell sizes in the RAMSES simulations due to the use of AMR, which degrades the spatial resolution in some regions of our radiative transfer maps. We make sure that the resolution achieved by the artificial ALMA observations on the models, \ie 100\,au, when observed at 0.87 mm and 140\,au at 1.3mm, is larger than the size of the AMR cells in the 2000 au central region of the synthetic observations. Beyond this central region, the AMR cell sizes are larger, and the statistics cannot be computed. However, in those regions the emission is anyway not recovered after spatial filtering, as it is shown in Fig. \ref{fig:polaris_canonical_filtered}. In the filtered maps, we apply the same criteria pixel selection as in \citet{LeGouellec2020}, which is based on a cutoff in Stokes $I$, whose value is obtained when the signal to noise ratio in polarized intensity is, on average, higher than 5. 

We present the evolution of $\StimesP$ values as a function of the normalized column density in our set of models III in Fig. \ref{fig:RT_maps_GE1}, sets IV and V in Fig. \ref{fig:RT_maps_GE2}, before and after filtering, for the two wavelengths of observation, \ie 0.87 and 1.3mm. Alongside those results, we plot for reference, in grey, the distributions obtained in \citet{LeGouellec2020} from several ALMA observations, where we split the 0.87 and 1.3mm observations. The initial distribution of \citet{LeGouellec2020} merged 0.87, 1.3, and 3 mm observations. However, given the results obtained in Fig. \ref{fig:RT_maps_vamax_diffB} in Appendix \ref{sec:app_RT_amax}, splitting the $\StimesP$ distribution by wavelength is necessary. Despite the fact that most of the ALMA observations were realized at 1.3mm, four cores were observed at 0.87\,mm (Serpens Emb 6, Emb 8, Emb 8(N), and NGC1333 IRAS4A). We note that, on average, the dust grain alignment efficiency, as traced by $\StimesP$, is higher at 1.3\,mm than at 0.87\,mm in our radiative transfer modeling (comparing the models implementing the same parameters) and in the ALMA observations.
This is most likely due to higher polarization fraction values at 1.3\,mm (this increase in polarization fraction with wavelengths for our different models can be seen in Fig. \ref{fig:RT_maps_vamax_diffB}). This evolution of polarization fraction with wavelength can be caused by several effects: temperature distribution in the line of sights, opacity effects, and largest aligned grains contributing more to the polarized dust emission at longer wavelengths.
In addition, the plots from Fig. \ref{fig:RT_maps_GE1} and \ref{fig:RT_maps_GE2} exhibit quite a large scatter in the distribution, and the column density maps in our models tend to be more peaked compared to ALMA observations\footnote{We present in Appendix \ref{sec:app_RT_alpha} a method attempting to take into account this later point.}. Therefore, we will tend to discuss the global evolution of a given $\StimesP$ distribution, as the radiative transfer parameters and the wavelength of observations vary. 

We find that increasing the ratio $f_{\rm{high}-J}$, the maximum grain size $a_{\rm{max}}$, and the central luminosity $L_\star$ implemented in POLARIS strongly influences the grain alignment efficiency in the envelope, typically causing the efficiency to increase (see also Appendices \ref{sec:app_RT_amax} and \ref{sec:app_RT_highJ}). Nevertheless, the grain-alignment efficiency values calculated from the simulations are on average lower than the values from ALMA observations of Class 0 cores, if luminosities $\leq$\,20 L$_\odot$ are considered. We note that our MHD model may exhibit higher inner volume gas density values (given the initial mass of the model of 30 $M_\odot$) compared with what is expected in typical low-mass Class 0 protostars, thus justifying the need of high luminosity to match the observed grain-alignment efficiency.

Thanks to the multiple sets of parameters we adopt to run our radiative transfer calculations (from I to VI), one can infer the optimal parameters that yield the values of grain alignment efficiency most similar to those observed in the ALMA observations. Assuming $f_{\rm{high}-J}\,=\,1$ and $a_{\rm{max}}\,=\,10\,\mu$m, Figure \ref{fig:RT_maps_GE1} shows that only the implementation of large values of $L_\star$, \ie 20, 50 and 100 L$_\odot$, produces mean values of $\StimesP$ at 0.87\,mm high enough that they are potentially able to match the values from ALMA observations. At 1.3mm, $\StimesP$ values from ALMA observations are notably higher than our models. 

The results presented in Fig. \ref{fig:RT_maps_GE2} argue that low values of $f_{\rm{high}-J}$ and $a_{\rm{max}}$ (\ie 0.25 and 2\,$\mu$m, respectively) hinder efficient grain alignment in the envelope. In other words, we cannot reproduce the average values of dust grain alignment efficiency obtained with ALMA observations if we do not implement a large fraction of  perfectly aligned dust grains (\ie $f_{\rm{high}-J}$ close to 1), and if we do not include a population of large dust grains in the core ($\ge\,10\,\mu$m). Figure \ref{fig:RT_maps_GE2} thus justifies the values of the two fixed parameters of $f_{\rm{high}-J}$ and $a_{\rm{max}}$ we have chosen in Fig. \ref{fig:RT_maps_GE1}. 

The sets of radiative transfer calculations presented here provide hints that relatively high irradiation \ie with central luminosities $L_\star$ higher than 20\,L$_\odot$ (but also large grains $\ge$$\,10\,\mu$m, and $f_{\rm{high}-J}$ close to 1), needs to be implemented in our model in order to increase the estimated values of grain alignment efficiency and match ALMA observations. Overall, the ALMA values $\StimesP$ remain larger than our models, suggesting that further investigations are necessary to improve our understanding of grain alignment theories and/or our knowledge of the physical conditions in the inner region (\ie $\sim\,10-1000$ au) of Class 0 protostellar cores.

\subsection{Dust grain rotational disruption}
\label{sec:RATD}

\begin{figure*}[!tbh]
\centering
\vspace{-0.2cm}
\includegraphics[scale=0.6,clip,trim= 2.9cm 2.5cm 2.2cm 0cm]{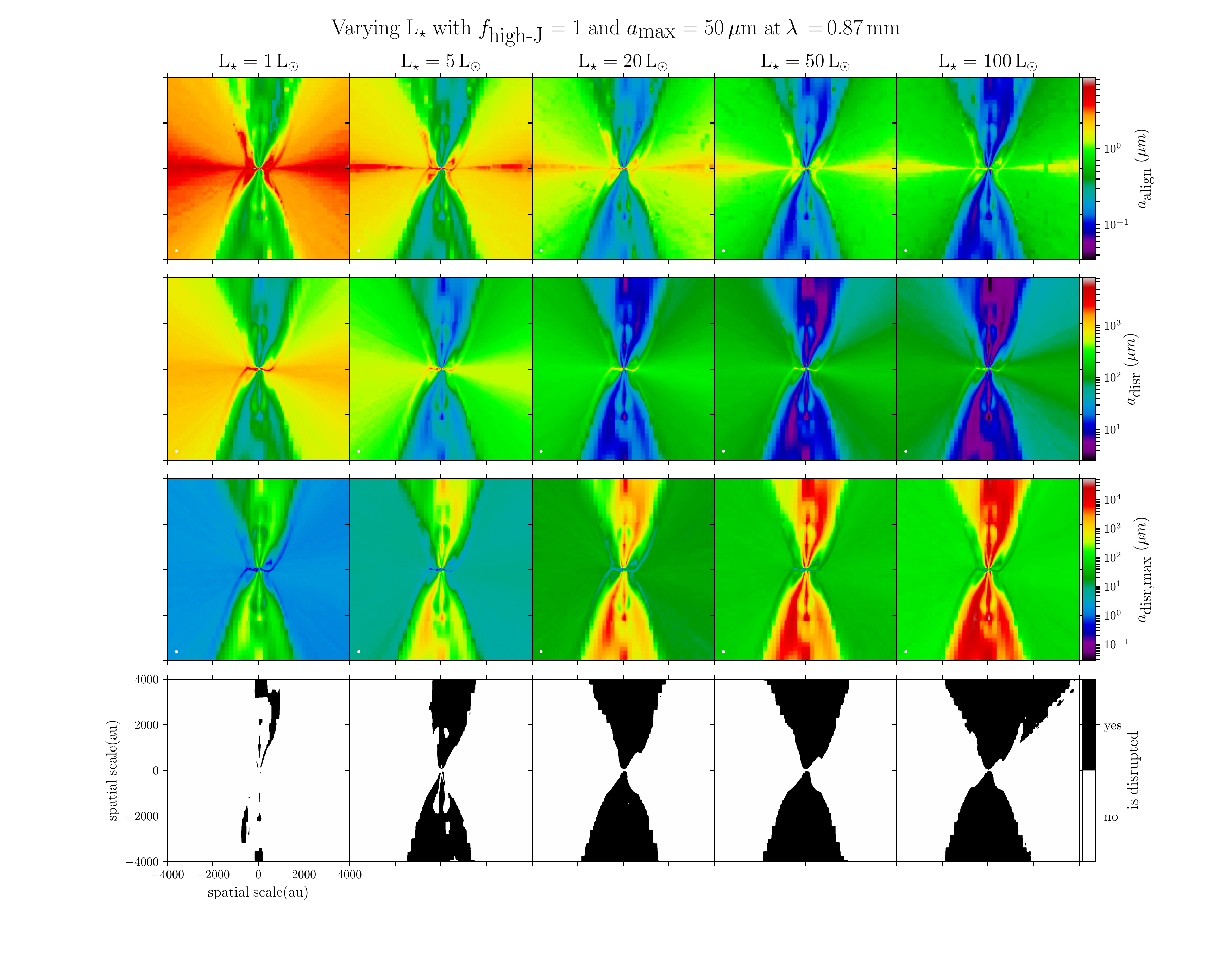}
\caption{\small Effect of the central luminosity $L_\star$ on the Radiative Torque Disruption of dust grains, at 0.87 mm, with the fixed parameters of $a_{\rm{max}}\,=\,50\,\mu$m, and $f_{\rm{high}-J}\,=\,1$. Each plot is a 2D-slice taken at the center of the core, in our radiative transfer results. Each column is one radiative transfer run of POLARIS, with $L_\star$\,=\, 1, 5, 20, 50, and 100\,L$_\odot$. The first row is the $a_{\rm{align}}$ parameter, \ie the dust grain size above which dust grains are considered aligned in our radiative transfer calculations. The second and third rows are the $a_{\rm{disr}}$ and $a_{\rm{disr,max}}$ parameters, respectively. In the theory of RATD, these two values correspond to the window in grain size, inside which dust grains are rotationally disrupted. The last row indicates whether dust grains are rotationally disrupted in our models for $S_{\rm{max}}\,=\,10^{5}\,\rm{erg}\,\rm{cm}^{-3}$, \ie if the intervals $[a_{\rm{align}}\,;\,a_{\rm{max}}]$ and $[a_{\rm{disr}}\,;\,a_{\rm{disr,max}}]$, overlap. For a given pixel, if there is an overlap, the pixel is black (\ie grains are disrupted). If it is not the case it is white (\ie grains are not disrupted).
}
\label{fig:RT_maps_RATD}
\end{figure*}

Figures \ref{fig:RT_maps_GE1} and \ref{fig:RT_maps_GE2} encourage us to consider relative high protostellar luminosity (with central luminosities $L_\star$ higher than 20\,L$_\odot$) so that the radiation field is sufficiently high in the core, in order to reproduce the level of grain alignment efficiency measured with ALMA. However, increasing the efficiency of radiative torques can in turn cause the rotational energy to overcome the tensile strength of dust grains, leading to the disruption of grains into fragments. Given that the large aligned grains are the ones that are efficiently aligned, dust grain disruption could deplete the large grains necessary to produce the polarized dust emission measured inside star forming cores with ALMA. 

This mechanism is called the RAdiative Torque Disruption (RATD; \citealt{Hoang2019NatAs}). Already detected in the star forming dense cores of Orion KL and $\rho$ Ophiuchus \citep{Tram2021,Tram2021b}, it can be triggered by RATs \citep{Hoang2019NatAs,Hoang2019,Hoang2020} or mechanical torques \citep{Hoang2019b,Hoang2019d}. \citet{Hoang2020b} derived an analytical model to investigate whether RATD is at work inside star forming cores. Here, we attempt to use the theoretical considerations of the RATD mechanism and compute them using the results of our radiative transfer modeling, in order to check if our claims, \ie the presence of large grains in the inner envelope and the need of a high central luminosity, may lead to the disruption of large grains. 

\begin{figure*}[!tbh]
\centering
% \vspace{-0.7cm}
\includegraphics[scale=0.58,clip,trim= 2.2cm 2.5cm 2.1cm 0.4cm]{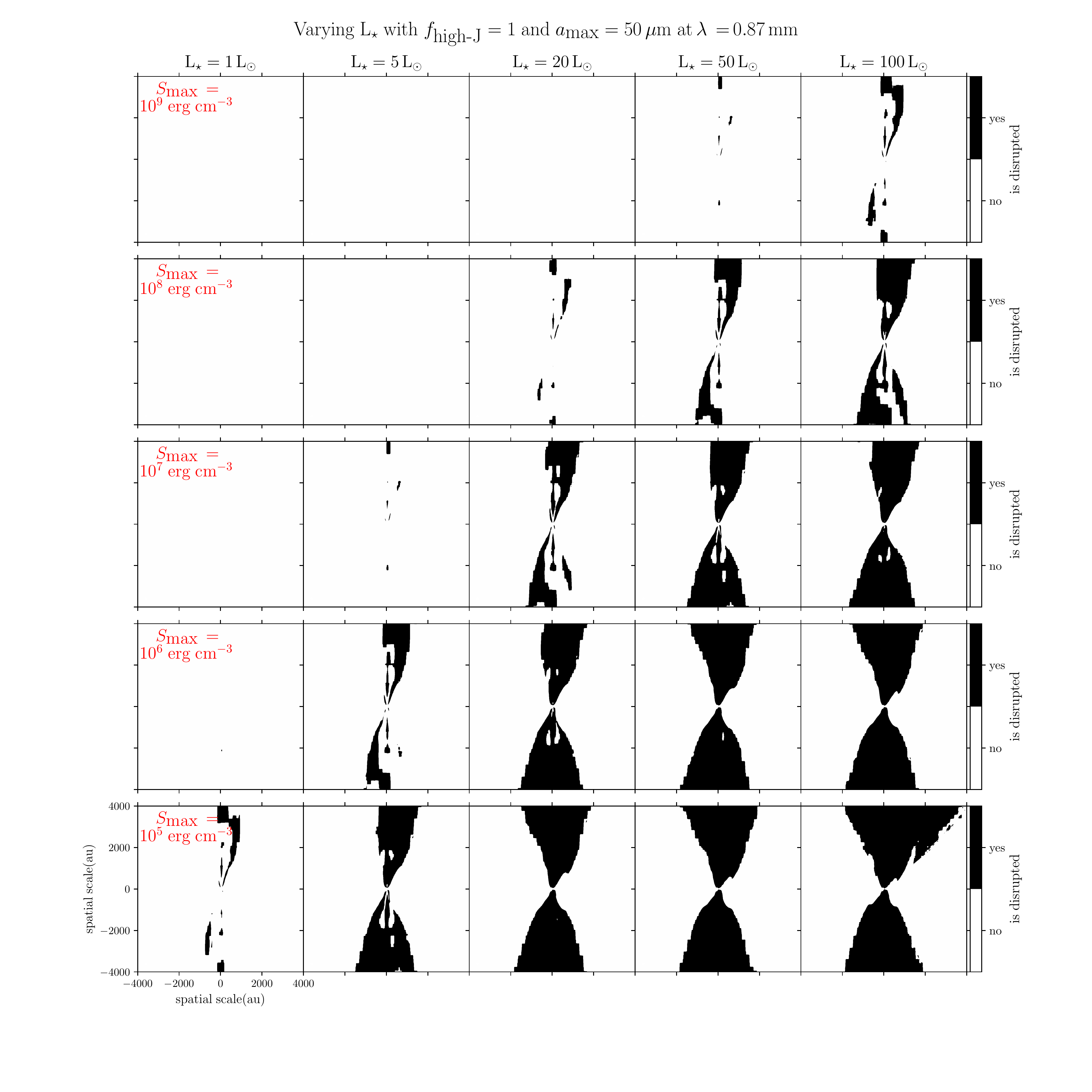}
\caption[Effect of the central luminosity $L_\star$ on the Radiative Torque Disruption of dust grains, at 0.87 mm]{\footnotesize Effects of the central luminosity $L_\star$ on the Radiative Torque Disruption of dust grains, at 0.87 mm, with the fixed parameters of $a_{\rm{max}}\,=\,50\,\mu$m, and $f_{\rm{high}-J}\,=\,1$. Each plot is a 2D-slice taken at the center of the core, in our radiative transfer results. Each column is one radiative transfer run of POLARIS, with $L_\star$\,=\, 1, 5, 20, 50, and 100\,L$_\odot$. Each row indicates whether dust grains are rotationally disrupted in our models, \ie if the intervals $[a_{\rm{align}}\,;\,a_{\rm{max}}]$ and $[a_{\rm{disr}}\,;\,a_{\rm{disr,max}}]$, overlap. For a given pixel, if there is an overlap, the pixel is black. If it is not the case it is white. The results from each row are obtained for a given value of grain tensile strength $S_{\rm{max}}$, \ie $10^{5}$, $10^{6}$, $10^{7}$ (the same plots as the bottom row of Figure \ref{fig:RT_maps_RATD}), $10^{8}$, and $10^{9}\,\rm{erg}\,\rm{cm}^{-3}$.
}
\label{fig:RT_maps_RATD2}
\end{figure*}

A dust grain of mass density $\rho$, of size $a$, rotating at the angular velocity $\omega$ experiences a tensile stress of $S\,=\,\rho a^2 \omega^2 /4$. We designate by $S_{\rm{max}}$ the tensile strength of dust grains. The critical angular velocity above which dust grains are rotationally disrupted is given by:
\begin{equation}
\omega_{\rm{disr}}\,=\,\frac{2}{a}\left(\frac{S_{\rm{max}}}{\rho}\right)^{1/2}\,\,.
\end{equation}
If the maximum rotation speed induced by RATs $\omega_{\rm{RAT}}$ exceeds $\omega_{\rm{disr}}$, dust grains are rotationally disrupted. Given the two regimes of the dust grain alignment efficiency $\bar{Q}_{\Gamma}$, defined by the relative values the grain size $a$ has with respect to the mean wavelength of the radiation field spectrum $\bar{\lambda}$ received by dust grains, one can define the grain size interval $[a_{\rm{disr}}\,;\,a_{\rm{disr,max}}]$, inside which the aligned dust grains are rotationally disrupted. The respective values of $a_{\rm{disr}}$ and $a_{\rm{disr,max}}$, are obtained developing the equation $\omega_{\rm{RAT}}\,=\,\omega_{\rm{disr}}$, in the two aforementioned regimes of the dust grain alignment efficiency. From \citet{Hoang2020b}, we have :
\begin{equation}
\label{equ:a_disr}
a_{\rm{disr}}\,=\,\left( \frac{ 0.8 n_{\rm{H}} \sqrt{2\pi m_{\rm{H}} k T_{\rm{gas}}} }{\gamma u_{\rm{rad}} \bar{\lambda}^{-2}} \right)^{1/2} \left(\frac{S_{\rm{max}}}{\rho}\right)^{1/4} (1+F_{\rm{IR}})^{1/2}\,\,,
\end{equation}
\rm{and}
\begin{equation}
\label{equ:a_disr_max}
a_{\rm{disr,max}}\,=\, \frac{ \gamma u_{\rm{rad}} \bar{\lambda} }{ 16 n_{\rm{H}} \sqrt{2\pi m_{\rm{H}} k T_{\rm{gas}}} } \left(\frac{S_{\rm{max}}}{\rho}\right)^{-1/2} (1+F_{\rm{IR}})^{-1}\,\,,
\end{equation}
where $\gamma$ is the anisotropy of the radiation field, $u_{\rm{rad}}$ is the radiation field, $\bar{\lambda}\,=\,\frac{\int^{\infty}_{0}  \lambda u_{\rm{rad}} d\lambda }{u_{\rm{rad}}}$ is the mean wavelength of the radiation field, $n_{\rm{H}}$ is the gas density, $T_{\rm{gas}}$ is the gas temperature, $m_{\rm{H}}$ is the mass of a hydrogen atom, and $F_{\rm{IR}}$ is the relative importance of rotational damping by infrared emission (due to the emission of infra-red photons emitted by the grain, which reduces the grain’s angular momentum, see \citealt{Draine1998}; we adopt the relation of \citealt{Hoang2020b}) with respect to the damping by gas collision\footnote{We note that in our calculation of the rotational disruption radii $a_{\rm{disr}}$ and $a_{\rm{disr,max}}$, we assume a maximum efficiency of the RAT mechanism (see  \citealt{Hoang2020b,Hoang2022b}).}.
% , \ie that the anisotropic component of the radiation field and the magnetic field vectors are collinear.}.
We adopt $\rho\,=\,3$\,g\,cm$^{-3}$ (this corresponds to the grain density of the dust model we adopt in POLARIS), and use the temperature, radiation field strength, and local gas density from each of our models. Both $\gamma$ and $\bar{\lambda}$ are derived in each cell of the grid during the radiative transfer calculations. However, $\bar{\lambda}$ is not present in the output files of POLARIS. Therefore, to calculate $\bar{\lambda}$, we use the relations in the Section 4 of \citet{Hoang2020b}, who derived, thanks to an analytical model, the mean wavelength as a function of the gas column density of a typical protostellar envelope. In the 2D slice of each of our models, we derive the column density ``seen'' by each cell by summing the gas density on the straight line separating the cell from the center (this apparent column density is dominated by the central $\sim$500 au region). We note that this is an approximation, as the photons received by a given cell have been reprocessed, and thus do not come ``straight'' from the protostar. Our values of $\bar{\lambda}$ are thus lower-limits. As both $a_{\rm{disr}}$ and $a_{\rm{disr,max}}$ are proportional to $\bar{\lambda}$, more realistic $\bar{\lambda}$ could thus shift the size range of rotationally disrupted grains toward larger values. However, the power-law dependence of $\bar{\lambda}$ with the apparent column density is weak, \ie $\sim$ 0.65.

We calculate the values of $a_{\rm{disr}}$ and $a_{\rm{disr,max}}$ for a given value of $S_{\rm{max}}$. The grains whose size falls within the size interval of aligned grains $[a_{\rm{align}}\,;\,a_{\rm{max}}]$ as well as the interval of rotationally disrupted grains $[a_{\rm{disr}}\,;\,a_{\rm{disr,max}}]$ can be considered as rotationally disrupted. Now only one parameter remains to be set for us to evaluate the equations \ref{equ:a_disr} and \ref{equ:a_disr_max}: the grains tensile strength $S_{\rm{max}}$. The tensile strength of interstellar dust is uncertain, because it depends on both the grain structure, \ie compact versus composite, and the grain composition, both of which are not strongly constrained quantities in Class 0 envelopes. We explore different grain structures, considering range $10^{5} - 10^{9}\,\rm{erg}\,\rm{cm}^{-3}$ for $S_{\rm{max}}$. However, given that the largest grains in protostellar envelopes that undergo collisions and coagulation during the collapse are expected to be aggregates \citep{Jones2016c}, our results will focus on the $S_{\rm{max}} \;=\;10^{5}\,\rm{erg}\,\rm{cm}^{-3}$ (see \cite{Hoang2019c,Hoang2020b}). The largest value for $a_{\textrm{max}}$ that we considering in our work is 50 $\mu$m, which implies that significant grain growth have occurred from the ISM-typical $a_{\textrm{max}}$ value of 0.5 $\mu$m. We note that the presence in protostellar envelopes of such large aggregates and underlying efficient grain growth, on which the resulting sub-millimeter polarization is highly dependent \citealt{Valdivia2019}, is still debated (see the recent works by \citealt{Guillet2020b,Bate2022,Silsbee2022} that put constraints on early dust grain growth given the initial conditions such as turbulent velocities, timescales, and gas density).

From our calculations, the intervals $[a_{\rm{align}}\,;\,a_{\rm{max}}]$ and $[a_{\rm{disr}}\,;\,a_{\rm{disr,max}}]$ can overlap for the largest grains, if the irradiation and tensile strength are sufficiently high and low, respectively. Figure \ref{fig:RT_maps_RATD} presents the results of the computation of $a_{\rm{disr}}$ and $a_{\rm{disr,max}}$, where we vary the protostar luminosity $L_\star$, for a tensile strength value of $10^{7}\,\rm{erg}\,\rm{cm}^{-3}$. The last row indicates whether both intervals overlap. As the radiation field strength increases, we notice that more grains get aligned, \ie $a_{\rm{align}}$ decreases, but at the same time $a_{\rm{disr}}$ decreases, and the region where $a_{\rm{disr}}$ is lower than $a_{\rm{max}}$ becomes larger. In other words, increasing the luminosity at the center causes larger and larger depletion of large grains. This depletion occurs mainly toward outflow cavities. However, as we saw earlier when implementing maximum grain size of 2$\mu$m for example, depletion of large grains considerably reduces the amount of polarized intensity in the core. Given that only grains larger than $\sim\,10\,\mu$m can produce a sufficient sub-millimeter polarized emission, our models require the rotational disruption of large grains to be limited in order to reproduce the high grain alignment efficiency revealed by ALMA observations. We also note that cells with high-density material along their lines of sight to the protostar, i.e., equatorial mid-planes (see Fig. \ref{fig:simu}), seem not to be affected by RATD. Therefore, they will be the last regions of the envelope to suffer depletion of large dust grains when the irradiation increases. 

In addition, we notice that there is a slight asymmetry, from the east to the west part of the core, in the plot displaying the region affected by RATD in Fig. \ref{fig:RT_maps_RATD}. This is due to the structure in density of the very inner part of the circumstellar disk, as can be seen in Fig. \ref{fig:simu}. When deriving the column density ``seen'' by a given cell, the densest cells located at the center govern the resulting column density. In the slice density map, small asymmetries toward the circumstellar disk result in differences in $\bar{\lambda}$, from the east to the west part of the map. At a given radius from the center, this, in turn, can cause the $a_{\rm{disr}}$ parameter to change from one side of the core to another. Therefore, inhomogeneities in density among the densest regions of the disk can be responsible for a region to be differently affected by RATD compared to one another. This phenomenon can be a potential explanation for asymmetric polarized intensity, such as that seen in B335, where strong polarization is only seen in the northern equatorial plane. One side might be more affected by RATD, and thus depleted in large grains that are needed to produce detectable polarized dust emission. This remains, however, an hypothesis, as different dust grains characteristics between the northern and southern portion of the equatorial plane have not been probed.

Finally, we investigate the locations affected by RATD in our models for several values of tensile strength $S_{\rm{max}}$, $10^{5}$, $10^{6}$, $10^{7}$, $10^{8}$, and $10^{9}\,\rm{erg}\,\rm{cm}^{-3}$, in Fig. \ref{fig:RT_maps_RATD2}. The smaller the implemented value of $S_{\rm{max}}$, the lower the luminosity required to disrupt large grains in a given region. 
The RATD mechanism clearly affects outflow cavities for $L_{\star}\,\ge\,$20L$_\odot$ if $S_{\rm{max}}\,\leq\,10^{7}\,\rm{erg}\,\rm{cm}^{-3}$, \ie if grains are composite, with grains $\geq\,5\,\mu$m being depleted. 
However, for $L_{\star}\,\ge\,$20L$_\odot$, if $S_{\rm{max}}\,\geq\,10^{8}\,\rm{erg}\,\rm{cm}^{-3}$, \ie if grains are extremely compact, $10-50\,\mu$m grains survive throughout all the envelope.

The RATD mechanism seems to affect one specific part of the envelope.
This mechanism, if occurring, could cause the depletion of large, moderately compact grains, \ie $a\,\geq\,5\,\mu$m and $S_{\rm{max}}\,\sim\,10^{5}\,\rm{erg}\,\rm{cm}^{-3}$, in outflow cavities that are directly exposed to the radiation field, if this latter is high enough ($L_{\star}\,\ge\,$20L$_\odot$). As we showed earlier, this may be problematic because both large grains and significant irradiation need to be implemented in order to reproduce the dust grain alignment efficiency values observed by ALMA. Assuming that RATs is the mechanism at play in the inner regions of Class 0 protostellar cores and that our implementation of the RAT theory is accurate enough, the combination of synthetically observed MHD models (with varying grain sizes and luminosities) alongside ALMA observations could constrain the tensile strength of the dust grains responsible for the polarized dust emission we observe.

\section{Discussion}
\label{sec:p3_disc}

Our radiative transfer analysis highlights that in the environment of Class 0 protostellar envelopes, the radiative feedback from the accretion/ejection processes has an important role in the efficiency of grain alignment, and potentially on dust grain evolution. One needs to reconcile the constraints we obtained from these results in terms of irradiation (and grain size, paramagneticity), with the phenomenon of grain rotational disruption, and any time variation of the accretion luminosity.

\subsection{The radiation field in a Class 0 core: the photons' sources and propagation}

The radiation field strength and spectrum is at the center of our study aiming at understanding the mechanism responsible for the dust polarization we observe. In the radiative transfer calculations we presented, the central protostellar embryo represented by a sink particle is the only source of high energy photons. We describe here below how this can represent a caveat.

\subsubsection{Accretion luminosity}

In the inner regions of a Class 0 protostellar core, the radiative energy come mainly from the accretion energy radiated away that is constantly reprocessed along its pathway. In reality, the densest regions of highest opacities in the core, whose typical size is smaller than the spatial resolution of our gridding pattern, are responsible for most of the reprocessing of the high energetic photons from the accretion shock. We note that, UV photons would be reprocessed faster than X-ray photons throughout the envelope \citep{Stauber2004,Stauber2005}. In our models, our implementation of the spectrum of the energy radiated away from the sink due to the mass accretion onto the protostellar embryo is incomplete. However, we notice that the high-energy photons are quickly reprocessed toward the central high density cells surrounding the sink particle, in such a way that changing the blackbody spectrum parameters (fixing the luminosity) does not affect the radiation field spectrum throughout the core.

\citet{Baraffe2009} developed an analytical model of the accretion thermal efficiency that accounts for the fraction of accretion luminosity that is radiated away, with respect to the energy absorbed by the protostar. This model also considers details of the accretion process, such as how much energy can be stored in viscous heating or rotational energy when accretion is happening via an accretion disk (\citealt{Hartmann1997}, see also \citealt{Jensen2018,Kuffmeier2018}). We attempted to use the smooth step function \citet{Baraffe2009} proposed to model the thermal efficiency of the accretion. In our case, using the method of \citet{Baraffe2009}, the accretion rate on the sink provides too high values of luminosity compared to what is tentatively measured in intermediate-mass protostars. This can be explained by the spatial resolution of our MHD simulation ($\sim\,$5au) being too low to properly take into account the storage of material in a disk, which can lower the mass accretion rate onto the protostellar embryo.

In order to reproduce the averaged grain alignment efficiency measured in ALMA observations with our model, our radiative transfer results show that a black body spectrum of luminosity $\geq$ 20\,L$_\odot$ is necessary. Indeed, the protostellar photospheric luminosity ($\leq$ 1\,L$_\odot$) alone cannot result in a significant polarized dust emission. On average, the luminosity of protostellar cores is, however, significantly below the expected values. This is the so-called luminosity problem \citep{Kenyon1990b,Kenyon1995,Young2005,Evans2009}. Therefore, requiring high radiative energy in all the cores that exhibit the high grain alignment efficiency quantified in \citet{LeGouellec2020} can be problematic because these values may be above the values observed in protostars. Nevertheless, \citet{Krumholz2012} proposed that a significant fraction of the accretion energy can escape through outflow cavities. This is also what we found when plotting the 2D slice of the radiation field in our models. The low values of measured bolometric luminosity, could thus be reconciled with the expected mass accretion rate onto Class 0 protostars, if the luminosity from the accretion is not isotropically distributed throughout the envelope.

\subsubsection{Shocks along outflow cavities}
\label{sec:shocks}

The only source of photons in our radiative transfer models is from the center, where the sink particle is located. However, from the temperature maps of the RAMSES simulations, we can see that the mechanical energy deposited by the jet is also responsible for heating the gas. 
This potential source of UV photons can directly heat the dense walls of the cavity, in contrast to the photons from the accretion that have already been reprocessed well before reaching the cavity walls. Therefore, the presence of shocks can modify the mean wavelength received by dust grains responsible for producing the polarized dust emission. In particular, if the radiation field spectrum contains more high energy photons thanks to these additional sources, smaller grains could get aligned; these grains are less likely to be affected by RATD. These suggestions are, however, not included in our models, as the gas dynamics (and associated shocks and radiative energy) are not taken into account in POLARIS (see the theoretical work of \citealt{Hoang2019d}). The required subsequent investigations would require the use of the modeling of shocks in cavities in order to identify the source location and spectrum of the photons produced in shocks. Estimating their energetics and penetration path-length could then inform us how they can contribute to the radiative torques applied on dust grains. One could also implement self-consistently in the radiative transfers additional sources of photons along the interaction zones between the jet and the envelope, based on 1D modeling of self-irradiated shocks calculated with the appropriate initial conditions \citep{Lehmann2020}. We also note that such shocks in outflow cavities may affect the size distribution of dust grains via grain shattering \citep{Jones1996,Guillet2009}.

\subsection{Will the dependence with radiation be canceled by RATD ?}

To reproduce the dust grain alignment efficiency inferred from ALMA observations of Class 0 cores, large grains ($\geq\,10\,\mu$m) need to be present in the irradiated inner envelope. However, while this irradiation increases the dust grain alignment efficiency, it can also trigger grain rotational disruption, and deplete large grains in the affected regions. 
This phenomenon is dependent on the density along the photons' pathway, the local gas density, and the radiation field. 
% This phenomenon is, however, dependent on the density of the medium through which photons need to pass, the denser regions of the inner core being the outflow cavity walls and equatorial planes. 
Therefore, the regions that are detected in dust polarization are also the regions ``resisting'' RATD (see Fig. \ref{fig:RT_maps_RATD}), that are irradiated and dense enough, such as cavity walls. Now, these considerations are made assuming a given tensile strength $S_{\rm{max}}$ of dust grains. As presented in Section \ref{sec:RATD}, grains which are more compact (higher $S_{\rm{max}}$) are better able to resist RATD. On the contrary, grains made of aggregates with low tensile strength are more easily destroyed by RATD. As significant grain growth is necessary to justify the observed polarized dust emission in Class 0 protostellar cores, one could constrain the tensile strength of dust grains studying in what conditions and via what processes they grew. 

Constrains on the grain internal structures, such as the tensile strength, could thus be obtained via observations of dust emission (constraining $a_{\textrm{max}}$) and dust polarization (constraining $a_{\textrm{disr}}$ with dust polarization models). In order to study the regime where RATD can disrupt dust grains within typical ISM irradiation and density conditions, \citet{Hoang2019c} explored the dynamical constraint for several interstellar dust models: a contact binary grain model made of a silicate particle sticked to a carbonaceous particle, a model of composite grains made of sub-fragments of silicate and carbonaceous materials, and a model of a silicate core with an amorphous carbon mantle. For a range of irradiation and density conditions, the authors obtained a variety of rotational disrupted size $a_{\textrm{disr}}$ given the internal structure, and tensile strength derived from these dust models, highlighting the importance of understanding grain internal structure to analyze the dust polarization observations.

In dense cores, large grains ($\geq\,1\,\mu$m) are expected to have grown via coagulation processes such as grain sticking, and are thought to have composite structures, or compact structures with a high tensile strength ($S_{\textrm{max}}\,=\,10^{7}\,\textrm{erg}\,\textrm{cm}^{-3}$) with potentially a core-thick ice mantle structure (\eg see \citealt{McClure2009,Andersen2014,Poteet2015} for observational supports of ice mantles). However, dust growth and the physical conditions of the environment where grains have grown are not completely observationally constrained, and neither is the grains' internal structure. It is likely that grains become fluffy aggregates via low-velocity collisions, but also can go through a phase of compaction at higher velocities \citep{Ormel2009}. In the interior of a protostellar core, dust grains from the high temperature and high density regions of the inner circumstellar disk may have different internal structures compared to the grains that have grown on longer timescales within dense and cold infalling structures. Recently \citet{Garcia2020} developed a dust model that follows the grain porosity (a direct proxy for the dust grain tensile strength) during grain growth processes (in this case the formation of porous aggregates) occurring in protoplanetary disks. The high density of the disk may cause the dust aggregates to be compressed by static compression due to ram pressure of the disk gas, decreasing the grains' porosity, and thus increasing the grains' tensile strength \citep{Kataoka2013a,Kataoka2013b}. Therefore, if these latter large dust grains are efficiently lifted-up to populate outflow cavity and cavity walls \citep{Wong2016,Tsukamoto2021}, their structure might allow them to resist more strongly RATD in such irradiated regions.

In other words, if RATs is the mechanism aligning dust grains in cores, and if our models, which require us to implement large dust grains and high irradiation conditions, are accurate, then in order to protect large grains from being depleted by RATD, we must explain the high tensile strength of the corresponding grains. To that end, it would be required to characterize the physical conditions experienced by grains both in the inner disk and in the cavities in order to constrain their compactness using dust grain evolution models (see \citealt{Garcia2020} for example).
 
\subsection{Effects of the radiation field strength on the grain precession timescales}
\label{sec:timescale_ana}

\subsubsection{Radiative precession timescales}

\begin{figure*}[!tbh]
\centering
% \vspace{-0.6cm}
%no entry for referee mode
\subfigure{\includegraphics[scale=0.6,clip,trim= 0.1cm 0cm 0cm 0.1cm]{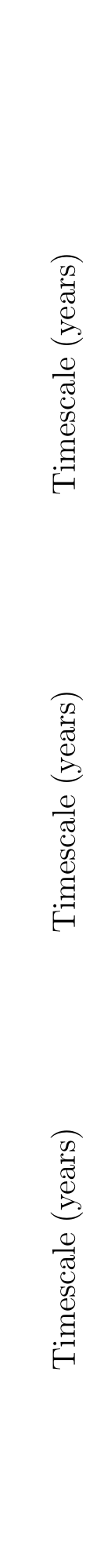}}
\subfigure{\includegraphics[scale=0.6,clip,trim= 3.5cm 1.2cm 3cm 0.2cm]{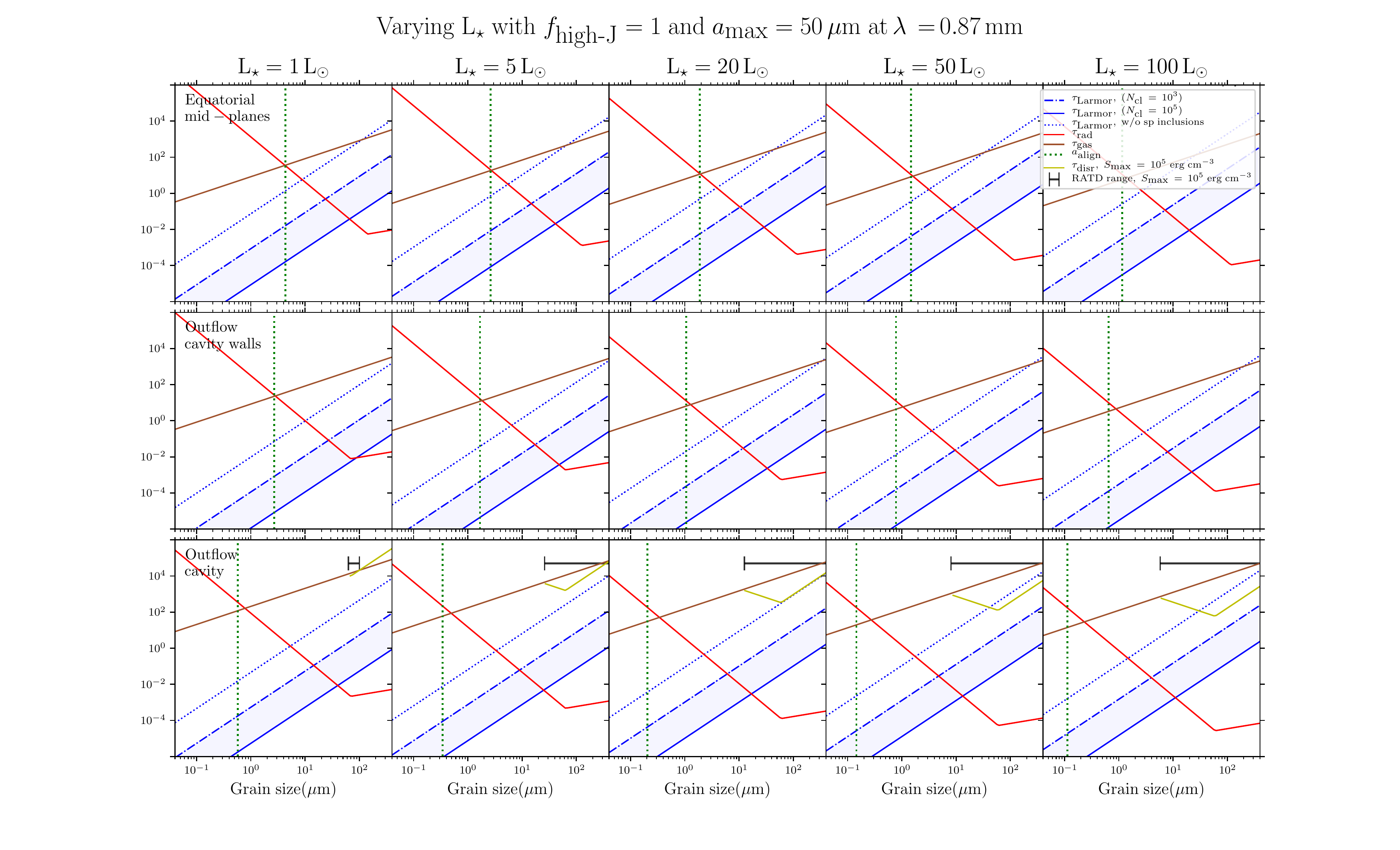}}
\subfigure{\includegraphics[scale=0.58,clip,trim= 2cm 0cm 2cm 1.1cm]{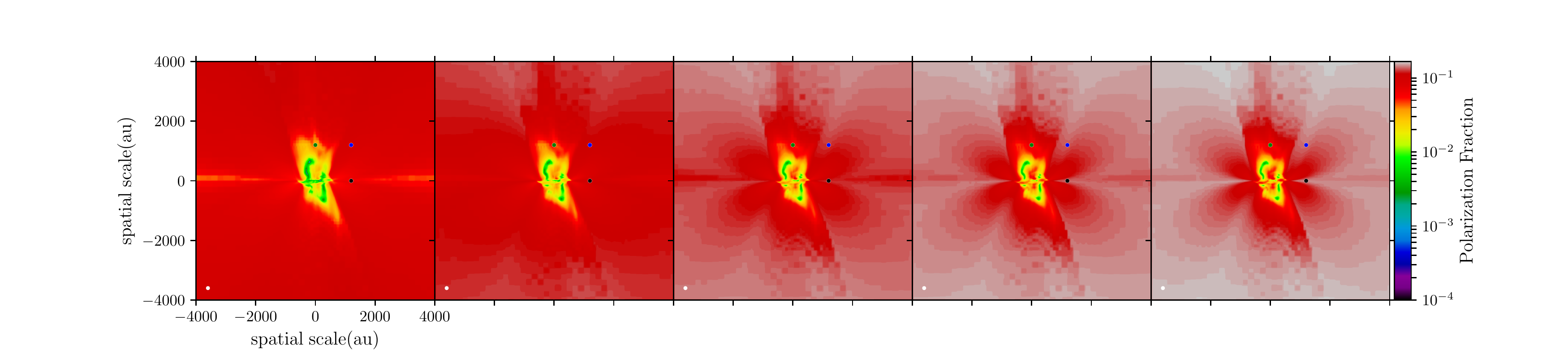}}
\vspace{-0.25in}
\caption{\footnotesize Effects of the central luminosity $L_\star$ on the alignment timescales in our radiative transfer models at 0.87 mm, with the fixed parameters of $a_{\textrm{max}}\,=\,50\,\mu$m, and $f_{\rm{high}-J}\,=\,1$. Each column is associated with a value of central luminosity $L_\star$. While the plots in three first rows represent the alignment timescales as a function of grain size, the plots in last row represent the polarization fraction.
The timescales are computed in specific locations, \ie toward the equatorial mid-plane (\textit{first row}), an outflow cavity wall (\textit{second row}), and in the middle of an outflow cavity (\textit{third row}), in the central 2D-slice of the model. These locations are indicated by the three colored circles in the polarization fraction maps, with the dark point for the first row (equatorial plane), the blue point for the second row (outflow cavity wall), and the green point for the third row (outflow cavity).
In each plot of the first three rows, we present the Larmor precession timescale $\tau_{\textrm{Larmor}}$ (for a grain without super paramagnetic inclusions, with the dotted line, and for two values of $N_{\textrm{cl}}$, being the number of atoms per cluster: 10$^3$ and 10$^5$, dot-dash and solid blue lines, respectively), the radiative precession timescale $\tau_{\textrm{rad}}$, the collisional gaseous damping time of the grain $\tau_{\textrm{gas}}$ (solid brown line), and the rotational disruption timescale $\tau_{\textrm{disr}}$ (for $S_{\textrm{max}}$ value of $10^{5}\,\textrm{erg}\,\textrm{cm}^{-3}$, solid dark yellow lines, respectively). We show the minimum grain size of aligned grains $a_{\textrm{align}}$ with the vertical dotted green line.
The range of grain size affected by rotationally disruption $a_{\textrm{disr}}-a_{\textrm{disr},\textrm{max}}$), for $S_{\textrm{max}}\,=\,10^{5}\,\textrm{erg}\,\textrm{cm}^{-3}$ is shown by the horizontal dark line.
At a given grain size, the shortest precession timescale between $\tau_{\rm{Larmor}}$ and $\tau_{\textrm{rad}}$ dictates the direction of alignment for the grains aligned at low-$J$, \ie with the magnetic field or the radiation field. 
% We select a region of 80\,au in size, located at $\sim\,$1200\,au from the center,
}
\label{fig:RT_timescales}
\end{figure*}

Finally, we attempt several timescale comparisons to determine whether our results are coherent with one another. The main part of the irradiation in the envelope that we measure in our models directly comes from the accretion shock onto the central protostellar embryo, which has been suggested to be episodic \citep{Vorobyov2005,Vorobyov2006,Baraffe2009,Vorobyov2010,Baraffe2012,Baraffe2017,Fischer2022}. During an accretion burst, the luminosity of a protostellar object can increase by a factor of few, up to $\sim\,10-100$. We need to determine how fast the temperature in the envelope adapts to a change in the central luminosity, how fast dust grain rotation, precession, and potential disruption, react to a change in irradiation and temperature. Figure \ref{fig:RT_timescales} presents the effects of the central luminosity value $L_\star$ on the different timescales concerning the precession, alignment, and disruption of dust grains. We present and discuss the results of the this timescale analysis shown in Fig. \ref{fig:RT_timescales} below.

In a core, the cooling time is commonly much shorter than the free-fall time (Equation 3 of \citealt{Hennebelle2020b}, which consist in calculating the radiative cooling time of a cell at a given distance, with its density, opacity, and temperature). We can consider that gas and dust temperature will become equal with one another in a few gaseous collisional timescale $\tau_{\rm{gas}}$ (see Fig. \ref{fig:RT_timescales}). In an outflow cavity wall of the fiducial radiative transfer, at $\sim\,$1000 au from the center, the cooling time is $\sim\,1\,$yr, and the free-fall time is $\sim\,6\times10^{3}\,$yr. As a consequence, it is reasonable to consider that the temperature adjusts instantaneously to a change of the central source luminosity, and that the gas dynamics is not affected. Therefore, if there is an accretion burst, one can expect the dust grains to immediately experience the different irradiation conditions. The accretion burst duration is thought to be $\leq\,$100-200 yr in protostars \citep{Vorobyov2013}, and the typical interval between two bursts varies within 10$^2$ and 10$^4$ years (different observational techniques have found a large variety of typical interval between bursts,\eg see \citealt{ContrerasPena2019,Jorgensen2020,LeeYH2021,Park2021,Fischer2022}). However, we note that in strongly irradiated regions the gas is likely warmer than the dust, even at high density (see for example the work of \citealt{Koumpia2015} who estimated a higher gas temperature by $\sim\,15$ K toward the PDR S 140). This may be due to shocks, and/or intense UV heating of the gas. Therefore, the temperature of the gas in outflow cavities and cavity walls may be higher than the dust temperature. This would affect our estimation of the gaseous damping timescale $\tau_{\textrm{gas}}$ (defined below), which scales with $T_{\textrm{gas}}^{-1/2}$. 

The timescale for a previously non-aligned grain to spin up via RATs up to a rotational velocity that would allow its alignment (at least three times the thermal velocity) is about the radiative precession time-scale $\tau_{\rm{rad}}$ if the grain is directly driven to a high-$J$ attractor point \citep{LazarianHoang2007}, given as:
\begin{equation}
\begin{split}
\tau_{\rm{rad}}\,\simeq\, 1.1\,\times\,10^{2}\, \hat{\rho}^{1/2}\hat{s}^{-1/3}a^{1/2}_{-5}\hat{T}_{\textrm{d}}^{1/2} \left( \frac{u_{\rm{rad}}}{u_{\rm{ISRF}}}\right)^{-1} \\
\times \left( \frac{\bar{\lambda}}{1.2\,\mu\rm{m}}\right)^{-1}
\left( \frac{\bar{\gamma} |\bar{\mathcal{Q}_{\Gamma}}|}{0.01}\right)^{-1}\,\rm{years}\,\,,
\end{split}
\label{equ:tau_rad}
\end{equation}
where $a_{-5}\,=\,a/10^{-5}$ cm, $\hat{\rho}\,=\,\rho/3\,$g cm$^{-3}$ the dust grain density, $\hat{s}\,=\,s/0.5$ the grains' aspect ratio, $\hat{T}_{\textrm{d}}\,=\,T_{\textrm{d}}/100$ K, $\bar{\gamma}$ is the anisotropy of the mean radiation field (see equation 20 of \citealt{Tazaki2017}, or equation 53 of \citealt{Hoang2022b}),
$\bar{\mathcal{Q}}_{\Gamma}$ is the RAT efficiency (see the relation 10 in \citealt{Hoang2020b}),
and $\bar{\lambda}$ is the mean wavelength of the radiation field\footnote{The values with a bar above result from the integration over the radiation field spectrum of the corresponding wavelength-dependent value, weighted by the radiation field $u_\lambda$}. The alignment timescale for a grain directly driven to a high-$J$ attractor point corresponds to a fast alignment, while the fraction $1-f_{\rm{high}-J}$ of grains that are driven to a low-$J$ attractor point experiences a slow alignment, as gaseous bombardments gradually move grains from the low-$J$ attractor toward a high-$J$ state \citep{Hoang2008,Hoang2016,Lazarian2021}. Such a slow alignment is typically achieved after $5-10\,\tau_{\rm{gas}}$, where $\tau_{\rm{gas}}$ is the collisional gaseous damping time of the grain (the infrared emission damping and plasma drag are neglected):
\begin{equation}
\tau_{\textrm{gas}}\,=\,\frac{3}{4\sqrt{\pi}} \frac{I_1}{n_{\textrm{H}} \mu m_{\textrm{H}} v_{\textrm{th}} a^4}\,\,,
\end{equation}
where $v_{\textrm{th}}\,=\,\sqrt{2 k_{\textrm{B}} T_{\textrm{gas}} / \mu m_{\textrm{H}}}$ is the thermal velocity, 
$I_1$ is the principal grain moment of inertia,
and $\mu$ is the mean molecular weight per hydrogen molecule. Nevertheless, the timescale for slow alignment is reduced when considering super-paramagnetic dust grains \citep{Hoang2008,Hoang2016,Lazarian2021,ChauGiang2022}, but can also increase for increasing radiation field strength because stronger radiative torques push grains to the low-$J$ attractor point \citep{Lazarian2021}. Therefore, the super-paramagneticity of dust grains plays an important role in the alignment timescale because (1) it governs the fraction of grains $f_{\rm{high}-J}$ that can be driven to a high-$J$ attractor point (and thus of grains that can experience a fast alignment), and (2) it diminishes the grain alignment timescale for grain experiencing slow alignment. The radiation field improves the strength of RATs and thus the maximum value of grain angular momentum (which can become high enough to trigger RATD), but forces grains to the low-$J$ attractor point that will experience slow alignment. An increase of the irradiation field during an accretion burst will thus generate an increase of the dust polarized emission on a timescale corresponding to the one of the slow alignment for the $1-f_{\rm{high}-J}$ fraction of grains driven at low-$J$. As the increase of irradiation also diminishes the minimum size of grains rotational disrupted $a_{\textrm{disr}}$, more grains could be rotationally disrupted. In the absence of high-$J$ attractor point for grains with inefficient magnetic relaxation (low $\delta_{\rm{m}}$) the increase of RATs via higher irradiation may thus cause slow disruption of grains (of the order of the slow alignment, \ie $5-10\,\tau_{\rm{gas}}$), because only low-$J$ attractors would exist. This makes RATD unlikely for low-$J$ aligned grains \citep{Hoang2019,Hoang2022b}.
In our model, the observed grain alignment efficiency strongly suggests that most grains are perfectly aligned with $f_{\rm{high}-J}\,\simeq\,1$ (see Appendix \ref{sec:app_RT_highJ} and \citealt{ChauGiang2022}). Therefore, in the type of environment we model, an increase of the radiation field strength can rapidly produce an increase of the dust polarized emission, \ie on a timescale of the order of $\tau_{\rm{rad}}$. 

The potential subsequent rotational disruption of grains that are supposed to be directly driven at high-$J$ also happens rapidly compared to the gaseous damping timescale, \ie of the order of the timescale $\tau_{\textrm{disr}}$ (see Fig. \ref{fig:RT_timescales}), which is the characteristic timescale for fast rotational grain disruption (\citealt{Hoang2019}; we note that this is much shorter than the shattering time by grain-grain collisions); it is derived as follows: 
\begin{equation}
\tau_{\textrm{disr}}\,=\,\frac{I_1\omega_{\textrm{disr}}}{\bar{\Gamma}_{\textrm{RAT}}}\,\,,
\end{equation}
where $\bar{\Gamma}_{\textrm{RAT}}$ is the radiative torque average over the incident radiation spectrum (as for $\tau_{\textrm{rad}}$, we adopt the value of $\bar{\mathcal{Q}}_{\Gamma}$ from \citealt{Hoang2020b}), and $\omega_{\textrm{disr}}$ is presented in Section \ref{sec:RATD}. If grains have super-paramagnetic (iron) inclusions such as fast alignment occurs, the dust polarization signal can have the time to increase during an accretion burst, because $\tau_{\rm{rad}}$ is generally much lower than the typical accretion burst duration. However, depending on the tensile strength of dust grains and luminosity reached during the burst, aligned dust grains could be rotationally disrupted if $\tau_{\rm{disr}}$ becomes comparable or lower than an accretion burst duration. If otherwise, slow alignment is occurring, slow disruption (on the order of the timescale for slow alignment, \ie $5-10\,\tau_{\rm{gas}}$) can happen in the steady-state regime where $a_{\textrm{disr}}\,\leq\,a_{\textrm{max}}$.
However, during an accretion burst the increase in grain alignment and rotational disruption efficiency may not happen in the case of slow alignment because $5-10\,\tau_{\rm{gas}}$ is of the order of an accretion burst duration. Even if only slow rotational disruption occurs, we note that RATD may remain an important process that can constrain early dust grain growth inside protostellar cores. Indeed, the typical timescale required to grow grains at sizes $\geq\,10\,\mu$m inside a medium of $n_{\textrm{H}}\,=\,10^{5}\,$cm$^{-3}$ is of the order the 7$-$10 times the free-fall timescale \citep{Ormel2009,Wong2016}, \ie $\sim\,7-10\times10^{5}$ years at such density (while the density of the envelope of our model is $n_{\textrm{H}}\,=\,10^{4}-10^{6}\,$cm$^{-3}$); this timescale is higher than the timescale of slow rotational disruption.

In addition, we note that an accretion burst may also influence the dust polarization position angle. Indeed, during such a burst the ionization increases and thus the coupling between the material and the magnetic field improves, such that a larger amount of material can dynamically affect magnetic field lines, which can be dragged more intensively inward. Such a change in dust polarization position angle given a change of magnetic field orientation would happen on a timescale of the order of the Larmor precession timescale, that is lower or of the order of the accretion burst duration, for grains $\geq\,1\,\mu$m.

\subsubsection{Radiative alignment}

Finally, if the radiation field is strong enough, grains can change their axis of alignment and become aligned with the axis of the anisotropic component of the radiation field (this is the $k$-RAT mechanism that we have already introduced; \citealt{LazarianHoang2007}). This occurs if the radiative precession timescale $\tau_{\rm{rad}}$ becomes shorter than the Larmor precession timescale, that illustrates the efficiency of the interaction between the grain magnetic moment with an external magnetic field. Following \citet{Hoang2014b,Tazaki2017}, we have:
\begin{equation}
\tau_{\rm{Larmor}}\,\simeq\,1.3\,\hat{\rho}\hat{s}^{-2/3}a^{2}_{-5}\hat{B}^{-1}\hat{\chi}^{-1} \,\,\rm{years}\,\,,
\end{equation}
where $\hat{B}\,=\,B/5\,\mu$G is the magnetic field, and $\hat{\chi}\,=\,\chi(0)/10^{-4}$ the grains' paramagnetic zero-frequency susceptibility. The zero-frequency susceptibility for a super-paramagnetic grain is given by Curie’s law \citep{Morrish2001}:
\begin{equation}
\chi(0)\,=\,\chi{_\textrm{sup}}(0)\,=\,1.2\times10^{-2} N_{\textrm{cl}} \phi_{\textrm{sp}} \left( \frac{15\,\textrm{K}}{T_\textrm{d}} \right)\,\,,
\end{equation}
where $\phi_{\rm{sp}}$ is the fraction of atoms that are super-paramagnetic, and $N_{\rm{cl}}$ is the number of atoms per cluster. From measurement of GEMS, the typical value  are $\phi_{\rm{sp}}$ = 0.03 \citep{Bradley1994,Martin1995,Goodman1995a}, and $N_{\rm{cl}}$ is expected to be $N_{\rm{cl}}$ = 10$^3$$-$10$^5$ \citep{Kneller1963,Jones1967}.
We highlight that the need to implement $f_{\rm{high}-J}\,\simeq\,1$ in our models argue toward super-paramagnetic grains (see also \citealt{ChauGiang2022}). We also consider an ordinary paramagnetic grain, with:
\begin{equation}
\chi(0)\,=\,4.2\,\times\,10^{-2} f_p \left( \frac{15\,\textrm{K}}{T_\textrm{d}} \right)\,\,,
\end{equation}
where $f_p$ is the fraction of atoms in the grain that are paramagnetic, evaluated at 10\% \citep{Draine1996,Tazaki2017}.  For reference, \citet{Draine1996} proposed $\chi(0)$ values in the range 4$\times10^{-5}-10^{-3}$ for ordinary paramagnetic grains. 
The evolution of $\tau_{\rm{Larmor}}$ as a function of grain size for a super-paramagnetic grain is shown in Fig. \ref{fig:RT_timescales}. We note that, while magnetic relaxation participates to the alignment of grains' angular momentum with magnetic field lines \citep{Weingartner2021}, grain alignment by magnetic relaxation alone is still inefficient in the absence of grain suprathermal rotation because of internal thermal fluctuations, even for super-paramagnetic grains \citep{Hoang2016,Hoang2016b}. For example, for typical protostellar envelope conditions \citet{Hoang2022b} calculated that super-paramagnetic relaxation is negligible for grains $\gtrsim\,1\,\mu$m inside protostellar envelope where $n\,\gtrsim\,10^{7}$ cm$^{-3}$. 
This makes Larmor precession the main agent in our models that controls the grain size parameter space describing which grains are aligned with magnetic field lines for grains $\gtrsim\,1\,\mu$m.
Therefore, at a given grain size the shortest precession timescale between $\tau_{\textrm{Larmor}}$ and $\tau_{\textrm{rad}}$ dictates the axis of alignment, \ie the magnetic field or the anisotropic component of the radiation field.

Equation \ref{equ:tau_rad} is valid for grains enduring slow-alignment, aligned at the low-$J$ attractor point. The radiative precession timescale for grains aligned in the high-$J$ state is several orders of magnitude higher. Indeed, at the high-$J$ attractor point the spin-up effect of RATs dominates, such as the Larmor precession is the strongest, and these grains are aligned with the magnetic field (see Section 5.3 of \citealt{Hoang2022b}).
As $L_\star$ increases, the grain size at which the transition of alignment axis must occur decreases, and can reach $\sim\,1-10\,\mu$m for paramagnetic grains when $L_\star\,\geq\,20\,$L$_\odot$, toward the most irradiated regions of the protostar, \ie inside outflow cavities.
% (this is dependent on the values chosen for the parameter $N_{\rm{cl}}$ that we vary for this timescale investigation).
In cavity walls and equatorial planes, where the dust polarization is detected, the transition would occur for paramagnetic grains $\geq\,10\,\mu$m. Such grain sizes are viable in these environments.
However, if aligned dust grains are super-paramagnetic with $f_{\rm{high}-J}\,\simeq\,1$, the grains whose size is eligible to the $k$-RAT mechanism would rotate rapidly enough to be aligned with the magnetic field, and be potentially affected by RATD where it is efficient, \ie in cavities (see Fig. \ref{fig:RT_maps_RATD2} and \ref{fig:RT_timescales}). The observed grain alignment efficiency is high, and points toward super-paramagnetic grains aligned with the magnetic field with $f_{\rm{high}-J}\,\simeq\,1$ \citep{LeGouellec2020,ChauGiang2022}.
Therefore, the $k$-RAT mechanism is unlikely in protostellar environments because the low-$J$ aligned grains eligible to $k$-RAT would produce too low polarization fraction values if they were to dominate the polarized dust emission.

With our approach, the most crucial model parameters that describe the alignment mechanism in a given protostellar envelope are thus $f_{\rm{high}-J}$, $L_\star$, and $S_{\textrm{max}}$. While $f_{\rm{high}-J}\,\simeq\,1$ and high irradiation (\ie $L_\star\,\geq\,20\,$L$_\odot$) are necessary to reproduce the observed grain alignment efficiency, one must also constrain the effect of RATD (the dependence of $a_{\textrm{align}}$ with the radiation field strength is weaker than the dependence of $a_{\textrm{disr}}$). The polarized dust emission observed in Class 0 protostars is thus produced in dense enough regions, and/or from compact enough aligned grains, where RATD does not disrupt the large grains that are responsible for the dust polarization.

\section{Conclusions and Summary}
\label{sec:p3_ccl}

While in paper I, an analysis of ALMA dust polarization and molecular line observations was presented, this paper explores in detail the physics of grain alignment occurring in Class 0 protostellar cores. To assess the role of the radiation field, we perform POLARIS synthetic observations of a MHD simulation of a collapsing protostellar core (that implements the radiation treatment method developed by \citet{Risse2020}, the protostellar jet implementation of \citealt{Verliat2022}). We compare the grain alignment efficiency that we derive from ALMA polarization observations of Class 0 protostellar cores, with the results from the radiative transfer modeling. The main results and conclusions of this paper are as follows:

\begin{enumerate}
\item We produce synthetic observations from a MHD models of the protostellar evolution for an intermediate mass core, using POLARIS. We show that these synthetic polarized dust emission maps reproduce the enhancement of polarized dust emission along outflow cavity walls that has been observed towards several objects.
\item In the radiative transfer computations, we vary the luminosity of the central object $L_\star$ and study how it favors dust grain alignment in the inner core. Comparing the dust grain alignment efficiency (as traced by $\StimesP$) derived in the radiative transfer calculations with those found in \citet{LeGouellec2020} with ALMA observations, we find that significant luminosity, \ie $L_{\star}\,\ge\,$20\,L$_{\odot}$, as well as super-paramagnetic grains (with a fraction supra-thermally rotating grains close to 1), and large grains (with $a_{\rm{max}}\,\geq\,10\,\mu$m), are required for the dust grain alignment efficiency of the synthetic observations to reach observed values. The values of central luminosity we implement are dependent on the characteristics of the model we use (core initial mass, inner density structures) to reproduce the dust polarization. 
\item One caveat of our modeling is the nature of the radiation field spectrum. The potential UV-photons produced in shocks in jets/outflows and along the cavity walls and the temperature increase triggered by the shocked-wind are not implemented in the radiative transfer calculations.
\item We implement the equations of the RAdiative Torque Disruption (RATD) phenomenon on the radiative transfer calculations. This phenomenon triggers the rotational disruption of the largest aligned dust grains, which can cause a depletion of large grains that are necessary to produce the observed polarized dust emission. Varying the irradiation and the tensile strength of dust grains, we find that in our models, RATD can occur in outflow cavities for central luminosities of $L_{\star}\,\gtrsim\,20$ L$_\odot$ and grain tensile strengths of $S_{\rm{max}}\,\lesssim\,10^{7}\,\rm{erg}\,\rm{cm}^{-3}$. The densest regions, \ie outflow cavity walls and more especially the equatorial mid-planes, seem more protected and are the last regions of the envelope to have their populations of large dust grains depleted by RATD.
\item Our models require both large grains and high irradiation, which raises two questions: where have large grains formed in Class 0 protostellar cores? And are their tensile strengths high enough to resist rotational disruption? These two questions are intrinsically linked, given that the environmental conditions (such as temperature) in which large grains have grown can dictate their structure (compact versus composite) and thus their tensile strength.
\item The physical conditions that favor the polarized dust emission detected toward equatorial planes and infalling structures such as accretion streamers still represent an open question. One needs to explain how large grains ($a\;\geq\;10\;\mu$m) form in the envelope, and how they resist disruption via RATD.
\item We study the characteristic timescales of the grain alignment physics and rotational disruption of grains, in an attempt to understand the effect that highly variable accretion rates would have on the dust grain alignment conditions.
In the environment we model, during an accretion burst or a steady-state phase of high luminosity from the protostellar embryo (with $L_{\star}\,\leq\,20\,$L$_\odot$), RATD could have enough time to disrupt the largest grains in irradiated regions, depending on their tensile strength.
In such luminous conditions, slowly rotating grains $\geq\,10\,\mu$m (at the low-$J$ attractor point) could in theory preferentially align themselves with the radiation field ($k$-RAT grain alignment). However, given the high grain alignment efficiency observed in protostellar envelopes, having mainly rapidly rotating grains (at the high-$J$ attractor point) aligned with the magnetic field is the most likely scenario. This suggests that the $k$-RAT grain alignment is not the main cause of the observed polarization dust emission in Class 0 objects.
\end{enumerate}

Assuming RATs are the predominant causes of dust polarization in Class 0 sources, the next step would be to investigate further the characteristics of the dust grains populating Class 0 envelopes: in particular, their composition and helicity (see \citealt{Hoang2022}). Right now we use standard MRN distribution of grains with ISM chemical grain properties (silicates and carbonaceous grains). We may have to revisit the picture of the dust grains in protostars, as their compositions and shape must favor the efficient grain alignment we observe (see for example the promising ``astrodust'' model \citealt{Draine2021,Hensley2022}). At this point, while RATs is favored by the investigations we present, one still needs to address the questions of grain growth, disruption, and structure. The radiation field strength seems to be an important agent to align grains in the inner core, but rotational disruption needs to be taken into account alongside grain structure to fully understand the role of irradiation in dust grain evolution. Another additional step in testing RAT theory involves the implementation of additional grain alignment theories in dust radiative transfer codes like POLARIS, \ie dust grains aligned with the radiation field, imperfect internal alignment \citep{Hoang2022,ChauGiang2022,Hoang2022b}, and mechanically aligned grains (\citealt{Hoang2018,Reissl2022,Hoang2022b}; requiring the implementation of dust-gas drift in 3D protostellar collapse simulations, see \citealt{Lebreuilly2019,Lebreuilly2020}). These other grain alignment mechanisms must also be further observationally characterized \citep{Pattle2021,Hull2022}.

~\\
{
\small
\textit{Acknowledgments}
V.J.M.L.G. and C.L.H.H. acknowledge the ESO Studentship Program, and the guidance and support of Eric Villard.
C.L.H.H. acknowledges the support of both the NAOJ Fellowship as well as JSPS KAKENHI grants 18K13586, 20K14527, and 22H01271. 
This work has received support from the European Research Council (ERC Starting Grant MagneticYSOs with grant agreement no. 679937).
We thank B-G Andersson and Thiem Hoang for providing helpful explanations.
ALMA is a partnership of ESO (representing its member states), NSF (USA) and NINS (Japan), together with NRC (Canada), MOST and ASIAA (Taiwan), and KASI (Republic of Korea), in cooperation with the Republic of Chile. The Joint ALMA Observatory is operated by ESO, AUI/NRAO and NAOJ.
The National Radio Astronomy Observatory is a facility of the National Science Foundation operated under cooperative agreement by Associated Universities, Inc. 

\textit{Facilities:} ALMA. 

\textit{Software:} APLpy, an open-source plotting package for Python hosted at \url{http://aplpy.github.com} \citep{Robitaille2012}. CASA \citep{McMullin2007}. Astropy \citep{Astropy2018}. POLARIS \url{https://portia.astrophysik.uni-kiel.de/polaris/} \citep{Reissl2016}

}

\bibliography{ms}

\begin{thebibliography}{}
\expandafter\ifx\csname natexlab\endcsname\relax\def\natexlab#1{#1}\fi

\bibitem[{{Agurto-Gangas} {et~al.}(2019){Agurto-Gangas}, {Pineda},
  {Sz{\H{u}}cs}, {Testi}, {Tazzari}, {Miotello}, {Caselli}, {Dunham},
  {Stephens}, \& {Bourke}}]{AgurtoGangas2019}
{Agurto-Gangas}, C., {Pineda}, J.~E., {Sz{\H{u}}cs}, L., {et~al.} 2019, \aap,
  623, A147

\bibitem[{{Anderl} {et~al.}(2016){Anderl}, {Maret}, {Cabrit}, {Belloche},
  {Maury}, {Andr{\'e}}, {Codella}, {Bacmann}, {Bontemps}, {Podio}, {Gueth}, \&
  {Bergin}}]{Anderl2016}
{Anderl}, S., {Maret}, S., {Cabrit}, S., {et~al.} 2016, \aap, 591, A3

\bibitem[{{Andersen} {et~al.}(2014){Andersen}, {Thi}, {Steinacker}, \&
  {Tothill}}]{Andersen2014}
{Andersen}, M., {Thi}, W.~F., {Steinacker}, J., \& {Tothill}, N. 2014, \aap,
  568, L3

\bibitem[{{Andersson} {et~al.}(2015){Andersson}, {Lazarian}, \&
  {Vaillancourt}}]{Andersson2015}
{Andersson}, B.-G., {Lazarian}, A., \& {Vaillancourt}, J.~E. 2015, \araa, 53,
  501

\bibitem[{{Andre} \& {Montmerle}(1994)}]{Andre1994}
{Andre}, P., \& {Montmerle}, T. 1994, \apj, 420, 837

\bibitem[{{Andre} {et~al.}(1993){Andre}, {Ward-Thompson}, \&
  {Barsony}}]{Andre1993}
{Andre}, P., {Ward-Thompson}, D., \& {Barsony}, M. 1993, \apj, 406, 122

\bibitem[{{Arce} \& {Goodman}(2001)}]{Arce2001}
{Arce}, H.~G., \& {Goodman}, A.~A. 2001, \apj, 554, 132

\bibitem[{{Astropy Collaboration} {et~al.}(2018){Astropy Collaboration},
  {Price-Whelan}, {Sip{\H o}cz}, {G{\"u}nther}, {Lim}, {Crawford}, {Conseil},
  {Shupe}, {Craig}, {Dencheva}, {Ginsburg}, {VanderPlas}, {Bradley},
  {P{\'e}rez-Su{\'a}rez}, {de Val-Borro}, {Aldcroft}, {Cruz}, {Robitaille},
  {Tollerud}, {Ardelean}, {Babej}, {Bach}, {Bachetti}, {Bakanov}, {Bamford},
  {Barentsen}, {Barmby}, {Baumbach}, {Berry}, {Biscani}, {Boquien}, {Bostroem},
  {Bouma}, {Brammer}, {Bray}, {Breytenbach}, {Buddelmeijer}, {Burke},
  {Calderone}, {Cano Rodr{\'{\i}}guez}, {Cara}, {Cardoso}, {Cheedella},
  {Copin}, {Corrales}, {Crichton}, {D'Avella}, {Deil}, {Depagne}, {Dietrich},
  {Donath}, {Droettboom}, {Earl}, {Erben}, {Fabbro}, {Ferreira}, {Finethy},
  {Fox}, {Garrison}, {Gibbons}, {Goldstein}, {Gommers}, {Greco}, {Greenfield},
  {Groener}, {Grollier}, {Hagen}, {Hirst}, {Homeier}, {Horton}, {Hosseinzadeh},
  {Hu}, {Hunkeler}, {Ivezi{\'c}}, {Jain}, {Jenness}, {Kanarek}, {Kendrew},
  {Kern}, {Kerzendorf}, {Khvalko}, {King}, {Kirkby}, {Kulkarni}, {Kumar},
  {Lee}, {Lenz}, {Littlefair}, {Ma}, {Macleod}, {Mastropietro}, {McCully},
  {Montagnac}, {Morris}, {Mueller}, {Mumford}, {Muna}, {Murphy}, {Nelson},
  {Nguyen}, {Ninan}, {N{\"o}the}, {Ogaz}, {Oh}, {Parejko}, {Parley}, {Pascual},
  {Patil}, {Patil}, {Plunkett}, {Prochaska}, {Rastogi}, {Reddy Janga},
  {Sabater}, {Sakurikar}, {Seifert}, {Sherbert}, {Sherwood-Taylor}, {Shih},
  {Sick}, {Silbiger}, {Singanamalla}, {Singer}, {Sladen}, {Sooley},
  {Sornarajah}, {Streicher}, {Teuben}, {Thomas}, {Tremblay}, {Turner},
  {Terr{\'o}n}, {van Kerkwijk}, {de la Vega}, {Watkins}, {Weaver}, {Whitmore},
  {Woillez}, {Zabalza}, \& {Astropy Contributors}}]{Astropy2018}
{Astropy Collaboration}, {Price-Whelan}, A.~M., {Sip{\H o}cz}, B.~M., {et~al.}
  2018, \aj, 156, 123

\bibitem[{{Bally}(2016)}]{Bally2016}
{Bally}, J. 2016, \araa, 54, 491

\bibitem[{{Baraffe} {et~al.}(2009){Baraffe}, {Chabrier}, \&
  {Gallardo}}]{Baraffe2009}
{Baraffe}, I., {Chabrier}, G., \& {Gallardo}, J. 2009, \apjl, 702, L27

\bibitem[{{Baraffe} {et~al.}(2017){Baraffe}, {Elbakyan}, {Vorobyov}, \&
  {Chabrier}}]{Baraffe2017}
{Baraffe}, I., {Elbakyan}, V.~G., {Vorobyov}, E.~I., \& {Chabrier}, G. 2017,
  \aap, 597, A19

\bibitem[{{Baraffe} {et~al.}(2012){Baraffe}, {Vorobyov}, \&
  {Chabrier}}]{Baraffe2012}
{Baraffe}, I., {Vorobyov}, E., \& {Chabrier}, G. 2012, \apj, 756, 118

\bibitem[{{Bate}(2022)}]{Bate2022}
{Bate}, M.~R. 2022, \mnras, 514, 2145

\bibitem[{{Bleuler} \& {Teyssier}(2014)}]{BleulerTeyssier2014}
{Bleuler}, A., \& {Teyssier}, R. 2014, \mnras, 445, 4015

\bibitem[{{Bonnell} \& {Bastien}(1992)}]{Bonnell1992}
{Bonnell}, I., \& {Bastien}, P. 1992, \apjl, 401, L31

\bibitem[{{Bradley}(1994)}]{Bradley1994}
{Bradley}, J.~P. 1994, Science, 265, 925

\bibitem[{{Chau Giang} {et~al.}(2022){Chau Giang}, {Hoang}, {Kim}, \&
  {Tram}}]{ChauGiang2022}
{Chau Giang}, N., {Hoang}, T., {Kim}, J.-G., \& {Tram}, L.~N. 2022, arXiv
  e-prints, arXiv:2210.01036

\bibitem[{{Commer{\c{c}}on} {et~al.}(2014){Commer{\c{c}}on}, {Debout}, \&
  {Teyssier}}]{Commercon2014}
{Commer{\c{c}}on}, B., {Debout}, V., \& {Teyssier}, R. 2014, \aap, 563, A11

\bibitem[{{Commer{\c{c}}on} {et~al.}(2022){Commer{\c{c}}on}, {Gonz{\'a}lez},
  {Mignon-Risse}, {Hennebelle}, \& {Vaytet}}]{Commercon2022}
{Commer{\c{c}}on}, B., {Gonz{\'a}lez}, M., {Mignon-Risse}, R., {Hennebelle},
  P., \& {Vaytet}, N. 2022, \aap, 658, A52

\bibitem[{{Commer{\c{c}}on} {et~al.}(2011{\natexlab{a}}){Commer{\c{c}}on},
  {Hennebelle}, \& {Henning}}]{Commercon2011b}
{Commer{\c{c}}on}, B., {Hennebelle}, P., \& {Henning}, T. 2011{\natexlab{a}},
  \apjl, 742, L9

\bibitem[{{Commer{\c{c}}on} {et~al.}(2011{\natexlab{b}}){Commer{\c{c}}on},
  {Teyssier}, {Audit}, {Hennebelle}, \& {Chabrier}}]{Commercon2011}
{Commer{\c{c}}on}, B., {Teyssier}, R., {Audit}, E., {Hennebelle}, P., \&
  {Chabrier}, G. 2011{\natexlab{b}}, \aap, 529, A35

\bibitem[{{Contreras Pe{\~n}a} {et~al.}(2019){Contreras Pe{\~n}a}, {Naylor}, \&
  {Morrell}}]{ContrerasPena2019}
{Contreras Pe{\~n}a}, C., {Naylor}, T., \& {Morrell}, S. 2019, \mnras, 486,
  4590

\bibitem[{{Cox} {et~al.}(2018){Cox}, {Harris}, {Looney}, {Li}, {Yang}, {Tobin},
  \& {Stephens}}]{Cox2018}
{Cox}, E.~G., {Harris}, R.~J., {Looney}, L.~W., {et~al.} 2018, \apj, 855, 92

\bibitem[{{D'Angelo} \& {Spruit}(2012)}]{DAngelo2012}
{D'Angelo}, C.~R., \& {Spruit}, H.~C. 2012, \mnras, 420, 416

\bibitem[{{Davis} \& {Greenstein}(1951)}]{DavisGreenstein1951}
{Davis}, Jr., L., \& {Greenstein}, J.~L. 1951, \apj, 114, 206

\bibitem[{{Dolginov} \& {Mitrofanov}(1976)}]{Dolginov1976}
{Dolginov}, A.~Z., \& {Mitrofanov}, I.~G. 1976, \apss, 43, 291

\bibitem[{{Draine} \& {Hensley}(2021)}]{Draine2021}
{Draine}, B.~T., \& {Hensley}, B.~S. 2021, \apj, 909, 94

\bibitem[{{Draine} \& {Lazarian}(1998)}]{Draine1998}
{Draine}, B.~T., \& {Lazarian}, A. 1998, \apj, 508, 157

\bibitem[{{Draine} \& {Weingartner}(1996)}]{Draine1996}
{Draine}, B.~T., \& {Weingartner}, J.~C. 1996, \apj, 470, 551

\bibitem[{{Draine} \& {Weingartner}(1997)}]{Draine1997}
---. 1997, \apj, 480, 633

\bibitem[{{Dunham} {et~al.}(2010){Dunham}, {Evans}, {Terebey}, {Dullemond}, \&
  {Young}}]{Dunham2010a}
{Dunham}, M.~M., {Evans}, Neal~J., I., {Terebey}, S., {Dullemond}, C.~P., \&
  {Young}, C.~H. 2010, \apj, 710, 470

\bibitem[{{Dunham} {et~al.}(2013){Dunham}, {Arce}, {Allen}, {Evans},
  {Broekhoven-Fiene}, {Chapman}, {Cieza}, {Gutermuth}, {Harvey}, {Hatchell},
  {Huard}, {Kirk}, {Matthews}, {Mer{\'{\i}}n}, {Miller}, {Peterson}, \&
  {Spezzi}}]{Dunham2013}
{Dunham}, M.~M., {Arce}, H.~G., {Allen}, L.~E., {et~al.} 2013, \aj, 145, 94

\bibitem[{Dunham {et~al.}(2015)Dunham, Allen, II, Broekhoven-Fiene, Cieza,
  Francesco, Gutermuth, Harvey, Hatchell, Heiderman, Huard, Johnstone, Kirk,
  Matthews, Miller, Peterson, \& Young}]{Dunham2015}
Dunham, M.~M., Allen, L.~E., II, N. J.~E., {et~al.} 2015, The Astrophysical
  Journal Supplement Series, 220, 11

\bibitem[{{Evans} {et~al.}(2009){Evans}, {Dunham}, {J{\o}rgensen}, {Enoch},
  {Mer{\'\i}n}, {van Dishoeck}, {Alcal{\'a}}, {Myers}, {Stapelfeldt}, {Huard},
  {Allen}, {Harvey}, {van Kempen}, {Blake}, {Koerner}, {Mundy}, {Padgett}, \&
  {Sargent}}]{Evans2009}
{Evans}, Neal~J., I., {Dunham}, M.~M., {J{\o}rgensen}, J.~K., {et~al.} 2009,
  \apjs, 181, 321

\bibitem[{{Fischer} {et~al.}(2022){Fischer}, {Hillenbrand}, {Herczeg},
  {Johnstone}, {K{\'o}sp{\'a}l}, \& {Dunham}}]{Fischer2022}
{Fischer}, W.~J., {Hillenbrand}, L.~A., {Herczeg}, G.~J., {et~al.} 2022, arXiv
  e-prints, arXiv:2203.11257

\bibitem[{{Fischer} {et~al.}(2019){Fischer}, {Safron}, \&
  {Megeath}}]{Fischer2019}
{Fischer}, W.~J., {Safron}, E., \& {Megeath}, S.~T. 2019, \apj, 872, 183

\bibitem[{{Frimann} {et~al.}(2016){Frimann}, {J{\o}rgensen}, {Padoan}, \&
  {Haugb{\o}lle}}]{Frimann2016}
{Frimann}, S., {J{\o}rgensen}, J.~K., {Padoan}, P., \& {Haugb{\o}lle}, T. 2016,
  \aap, 587, A60

\bibitem[{{Fromang} {et~al.}(2006){Fromang}, {Hennebelle}, \&
  {Teyssier}}]{Fromang2006}
{Fromang}, S., {Hennebelle}, P., \& {Teyssier}, R. 2006, \aap, 457, 371

\bibitem[{{Galametz} {et~al.}(2019){Galametz}, {Maury}, {Valdivia}, {Testi},
  {Belloche}, \& {Andr{\'e}}}]{Galametz2019}
{Galametz}, M., {Maury}, A.~J., {Valdivia}, V., {et~al.} 2019, \aap, 632, A5

\bibitem[{{Garcia} \& {Gonzalez}(2020)}]{Garcia2020}
{Garcia}, A. J.~L., \& {Gonzalez}, J.-F. 2020, \mnras, 493, 1788

\bibitem[{{Gerrard} {et~al.}(2019){Gerrard}, {Federrath}, \&
  {Kuruwita}}]{Gerrard2019}
{Gerrard}, I.~A., {Federrath}, C., \& {Kuruwita}, R. 2019, \mnras, 485, 5532

\bibitem[{{Goodman} \& {Whittet}(1995)}]{Goodman1995a}
{Goodman}, A.~A., \& {Whittet}, D.~C.~B. 1995, \apjl, 455, L181

\bibitem[{{Guillet} {et~al.}(2020){Guillet}, {Hennebelle}, {Pineau des
  For{\^e}ts}, {Marcowith}, {Commer{\c{c}}on}, \& {Marchand}}]{Guillet2020b}
{Guillet}, V., {Hennebelle}, P., {Pineau des For{\^e}ts}, G., {et~al.} 2020,
  \aap, 643, A17

\bibitem[{{Guillet} {et~al.}(2009){Guillet}, {Jones}, \& {Pineau Des
  For{\^e}ts}}]{Guillet2009}
{Guillet}, V., {Jones}, A.~P., \& {Pineau Des For{\^e}ts}, G. 2009, \aap, 497,
  145

\bibitem[{{Hartmann} {et~al.}(1997){Hartmann}, {Cassen}, \&
  {Kenyon}}]{Hartmann1997}
{Hartmann}, L., {Cassen}, P., \& {Kenyon}, S.~J. 1997, \apj, 475, 770

\bibitem[{{Hennebelle}(2018)}]{Hennebelle2018}
{Hennebelle}, P. 2018, \aap, 611, A24

\bibitem[{{Hennebelle} {et~al.}(2020){Hennebelle}, {Commer{\c{c}}on}, {Lee}, \&
  {Chabrier}}]{Hennebelle2020b}
{Hennebelle}, P., {Commer{\c{c}}on}, B., {Lee}, Y.-N., \& {Chabrier}, G. 2020,
  \apj, 904, 194

\bibitem[{{Hensley} \& {Draine}(2022)}]{Hensley2022}
{Hensley}, B.~S., \& {Draine}, B.~T. 2022, arXiv e-prints, arXiv:2208.12365

\bibitem[{{Hildebrand} \& {Dragovan}(1995)}]{Hildebrand_Dragovan1995}
{Hildebrand}, R.~H., \& {Dragovan}, M. 1995, \apj, 450, 663

\bibitem[{{Hoang}(2019)}]{Hoang2019c}
{Hoang}, T. 2019, \apj, 876, 13

\bibitem[{{Hoang}(2020)}]{Hoang2020}
---. 2020, Galaxies, 8, 52

\bibitem[{{Hoang}(2022)}]{Hoang2022}
---. 2022, \apj, 928, 102

\bibitem[{{Hoang} {et~al.}(2018){Hoang}, {Cho}, \& {Lazarian}}]{Hoang2018}
{Hoang}, T., {Cho}, J., \& {Lazarian}, A. 2018, \apj, 852, 129

\bibitem[{{Hoang} \& {Lazarian}(2008)}]{Hoang2008}
{Hoang}, T., \& {Lazarian}, A. 2008, \mnras, 388, 117

\bibitem[{{Hoang} \& {Lazarian}(2009)}]{Hoang2009}
---. 2009, \apj, 697, 1316

\bibitem[{{Hoang} \& {Lazarian}(2014)}]{Hoang2014b}
---. 2014, \mnras, 438, 680

\bibitem[{{Hoang} \& {Lazarian}(2016{\natexlab{a}})}]{Hoang2016}
---. 2016{\natexlab{a}}, \apj, 831, 159

\bibitem[{{Hoang} \& {Lazarian}(2016{\natexlab{b}})}]{Hoang2016b}
---. 2016{\natexlab{b}}, \apj, 821, 91

\bibitem[{{Hoang} \& {Lee}(2020)}]{Hoang2019b}
{Hoang}, T., \& {Lee}, H. 2020, \apj, 896, 144

\bibitem[{{Hoang} \& {Tram}(2019)}]{Hoang2019d}
{Hoang}, T., \& {Tram}, L.~N. 2019, \apj, 877, 36

\bibitem[{{Hoang} \& {Tram}(2020)}]{Hoang2019}
---. 2020, \apj, 891, 38

\bibitem[{{Hoang} {et~al.}(2019){Hoang}, {Tram}, {Lee}, \&
  {Ahn}}]{Hoang2019NatAs}
{Hoang}, T., {Tram}, L.~N., {Lee}, H., \& {Ahn}, S.-H. 2019, Nature Astronomy,
  3, 766

\bibitem[{{Hoang} {et~al.}(2021){Hoang}, {Tram}, {Lee}, {Diep}, \&
  {Ngoc}}]{Hoang2020b}
{Hoang}, T., {Tram}, L.~N., {Lee}, H., {Diep}, P.~N., \& {Ngoc}, N.~B. 2021,
  \apj, 908, 218

\bibitem[{{Hoang} {et~al.}(2022){Hoang}, {Tram}, {Minh Phan}, {Giang},
  {Phuong}, \& {Dieu}}]{Hoang2022b}
{Hoang}, T., {Tram}, L.~N., {Minh Phan}, V.~H., {et~al.} 2022, \aj, 164, 248

\bibitem[{{Hosokawa} \& {Omukai}(2009)}]{Hosokawa2009}
{Hosokawa}, T., \& {Omukai}, K. 2009, \apj, 691, 823

\bibitem[{{Hull} {et~al.}(2020){Hull}, {Le Gouellec}, {Girart}, {Tobin}, \&
  {Bourke}}]{Hull2020a}
{Hull}, C. L.~H., {Le Gouellec}, V. J.~M., {Girart}, J.~M., {Tobin}, J.~J., \&
  {Bourke}, T.~L. 2020, \apj, 892, 152

\bibitem[{{Hull} {et~al.}(2017){Hull}, {Girart}, {Tychoniec}, {Rao},
  {Cort{\'e}s}, {Pokhrel}, {Zhang}, {Houde}, {Dunham}, {Kristensen}, {Lai},
  {Li}, \& {Plambeck}}]{Hull2017b}
{Hull}, C.~L.~H., {Girart}, J.~M., {Tychoniec}, {\L}., {et~al.} 2017, \apj,
  847, 92

\bibitem[{{Hull} {et~al.}(2022){Hull}, {Yang}, {Cort{\'e}s}, {Dent}, {Kral},
  {Li}, {Le Gouellec}, {Hughes}, {Milli}, {Teague}, \& {Wyatt}}]{Hull2022}
{Hull}, C. L.~H., {Yang}, H., {Cort{\'e}s}, P.~C., {et~al.} 2022, \apj, 930, 49

\bibitem[{{Jensen} \& {Haugb{\o}lle}(2018)}]{Jensen2018}
{Jensen}, S.~S., \& {Haugb{\o}lle}, T. 2018, \mnras, 474, 1176

\bibitem[{{Jones}(2016)}]{Jones2016c}
{Jones}, A.~P. 2016, Royal Society Open Science, 3, 160224

\bibitem[{{Jones} {et~al.}(1996){Jones}, {Tielens}, \&
  {Hollenbach}}]{Jones1996}
{Jones}, A.~P., {Tielens}, A.~G.~G.~M., \& {Hollenbach}, D.~J. 1996, \apj, 469,
  740

\bibitem[{{Jones} \& {Spitzer}(1967)}]{Jones1967}
{Jones}, R.~V., \& {Spitzer}, Lyman, J. 1967, \apj, 147, 943

\bibitem[{{J{\o}rgensen} {et~al.}(2020){J{\o}rgensen}, {Belloche}, \&
  {Garrod}}]{Jorgensen2020}
{J{\o}rgensen}, J.~K., {Belloche}, A., \& {Garrod}, R.~T. 2020, \araa, 58, 727

\bibitem[{{J{\o}rgensen} {et~al.}(2013){J{\o}rgensen}, {Visser}, {Sakai},
  {Bergin}, {Brinch}, {Harsono}, {Lindberg}, {van Dishoeck}, {Yamamoto},
  {Bisschop}, \& {Persson}}]{Jorgensen2013}
{J{\o}rgensen}, J.~K., {Visser}, R., {Sakai}, N., {et~al.} 2013, \apjl, 779,
  L22

\bibitem[{{Kataoka} {et~al.}(2013{\natexlab{a}}){Kataoka}, {Tanaka}, {Okuzumi},
  \& {Wada}}]{Kataoka2013b}
{Kataoka}, A., {Tanaka}, H., {Okuzumi}, S., \& {Wada}, K. 2013{\natexlab{a}},
  \aap, 557, L4

\bibitem[{{Kataoka} {et~al.}(2013{\natexlab{b}}){Kataoka}, {Tanaka}, {Okuzumi},
  \& {Wada}}]{Kataoka2013a}
---. 2013{\natexlab{b}}, \aap, 554, A4

\bibitem[{{Kenyon} {et~al.}(1994){Kenyon}, {Dobrzycka}, \&
  {Hartmann}}]{Kenyon1994}
{Kenyon}, S.~J., {Dobrzycka}, D., \& {Hartmann}, L. 1994, \aj, 108, 1872

\bibitem[{{Kenyon} \& {Hartmann}(1995)}]{Kenyon1995}
{Kenyon}, S.~J., \& {Hartmann}, L. 1995, \apjs, 101, 117

\bibitem[{{Kenyon} {et~al.}(1990){Kenyon}, {Hartmann}, {Strom}, \&
  {Strom}}]{Kenyon1990b}
{Kenyon}, S.~J., {Hartmann}, L.~W., {Strom}, K.~M., \& {Strom}, S.~E. 1990,
  \aj, 99, 869

\bibitem[{{Kneller} \& {Luborsky}(1963)}]{Kneller1963}
{Kneller}, E.~F., \& {Luborsky}, F.~E. 1963, Journal of Applied Physics, 34,
  656

\bibitem[{{Ko} {et~al.}(2020){Ko}, {Liu}, {Lai}, {Ching}, {Rao}, \&
  {Girart}}]{Ko2019}
{Ko}, C.-L., {Liu}, H.~B., {Lai}, S.-P., {et~al.} 2020, \apj, 889, 172

\bibitem[{{Koumpia} {et~al.}(2015){Koumpia}, {Harvey}, {Ossenkopf}, {van der
  Tak}, {Mookerjea}, {Fuente}, \& {Kramer}}]{Koumpia2015}
{Koumpia}, E., {Harvey}, P.~M., {Ossenkopf}, V., {et~al.} 2015, \aap, 580, A68

\bibitem[{{Kristensen} \& {Dunham}(2018)}]{Kristensen2018}
{Kristensen}, L.~E., \& {Dunham}, M.~M. 2018, \aap, 618, A158

\bibitem[{{Krumholz} {et~al.}(2012){Krumholz}, {Klein}, \&
  {McKee}}]{Krumholz2012}
{Krumholz}, M.~R., {Klein}, R.~I., \& {McKee}, C.~F. 2012, \apj, 754, 71

\bibitem[{{Krumholz} {et~al.}(2004){Krumholz}, {McKee}, \&
  {Klein}}]{Krumholz2004}
{Krumholz}, M.~R., {McKee}, C.~F., \& {Klein}, R.~I. 2004, \apj, 611, 399

\bibitem[{{Kuffmeier} {et~al.}(2018){Kuffmeier}, {Frimann}, {Jensen}, \&
  {Haugb{\o}lle}}]{Kuffmeier2018}
{Kuffmeier}, M., {Frimann}, S., {Jensen}, S.~S., \& {Haugb{\o}lle}, T. 2018,
  \mnras, 475, 2642

\bibitem[{{Kuffmeier} {et~al.}(2020){Kuffmeier}, {Reissl}, {Wolf}, {Stephens},
  \& {Calcutt}}]{Kuffmeier2020}
{Kuffmeier}, M., {Reissl}, S., {Wolf}, S., {Stephens}, I., \& {Calcutt}, H.
  2020, \aap, 639, A137

\bibitem[{{Kuiper} \& {Yorke}(2013)}]{Kuiper2013}
{Kuiper}, R., \& {Yorke}, H.~W. 2013, \apj, 772, 61

\bibitem[{{Kulkarni} \& {Romanova}(2008)}]{Kulkarni2008}
{Kulkarni}, A.~K., \& {Romanova}, M.~M. 2008, \mnras, 386, 673

\bibitem[{{Kwon} {et~al.}(2019){Kwon}, {Stephens}, {Tobin}, {Looney}, {Li},
  {van der Tak}, \& {Crutcher}}]{Kwon2019}
{Kwon}, W., {Stephens}, I.~W., {Tobin}, J.~J., {et~al.} 2019, \apj, 879, 25

\bibitem[{{Lazarian}(2020)}]{Lazarian2020}
{Lazarian}, A. 2020, \apj, 902, 97

\bibitem[{{Lazarian} \& {Draine}(1999)}]{Lazarian1999a}
{Lazarian}, A., \& {Draine}, B.~T. 1999, \apjl, 520, L67

\bibitem[{{Lazarian} \& {Efroimsky}(1999)}]{Lazarian1999c}
{Lazarian}, A., \& {Efroimsky}, M. 1999, \mnras, 303, 673

\bibitem[{{Lazarian} \& {Hoang}(2007)}]{LazarianHoang2007}
{Lazarian}, A., \& {Hoang}, T. 2007, \mnras, 378, 910

\bibitem[{{Lazarian} \& {Hoang}(2008)}]{Lazarian2008}
---. 2008, \apjl, 676, L25

\bibitem[{{Lazarian} \& {Hoang}(2021)}]{Lazarian2021}
---. 2021, \apj, 908, 12

\bibitem[{{Lazarian} \& {Roberge}(1997)}]{Lazarian1997}
{Lazarian}, A., \& {Roberge}, W.~G. 1997, \apj, 484, 230

\bibitem[{{Le Gouellec} {et~al.}(2022){Le Gouellec}, {Maury}, \&
  {Hull}}]{LeGouellec2023}
{Le Gouellec}, V. J.~M., {Maury}, A.~J., \& {Hull}, C. L.~H. 2022, arXiv
  e-prints, arXiv:2212.11899

\bibitem[{{Le Gouellec} {et~al.}(2019){Le Gouellec}, {Hull}, {Maury}, {Girart},
  {Tychoniec}, {Kristensen}, {Li}, {Louvet}, {Cortes}, \&
  {Rao}}]{LeGouellec2019a}
{Le Gouellec}, V. J.~M., {Hull}, C. L.~H., {Maury}, A.~J., {et~al.} 2019, \apj,
  885, 106

\bibitem[{{Le Gouellec} {et~al.}(2020){Le Gouellec}, {Maury}, {Guillet},
  {Hull}, {Girart}, {Verliat}, {Mignon-Risse}, {Valdivia}, {Hennebelle},
  {Gonz{\'a}lez}, \& {Louvet}}]{LeGouellec2020}
{Le Gouellec}, V.~J.~M., {Maury}, A.~J., {Guillet}, V., {et~al.} 2020, \aap,
  644, A11

\bibitem[{{Lebreuilly} {et~al.}(2019){Lebreuilly}, {Commer{\c{c}}on}, \&
  {Laibe}}]{Lebreuilly2019}
{Lebreuilly}, U., {Commer{\c{c}}on}, B., \& {Laibe}, G. 2019, \aap, 626, A96

\bibitem[{{Lebreuilly} {et~al.}(2020){Lebreuilly}, {Commer{\c{c}}on}, \&
  {Laibe}}]{Lebreuilly2020}
---. 2020, \aap, 641, A112

\bibitem[{{Lee} {et~al.}(2021{\natexlab{a}}){Lee}, {Johnstone}, {Lee},
  {Herczeg}, {Mairs}, {Contreras-Pe{\~n}a}, {Hatchell}, {Naylor}, {Bell},
  {Bourke}, {Broughton}, {Francis}, {Gupta}, {Harsono}, {Liu}, {Park},
  {Plovie}, {Moriarty-Schieven}, {Scholz}, {Sharma}, {Teixeira}, {Wang},
  {Aikawa}, {Bower}, {Vivien Chen}, {Bae}, {Baek}, {Chapman}, {Ping Chen},
  {Du}, {Dutta}, {Forbrich}, {Guo}, {Inutsuka}, {Kang}, {Kirk}, {Kuan}, {Kwon},
  {Lai}, {Lalchand}, {Lane}, {Lee}, {Liu}, {Morata}, {Pearson}, {Pon}, {Sahu},
  {Shang}, {Stamatellos}, {Tang}, {Xu}, {Yoo}, \& {Rawlings}}]{LeeYH2021}
{Lee}, Y.-H., {Johnstone}, D., {Lee}, J.-E., {et~al.} 2021{\natexlab{a}}, \apj,
  920, 119

\bibitem[{{Lee} {et~al.}(2021{\natexlab{b}}){Lee}, {Charnoz}, \&
  {Hennebelle}}]{LeeYN2021}
{Lee}, Y.-N., {Charnoz}, S., \& {Hennebelle}, P. 2021{\natexlab{b}}, \aap, 648,
  A101

\bibitem[{{Lehmann} {et~al.}(2020){Lehmann}, {Godard}, {Pineau des For{\^e}ts},
  \& {Falgarone}}]{Lehmann2020}
{Lehmann}, A., {Godard}, B., {Pineau des For{\^e}ts}, G., \& {Falgarone}, E.
  2020, \aap, 643, A101

\bibitem[{{Levermore}(1984)}]{Levermore1984}
{Levermore}, C.~D. 1984, \jqsrt, 31, 149

\bibitem[{{Levermore} \& {Pomraning}(1981)}]{Levermore1981}
{Levermore}, C.~D., \& {Pomraning}, G.~C. 1981, \apj, 248, 321

\bibitem[{{Martin}(1995)}]{Martin1995}
{Martin}, P.~G. 1995, \apjl, 445, L63

\bibitem[{{Masson} {et~al.}(2012){Masson}, {Teyssier}, {Mulet-Marquis},
  {Hennebelle}, \& {Chabrier}}]{Masson2012}
{Masson}, J., {Teyssier}, R., {Mulet-Marquis}, C., {Hennebelle}, P., \&
  {Chabrier}, G. 2012, \apjs, 201, 24

\bibitem[{{Mathis} {et~al.}(1983){Mathis}, {Mezger}, \& {Panagia}}]{Mathis1983}
{Mathis}, J.~S., {Mezger}, P.~G., \& {Panagia}, N. 1983, \aap, 500, 259

\bibitem[{{Mathis} {et~al.}(1977){Mathis}, {Rumpl}, \&
  {Nordsieck}}]{Mathis1977}
{Mathis}, J.~S., {Rumpl}, W., \& {Nordsieck}, K.~H. 1977, \apj, 217, 425

\bibitem[{{Maury} {et~al.}(2018){Maury}, {Girart}, {Zhang}, {Hennebelle},
  {Keto}, {Rao}, {Lai}, {Ohashi}, \& {Galametz}}]{Maury2018}
{Maury}, A.~J., {Girart}, J.~M., {Zhang}, Q., {et~al.} 2018, \mnras, 477, 2760

\bibitem[{{McClure}(2009)}]{McClure2009}
{McClure}, M. 2009, \apjl, 693, L81

\bibitem[{{McMullin} {et~al.}(2007){McMullin}, {Waters}, {Schiebel}, {Young},
  \& {Golap}}]{McMullin2007}
{McMullin}, J.~P., {Waters}, B., {Schiebel}, D., {Young}, W., \& {Golap}, K.
  2007, in Astronomical Society of the Pacific Conference Series, Vol. 376,
  Astronomical Data Analysis Software and Systems XVI, ed. R.~A. {Shaw},
  F.~{Hill}, \& D.~J. {Bell}, 127

\bibitem[{{Mignon-Risse} {et~al.}(2020){Mignon-Risse}, {Gonz{\'a}lez},
  {Commer{\c{c}}on}, \& {Rosdahl}}]{Risse2020}
{Mignon-Risse}, R., {Gonz{\'a}lez}, M., {Commer{\c{c}}on}, B., \& {Rosdahl}, J.
  2020, \aap, 635, A42

\bibitem[{{Miotello} {et~al.}(2014){Miotello}, {Testi}, {Lodato}, {Ricci},
  {Rosotti}, {Brooks}, {Maury}, \& {Natta}}]{Miotello2014}
{Miotello}, A., {Testi}, L., {Lodato}, G., {et~al.} 2014, \aap, 567, A32

\bibitem[{{Morrish}(2001)}]{Morrish2001}
{Morrish}, A.~H. 2001, {The Physical Principles of Magnetism}

\bibitem[{{Myers}(2011)}]{Myers2011}
{Myers}, P.~C. 2011, \apj, 743, 98

\bibitem[{{Myers}(2014)}]{Myers2014}
---. 2014, \apj, 781, 33

\bibitem[{{Myers} {et~al.}(1998){Myers}, {Adams}, {Chen}, \&
  {Schaff}}]{Myers1998}
{Myers}, P.~C., {Adams}, F.~C., {Chen}, H., \& {Schaff}, E. 1998, \apj, 492,
  703

\bibitem[{{Nakatani} {et~al.}(2020){Nakatani}, {Liu}, {Ohashi}, {Zhang},
  {Hanawa}, {Chandler}, {Oya}, \& {Sakai}}]{Nakatani2020}
{Nakatani}, R., {Liu}, H.~B., {Ohashi}, S., {et~al.} 2020, \apjl, 895, L2

\bibitem[{{Offner} {et~al.}(2009){Offner}, {Klein}, {McKee}, \&
  {Krumholz}}]{Offner2009}
{Offner}, S.~S.~R., {Klein}, R.~I., {McKee}, C.~F., \& {Krumholz}, M.~R. 2009,
  \apj, 703, 131

\bibitem[{{Offner} \& {McKee}(2011)}]{Offner2011b}
{Offner}, S. S.~R., \& {McKee}, C.~F. 2011, \apj, 736, 53

\bibitem[{{Ohashi} {et~al.}(2021){Ohashi}, {Kobayashi}, {Nakatani}, {Okuzumi},
  {Tanaka}, {Murakawa}, {Zhang}, {Liu}, \& {Sakai}}]{Ohashi2021}
{Ohashi}, S., {Kobayashi}, H., {Nakatani}, R., {et~al.} 2021, \apj, 907, 80

\bibitem[{{Ormel} {et~al.}(2009){Ormel}, {Paszun}, {Dominik}, \&
  {Tielens}}]{Ormel2009}
{Ormel}, C.~W., {Paszun}, D., {Dominik}, C., \& {Tielens}, A.~G.~G.~M. 2009,
  \aap, 502, 845

\bibitem[{{Ostriker} \& {Shu}(1995)}]{Ostriker1995}
{Ostriker}, E.~C., \& {Shu}, F.~H. 1995, \apj, 447, 813

\bibitem[{{Park} {et~al.}(2021){Park}, {Lee}, {Contreras Pe{\~n}a},
  {Johnstone}, {Herczeg}, {Lee}, {Lee}, {Bhardwaj}, \&
  {Moriarty-Schieven}}]{Park2021}
{Park}, W., {Lee}, J.-E., {Contreras Pe{\~n}a}, C., {et~al.} 2021, \apj, 920,
  132

\bibitem[{{Pattle} {et~al.}(2021){Pattle}, {Lai}, {Wright}, {Coud{\'e}},
  {Plambeck}, {Hoang}, {Tang}, {Bastien}, {Eswaraiah}, {Furuya}, {Hwang},
  {Inutsuka}, {Kim}, {Kirchschlager}, {Kwon}, {Lee}, {Liu}, {Lyo}, {Ohashi},
  {Rawlings}, {Tahani}, {Tamura}, {Soam}, {Wang}, \&
  {Ward-Thompson}}]{Pattle2021}
{Pattle}, K., {Lai}, S.-P., {Wright}, M., {et~al.} 2021, \mnras, 503, 3414

\bibitem[{{Pillai} {et~al.}(2020){Pillai}, {Clemens}, {Reissl}, {Myers},
  {Kauffmann}, {Lopez-Rodriguez}, {Alves}, {Franco}, {Henshaw}, {Menten},
  {Nakamura}, {Seifried}, {Sugitani}, \& {Wiesemeyer}}]{Pillai2020}
{Pillai}, T. G.~S., {Clemens}, D.~P., {Reissl}, S., {et~al.} 2020, Nature
  Astronomy, 4, 1195

\bibitem[{{Planck Collaboration} {et~al.}(2020){Planck Collaboration},
  {Aghanim}, {Akrami}, {Alves}, {Ashdown}, {Aumont}, {Baccigalupi},
  {Ballardini}, {Banday}, {Barreiro}, {Bartolo}, {Basak}, {Benabed}, {Bernard},
  {Bersanelli}, {Bielewicz}, {Bock}, {Bond}, {Borrill}, {Bouchet}, {Boulanger},
  {Bracco}, {Bucher}, {Burigana}, {Calabrese}, {Cardoso}, {Carron}, {Chary},
  {Chiang}, {Colombo}, {Combet}, {Crill}, {Cuttaia}, {de Bernardis}, {de
  Zotti}, {Delabrouille}, {Delouis}, {Di Valentino}, {Dickinson}, {Diego},
  {Dor{\'e}}, {Douspis}, {Ducout}, {Dupac}, {Efstathiou}, {Elsner},
  {En{\ss}lin}, {Eriksen}, {Falgarone}, {Fantaye}, {Fernandez-Cobos},
  {Ferri{\`e}re}, {Finelli}, {Forastieri}, {Frailis}, {Fraisse}, {Franceschi},
  {Frolov}, {Galeotta}, {Galli}, {Ganga}, {G{\'e}nova-Santos}, {Gerbino},
  {Ghosh}, {Gonz{\'a}lez-Nuevo}, {G{\'o}rski}, {Gratton}, {Green}, {Gruppuso},
  {Gudmundsson}, {Guillet}, {Handley}, {Hansen}, {Helou}, {Herranz}, {Hivon},
  {Huang}, {Jaffe}, {Jones}, {Keih{\"a}nen}, {Keskitalo}, {Kiiveri}, {Kim},
  {Krachmalnicoff}, {Kunz}, {Kurki-Suonio}, {Lagache}, {Lamarre}, {Lasenby},
  {Lattanzi}, {Lawrence}, {Le Jeune}, {Levrier}, {Liguori}, {Lilje},
  {Lindholm}, {L{\'o}pez-Caniego}, {Lubin}, {Ma}, {Mac{\'\i}as-P{\'e}rez},
  {Maggio}, {Maino}, {Mandolesi}, {Mangilli}, {Marcos-Caballero}, {Maris},
  {Martin}, {Mart{\'\i}nez-Gonz{\'a}lez}, {Matarrese}, {Mauri}, {McEwen},
  {Melchiorri}, {Mennella}, {Migliaccio}, {Miville-Desch{\^e}nes}, {Molinari},
  {Moneti}, {Montier}, {Morgante}, {Moss}, {Natoli}, {Pagano}, {Paoletti},
  {Patanchon}, {Perrotta}, {Pettorino}, {Piacentini}, {Polastri}, {Polenta},
  {Puget}, {Rachen}, {Reinecke}, {Remazeilles}, {Renzi}, {Ristorcelli},
  {Rocha}, {Rosset}, {Roudier}, {Rubi{\~n}o-Mart{\'\i}n}, {Ruiz-Granados},
  {Salvati}, {Sandri}, {Savelainen}, {Scott}, {Sirignano}, {Sunyaev},
  {Suur-Uski}, {Tauber}, {Tavagnacco}, {Tenti}, {Toffolatti}, {Tomasi},
  {Trombetti}, {Valiviita}, {Vansyngel}, {Van Tent}, {Vielva}, {Villa},
  {Vittorio}, {Wandelt}, {Wehus}, {Zacchei}, \& {Zonca}}]{Planck2018XII}
{Planck Collaboration}, {Aghanim}, N., {Akrami}, Y., {et~al.} 2020, \aap, 641,
  A12

\bibitem[{{Plunkett} {et~al.}(2015){Plunkett}, {Arce}, {Corder}, {Dunham},
  {Garay}, \& {Mardones}}]{Plunkett2015}
{Plunkett}, A.~L., {Arce}, H.~G., {Corder}, S.~A., {et~al.} 2015, \apj, 803, 22

\bibitem[{{Poteet} {et~al.}(2015){Poteet}, {Whittet}, \& {Draine}}]{Poteet2015}
{Poteet}, C.~A., {Whittet}, D. C.~B., \& {Draine}, B.~T. 2015, \apj, 801, 110

\bibitem[{{Purcell}(1979)}]{Purcell1979}
{Purcell}, E.~M. 1979, \apj, 231, 404

\bibitem[{{Ray} \& {Ferreira}(2021)}]{Ray2021}
{Ray}, T.~P., \& {Ferreira}, J. 2021, \nar, 93, 101615

\bibitem[{{Reissl} {et~al.}(2020){Reissl}, {Guillet}, {Brauer}, {Levrier},
  {Boulanger}, \& {Klessen}}]{Reissl2020}
{Reissl}, S., {Guillet}, V., {Brauer}, R., {et~al.} 2020, \aap, 640, A118

\bibitem[{{Reissl} {et~al.}(2022){Reissl}, {Meehan}, \& {Klessen}}]{Reissl2022}
{Reissl}, S., {Meehan}, P., \& {Klessen}, R.~S. 2022, arXiv e-prints,
  arXiv:2201.03694

\bibitem[{{Reissl} {et~al.}(2017){Reissl}, {Seifried}, {Wolf}, {Banerjee}, \&
  {Klessen}}]{Reissl2017}
{Reissl}, S., {Seifried}, D., {Wolf}, S., {Banerjee}, R., \& {Klessen}, R.~S.
  2017, \aap, 603, A71

\bibitem[{{Reissl} {et~al.}(2016){Reissl}, {Wolf}, \& {Brauer}}]{Reissl2016}
{Reissl}, S., {Wolf}, S., \& {Brauer}, R. 2016, \aap, 593, A87

\bibitem[{{Robitaille} \& {Bressert}(2012)}]{Robitaille2012}
{Robitaille}, T., \& {Bressert}, E. 2012, {APLpy: Astronomical Plotting Library
  in Python}, Astrophysics Source Code Library, ascl:1208.017

\bibitem[{{Rosdahl} {et~al.}(2013){Rosdahl}, {Blaizot}, {Aubert}, {Stranex}, \&
  {Teyssier}}]{Rosdahl2013}
{Rosdahl}, J., {Blaizot}, J., {Aubert}, D., {Stranex}, T., \& {Teyssier}, R.
  2013, \mnras, 436, 2188

\bibitem[{{Rosdahl} \& {Teyssier}(2015)}]{Rosdahl2015}
{Rosdahl}, J., \& {Teyssier}, R. 2015, \mnras, 449, 4380

\bibitem[{{Rosen} \& {Krumholz}(2020)}]{Rosen2020}
{Rosen}, A.~L., \& {Krumholz}, M.~R. 2020, \aj, 160, 78

\bibitem[{{Sadavoy} {et~al.}(2018{\natexlab{a}}){Sadavoy}, {Myers}, {Stephens},
  {Tobin}, {Commer{\c{c}}on}, {Henning}, {Looney}, {Kwon}, {Segura-Cox}, \&
  {Harris}}]{Sadavoy2018a}
{Sadavoy}, S.~I., {Myers}, P.~C., {Stephens}, I.~W., {et~al.}
  2018{\natexlab{a}}, \apj, 859, 165

\bibitem[{{Sadavoy} {et~al.}(2018{\natexlab{b}}){Sadavoy}, {Myers}, {Stephens},
  {Tobin}, {Kwon}, {Segura-Cox}, {Henning}, {Commer{\c c}on}, \&
  {Looney}}]{Sadavoy2018b}
---. 2018{\natexlab{b}}, \apj, 869, 115

\bibitem[{{Sadavoy} {et~al.}(2019){Sadavoy}, {Stephens}, {Myers}, {Looney},
  {Tobin}, {Kwon}, {Commer{\c{c}}on}, {Segura-Cox}, {Henning}, \&
  {Hennebelle}}]{Sadavoy2019}
{Sadavoy}, S.~I., {Stephens}, I.~W., {Myers}, P.~C., {et~al.} 2019, \apjs, 245,
  2

\bibitem[{{Silsbee} {et~al.}(2022){Silsbee}, {Akimkin}, {Ivlev}, {Testi},
  {Gong}, \& {Caselli}}]{Silsbee2022}
{Silsbee}, K., {Akimkin}, V., {Ivlev}, A.~V., {et~al.} 2022, \apj, 940, 188

\bibitem[{{St{\"a}uber} {et~al.}(2005){St{\"a}uber}, {Doty}, {van Dishoeck}, \&
  {Benz}}]{Stauber2005}
{St{\"a}uber}, P., {Doty}, S.~D., {van Dishoeck}, E.~F., \& {Benz}, A.~O. 2005,
  \aap, 440, 949

\bibitem[{{St{\"a}uber} {et~al.}(2004){St{\"a}uber}, {Doty}, {van Dishoeck},
  {J{\o}rgensen}, \& {Benz}}]{Stauber2004}
{St{\"a}uber}, P., {Doty}, S.~D., {van Dishoeck}, E.~F., {J{\o}rgensen}, J.~K.,
  \& {Benz}, A.~O. 2004, \aap, 425, 577

\bibitem[{{Takahashi} {et~al.}(2019){Takahashi}, {Machida}, {Tomisaka}, {Ho},
  {Fomalont}, {Nakanishi}, \& {Girart}}]{Takahashi2019}
{Takahashi}, S., {Machida}, M.~N., {Tomisaka}, K., {et~al.} 2019, \apj, 872, 70

\bibitem[{{Takasao} {et~al.}(2019){Takasao}, {Tomida}, {Iwasaki}, \&
  {Suzuki}}]{Takasao2019}
{Takasao}, S., {Tomida}, K., {Iwasaki}, K., \& {Suzuki}, T.~K. 2019, \apjl,
  878, L10

\bibitem[{{Tazaki} {et~al.}(2017){Tazaki}, {Lazarian}, \&
  {Nomura}}]{Tazaki2017}
{Tazaki}, R., {Lazarian}, A., \& {Nomura}, H. 2017, \apj, 839, 56

\bibitem[{{Teyssier, R.}(2002)}]{Teyssier2002}
{Teyssier, R.} 2002, A\&A, 385, 337

\bibitem[{{Tram} {et~al.}(2021{\natexlab{a}}){Tram}, {Hoang}, {Lee}, {Santos},
  {Soam}, {Lesaffre}, {Gusdorf}, \& {Reach}}]{Tram2021}
{Tram}, L.~N., {Hoang}, T., {Lee}, H., {et~al.} 2021{\natexlab{a}}, \apj, 906,
  115

\bibitem[{{Tram} {et~al.}(2021{\natexlab{b}}){Tram}, {Lee}, {Hoang}, {Michail},
  {Chuss}, {Nickerson}, {Rangwala}, \& {Reach}}]{Tram2021b}
{Tram}, L.~N., {Lee}, H., {Hoang}, T., {et~al.} 2021{\natexlab{b}}, \apj, 908,
  159

\bibitem[{{Tsukamoto} {et~al.}(2021){Tsukamoto}, {Machida}, \&
  {Inutsuka}}]{Tsukamoto2021}
{Tsukamoto}, Y., {Machida}, M.~N., \& {Inutsuka}, S.-i. 2021, \apjl, 920, L35

\bibitem[{{Valdivia} {et~al.}(2019){Valdivia}, {Maury}, {Brauer}, {Hennebelle},
  {Galametz}, {Guillet}, \& {Reissl}}]{Valdivia2019}
{Valdivia}, V., {Maury}, A., {Brauer}, R., {et~al.} 2019, \mnras, 488, 4897

\bibitem[{{Valdivia} {et~al.}(2022){Valdivia}, {Maury}, \&
  {Hennebelle}}]{Valdivia2022}
{Valdivia}, V., {Maury}, A., \& {Hennebelle}, P. 2022, \aap, 668, A83

\bibitem[{{Vaytet} {et~al.}(2018){Vaytet}, {Commer{\c{c}}on}, {Masson},
  {Gonz{\'a}lez}, \& {Chabrier}}]{Vaytet2018}
{Vaytet}, N., {Commer{\c{c}}on}, B., {Masson}, J., {Gonz{\'a}lez}, M., \&
  {Chabrier}, G. 2018, \aap, 615, A5

\bibitem[{{Verliat} {et~al.}(2022){Verliat}, {Hennebelle}, {Gonz{\'a}lez},
  {Lee}, \& {Geen}}]{Verliat2022}
{Verliat}, A., {Hennebelle}, P., {Gonz{\'a}lez}, M., {Lee}, Y.-N., \& {Geen},
  S. 2022, \aap, 663, A6

\bibitem[{{Visser} {et~al.}(2015){Visser}, {Bergin}, \&
  {J{\o}rgensen}}]{Visser2015}
{Visser}, R., {Bergin}, E.~A., \& {J{\o}rgensen}, J.~K. 2015, \aap, 577, A102

\bibitem[{{Visser} {et~al.}(2012){Visser}, {Kristensen}, {Bruderer}, {van
  Dishoeck}, {Herczeg}, {Brinch}, {Doty}, {Harsono}, \& {Wolfire}}]{Visser2012}
{Visser}, R., {Kristensen}, L.~E., {Bruderer}, S., {et~al.} 2012, \aap, 537,
  A55

\bibitem[{{Vorobyov} {et~al.}(2013){Vorobyov}, {Baraffe}, {Harries}, \&
  {Chabrier}}]{Vorobyov2013}
{Vorobyov}, E.~I., {Baraffe}, I., {Harries}, T., \& {Chabrier}, G. 2013, \aap,
  557, A35

\bibitem[{{Vorobyov} \& {Basu}(2005)}]{Vorobyov2005}
{Vorobyov}, E.~I., \& {Basu}, S. 2005, \apjl, 633, L137

\bibitem[{{Vorobyov} \& {Basu}(2006)}]{Vorobyov2006}
---. 2006, \apj, 650, 956

\bibitem[{{Vorobyov} \& {Basu}(2010)}]{Vorobyov2010}
---. 2010, \apj, 719, 1896

\bibitem[{{Weingartner} \& {Draine}(2001)}]{Weingartner2001}
{Weingartner}, J.~C., \& {Draine}, B.~T. 2001, \apj, 548, 296

\bibitem[{{Weingartner} {et~al.}(2021){Weingartner}, {Kolasi}, \&
  {Woods}}]{Weingartner2021}
{Weingartner}, J.~C., {Kolasi}, E., \& {Woods}, C. 2021, \mnras, 504, 1164

\bibitem[{{Wong} {et~al.}(2016){Wong}, {Hirashita}, \& {Li}}]{Wong2016}
{Wong}, Y. H.~V., {Hirashita}, H., \& {Li}, Z.-Y. 2016, \pasj, 68, 67

\bibitem[{{Yang}(2021)}]{Yang2021}
{Yang}, H. 2021, \apj, 911, 125

\bibitem[{{Young} \& {Evans}(2005)}]{Young2005}
{Young}, C.~H., \& {Evans}, Neal~J., I. 2005, \apj, 627, 293

\bibitem[{{Zakri} {et~al.}(2022){Zakri}, {Megeath}, {Fischer}, {Gutermuth},
  {Furlan}, {Hartmann}, {Karnath}, {Osorio}, {Safron}, {Stanke}, {Stutz},
  {Tobin}, {Allen}, {Federman}, {Habel}, {Manoj}, {Narang}, {Pokhrel},
  {Rebull}, {Sheehan}, \& {Watson}}]{Zakri222}
{Zakri}, W., {Megeath}, S.~T., {Fischer}, W.~J., {et~al.} 2022, \apjl, 924, L23

\end{thebibliography}
\bibliographystyle{apj}

%\cleardoublepage
%\newpage

\begin{appendix}
% \addcontentsline{toc}{section}{Appendix}
% \renewcommand{\thesection}{\Alph{section}}

% \clearpage
\section{Details about our synthetic observations and MHD model}
\label{sec:app_MHD_RT_details}

This simulation belongs to a set of simulations aiming to study the evolution of protostellar cores, and especially the role of jets, initial turbulent and magnetic energy, accretion efficiency and luminosity, and protostellar outflow feedback. Constructing this model was motivated by our quest to reproduce the observable features of Class 0 cores, observed at high angular resolution with ALMA. While outflows and jets are ubiquitous in young protostellar cores \citep{Bally2016,Ray2021}, their complete modeling remains difficult. The main reason is that they are launched at sub-au scales, while the largest scale we need to implement to model an isolated collapsing core is $\sim$0.1\,pc (the tiniest resolution element of the simulation, that uses the AMR, is $\sim$5 au). In addition, the physics occurring at the sub-au scales of the accretion demands very small time-steps in the calculations, which in turn, correspond to unrealistic computational times in current simulations. Magneto-centrifugal outflows can be launched \citep{Gerrard2019}, however, their energetics do not correspond to the protostellar jets observed in Class 0 cores. Therefore, a jet is implemented by hand at the creation of the sink particle. This jet ejects one third of the accreted material, with a speed of 66\% of the escape speed, and with an opening angle of 30$^\circ$.

The ALMA observations of Class 0 protostars used in \citet{LeGouellec2020} to estimate an average grain alignment efficiency correspond to objects located in close-by low- and intermediate mass star forming regions. They span a range of total envelope mass of $\sim$1-40$M_\odot$, and are likely of heterogeneous evolutionary ages. Not all of these cores exhibit pristine and symmetric envelope structures that we observe in our simulation. This suggests that the cores observed by ALMA have been formed in a variety of environments, \ie with different relative amount of large-scale turbulent, gravitational, and magnetic energies. We chose not to implement initial turbulence in order to obtain a pristine case of protostellar collapse, and understand more easily the propagation of photons in the radiative transfer calculations. If turbulence is implemented, significant precession of the outflow axis is caused by the sink particle' spin axis changes, that are due to the turbulent accretion flow. This causes outflow broadening and additional entrained material \citep{Rosen2020}. This could have limited our interpretations, especially as we aim to identify specific grain alignment environmental conditions, \eg toward outflow cavities and equatorial mid-planes. The turbulence in a core is also responsible for delaying the formation of the sink in simulations. The simulation snapshot we have chosen corresponds to a relatively young star forming core, compared to what is derived from observations for Class 0/I objects. The absence of initial turbulence (and thus the more rapid formation of the sink) can thus explain this young age. The initial mass-to-flux ratio, which varies significantly among star forming cores \citep{Hennebelle2018}, also plays a role in regulating the core dynamics during the collapse. We chose a reasonably magnetized core with $\mu\,=\,$5.

The luminosity from the accretion and the associated radiative pressure \citep{Hennebelle2020b,Rosen2020}, alongside the mechanical energy of the outflow/jet system, provide a radiative feedback in star forming regions, and can be important to reproduce the star forming activity of a given region. However, the effect of the accretion luminosity and subsequent radiative pressure on the evolution of the envelope are not implemented in this MHD simulation. We investigate the heating from the total protostellar luminosity (photospheric and accretion luminosities) in our radiative transfer calculations, and look at the resulting radiative field. From the jet and outflow launching models, the fraction of accretion energy that translates into radiative energy is still a free parameter in models \citep{Ostriker1995,Offner2009}. Therefore, we chose to explore a broad range of protostar luminosities in Section \ref{sec:p3_RT}.

The radiation field that escapes from the surface of the central protostellar embryo and the accretion shock are different in nature, compared to the radiation field escaping out of the sink particle. The radiation field escaping out of the sink particle is supposed to have already been reprocessed by high density/opacity material. 
% This causes the spectrum of the blackbody radiation field to evolve. 
In other words, at a given luminosity $\textrm{L}_\star \,\propto\, \textrm{T}_\star^4 \textrm{R}_\star^2$, the two parameters $\textrm{T}_\star$ and $\textrm{R}_\star$ evolve in such a way that the spectrum becomes less rich in high energy photons as they propagate within this high density material.
However, the radiation field is expected to be reprocessed toward longer wavelength photons very quickly, \ie at scales $\lesssim$ 1 au. Given the spatial scales we study, the shape of the blackbody spectrum thus does not have a significant impact on our results.
Therefore, in the work presented here we fix $\textrm{R}_\star$ and vary $\textrm{T}_\star$ in order to vary $\textrm{L}_\star$. At given luminosities, we have investigated the impact of varying $(\textrm{R}_\star,\textrm{T}_\star)$, and found that indeed, no effects are seen on the grain alignment parameters, and the dust polarization maps.

%Mention the recent Wruster works of core collapse simulation?

\clearpage
\section{The effects of the maximum grain size on the radiative transfer results}
\label{sec:app_RT_amax}

\citet{Valdivia2019} first presented the impact of the maximum dust grain size on polarized dust emission with POLARIS radiative transfers. In this Appendix, we present such effects on our model, exploring $a_{\rm{max}}\,=\,$0.5, 2, 10, 30, and 50\,$\mu$m. This parameter is of great importance on the grain alignment efficiency within the dense regions of protostars. Because photons are rapidly reprocessed toward longer wavelength in the envelope of protostars, and radiative torques can only spin up dust grains whose size are larger than the wavelength of the anisotropic component of the radiation field, dust grains larger than $\sim$ 10 $\mu$m must populate protostellar envelopes. Otherwise, the polarized dust emission would not be observable. We note that our study does not implement the maximum size of internally aligned grains. For the densest conditions of our model (\ie $n_{\textrm{H}}\,\geq\,10^{9}$ cm$^{-3}$), grains larger than $\sim\,$50 $\mu$m should not have efficient enough Barnett and inelastic relaxation processes to resist gaseous de-alignment. However, considering high irradiation conditions and a high level of iron inclusions, \ie $u_{\rm{rad}}$/$u_{\rm{ISRF}}\,\geq\,10^{6}$ and $N_{\textrm{cl}}\,\sim\,10^{5}$, can increase this upper limit (see Section 4 of \citealt{Hoang2022b}).

\begin{figure*}[!tbh]
%\noindent\begin{minipage}{\textwidth}
\centering
% \vspace{-0.6cm}
\includegraphics[scale=0.6,clip,trim= 3.0cm 3.0cm 2cm 0.5cm]{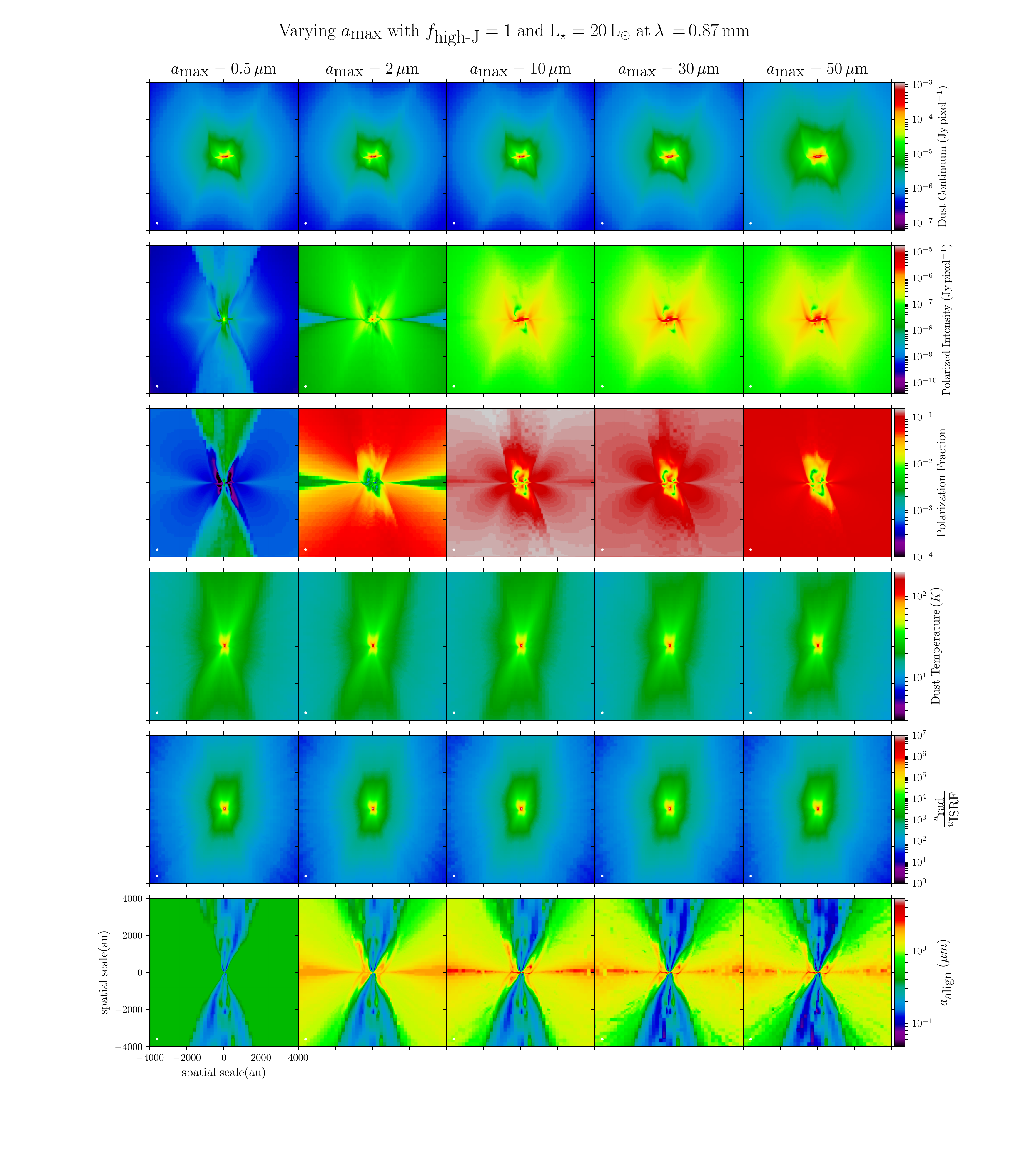}
\vspace{-0.3cm}
\caption{\footnotesize Effects of the maximum dust grain size $a_{\rm{max}}$ on the radiative transfer results at 0.87 mm, with the fixed parameters of $f_{\rm{high}-J}\,=\,1$, and $L_\star$\,=\,20\,L$_\odot$. This correspond to the set I of Table \ref{t.RT_details}. Each column is one radiative transfer run of POLARIS, with $a_{\rm{max}}\,=\,$0.5, 2, 10, 30, and 50\,$\mu$m. Each row is a quantity provided by the radiative transfer, from the first to the sixth row: total intensity Stokes $I$, polarized intensity $P$, polarization fraction, 2D temperature slice obtained at the center, 2D radiation field slice obtained at the center u$_{\rm{rad}}$/u$_{\rm{ISRF}}$, and the 2D slice of the $a_{\rm{align}}$ parameter obtained at the center. 
}
\label{fig:RT_maps_vamax}
%\end{minipage}
\end{figure*}

Figure \ref{fig:RT_maps_vamax} shows that changing the maximum grain size from 0.5 to 50\,$\mu$m, with $f_{\rm{high}-J}\,=\,1$ and $L_\star$\,=\,20\,L$_\odot$, involves significant changes in the resulting dust polarization maps (see also \citealt{ChauGiang2022}). The heating of the envelope remains the same, \ie changing the last bins of the MRN distribution does not affect the temperature or the radiation field in our calculations. The maximum grain size has, however, a large impact on the emissivity properties of dust grains, as Stokes $I$ derived at 0.87 mm increases with increasing $a_{\rm{max}}$ from 10 to 50\,$\mu$m. However the total intensity remain on average constant between 0.5 and 10\,$\mu$m. The polarized intensity, that depends on how evolves the fraction of aligned dust grains, evolves differently than Stokes $I$ does. While the irradiation remains the same in the cases of this figure, the evolution of the mean wavelength of the photons impinging onto dust grains could explain the evolution of the $a_{\rm{align}}$. When the $a_{\rm{max}}$ increases, $a_{\rm{align}}$ also increases. However, the more the maximum grain size increases, the larger is the fraction of aligned grains, and the more numerous are the grains that contributes to the polarized dust emission. This explain why the polarized intensity increases with increasing $a_{\rm{max}}$. Finally, the increase in polarized intensity is different from the increase in Stokes $I$. At 0.87 mm, this leads to a maximum polarization fraction in the case implementing $a_{\rm{max}}\,=\,10\,\mu$m.

\begin{figure*}[!tbh]
%\noindent\begin{minipage}{\textwidth}
\centering
% \vspace{-0.5cm}
\includegraphics[scale=0.65,clip,trim= 3.9cm 0.5cm 3cm 0cm]{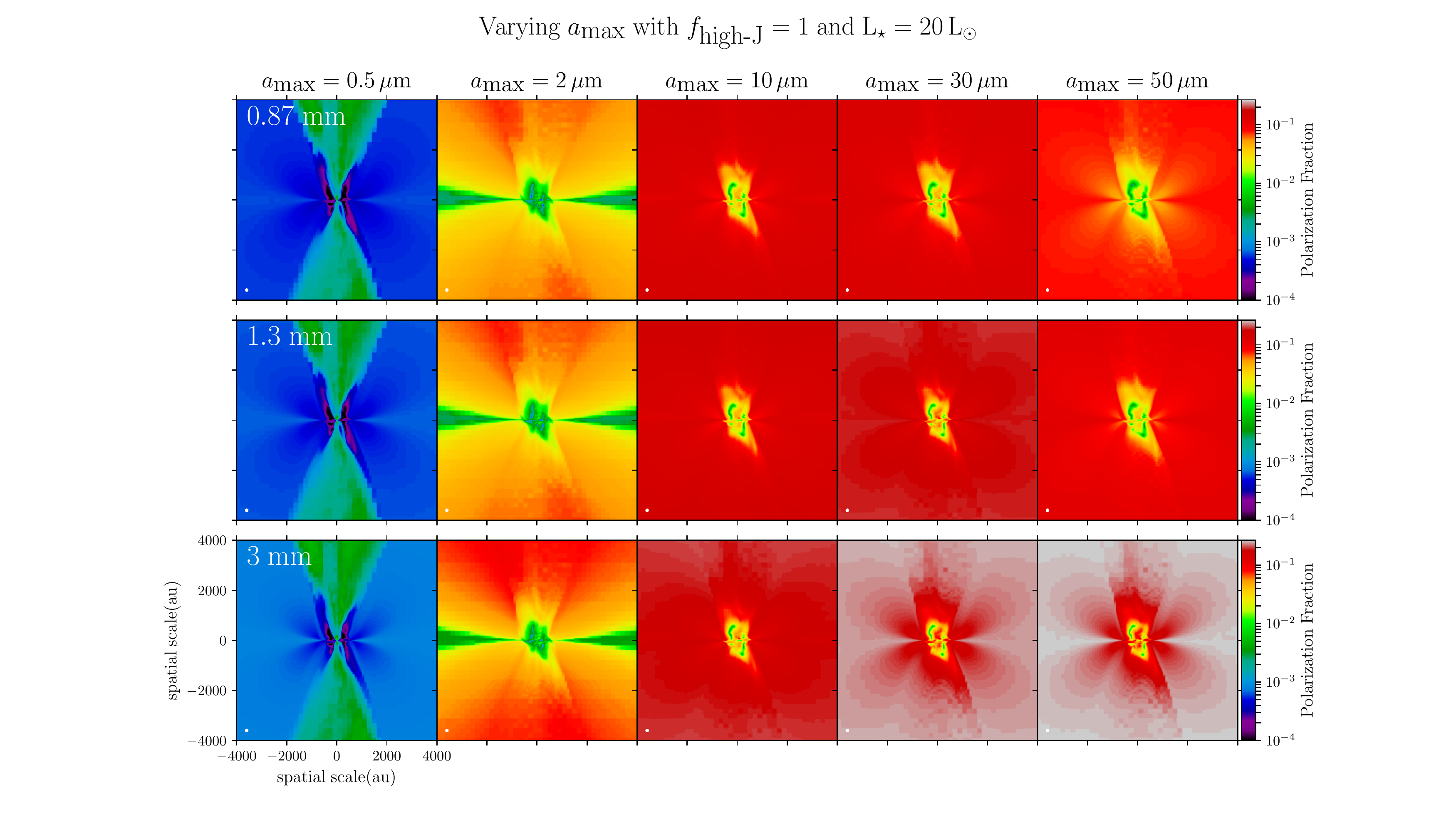}
\vspace{-0.2cm}
\caption{\small Effects of the maximum dust grain size $a_{\rm{max}}$ on the radiative transfer results at 0.87 mm, 1.3mm, and 3mm, with the fixed parameters of $f_{\rm{high}-J}\,=\,1$, and $L_\star$\,=\,20\,L$_\odot$. Each column is one radiative transfer run of POLARIS, with $a_{\rm{max}}\,=\,$0.5, 2, 10, 30, and 50\,$\mu$m, while each row corresponds to a wavelength of observation for the radiative transfer, \ie 0.87 mm, 1.3mm, and 3mm, as indicated on the first plot of each row. We only show the polarization fraction.
}
\label{fig:RT_maps_vamax_diffB}
%\end{minipage}
\end{figure*}

We show in Fig. \ref{fig:RT_maps_vamax_diffB} the impact that the wavelength of observation (0.87, 1.3, and 3mm) has on the polarization fraction maps, for the different values of maximum grain sizes of $a_{\rm{max}}\,=\,$0.5, 2, 10, 30, and 50\,$\mu$m. Within this range of maximum grain sizes, the maximum average polarization fraction in the core is obtained at $a_{\rm{max}}\,=\,10\,\mu$m when observing at 0.87 mm, at $a_{\rm{max}}\,=\,30\,\mu$m when observing at 1.3 mm, and at $a_{\rm{max}}\,=\,50\,\mu$m when observing at 3mm.
This is consistent with \citet{Valdivia2019}, who mentioned that larger dust grains behave more as spherical grains, producing a smaller difference between Stokes $I$ and the polarized intensity, thus producing a smaller polarization fraction.
In other words, when observed as a function of maximum grain size $a_{\rm{max}}$, the polarization fraction peak shifts to a higher value of $a_{\rm{max}}$ when increasing the wavelength of observations.
In addition, at a given maximum grain size, the polarization fraction increases with increasing wavelength of observation, especially for $a_{\rm{max}}\,=\,30$ and 50\,$\mu$m.
This highlights the importance of multi-wavelength observations to target the evolution of the polarization fraction. Given that the magnetic field morphology does not change, multi-wavelength observations of a given core offer the possibilities to constrain the properties of the aligned dust grains. If our implementation of RATs accurately reproduce the alignment of grains in the inner core, one could fit the evolution of the polarization fraction with wavelength and constrain the maximum grain sizes in the region exhibiting polarized dust emission. This could in turn bring constrains on the minimum size of rotationally disrupted grains. However, one needs to mind that the aligned grains contributing to the polarized dust emission may come from different regions when observing at different wavelength. How vary the magnetic field morphology across these different regions of emission can also affect the multi-wavelength evolution of the polarization fraction (see \citealt{Valdivia2022}). 

\begin{figure*}[!tbh]
%\noindent\begin{minipage}{\textwidth}
\centering
% \vspace{-0.6cm}
\subfigure{\includegraphics[scale=\scaleGE,clip,trim= 1.6cm 0.5cm 2cm 2cm]{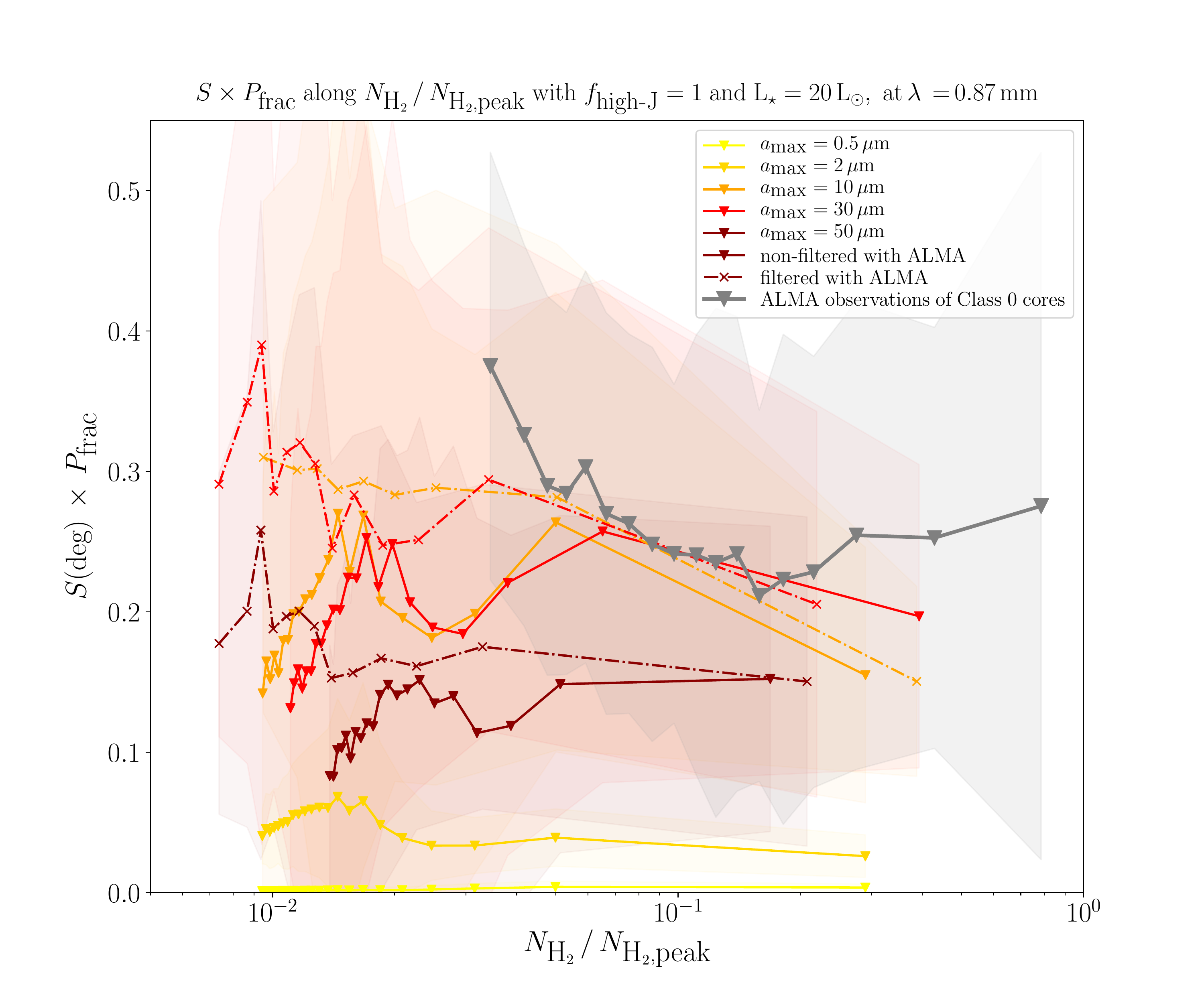}}
\subfigure{\includegraphics[scale=\scaleGE,clip,trim= 3.58cm 0.5cm 2cm 2cm]{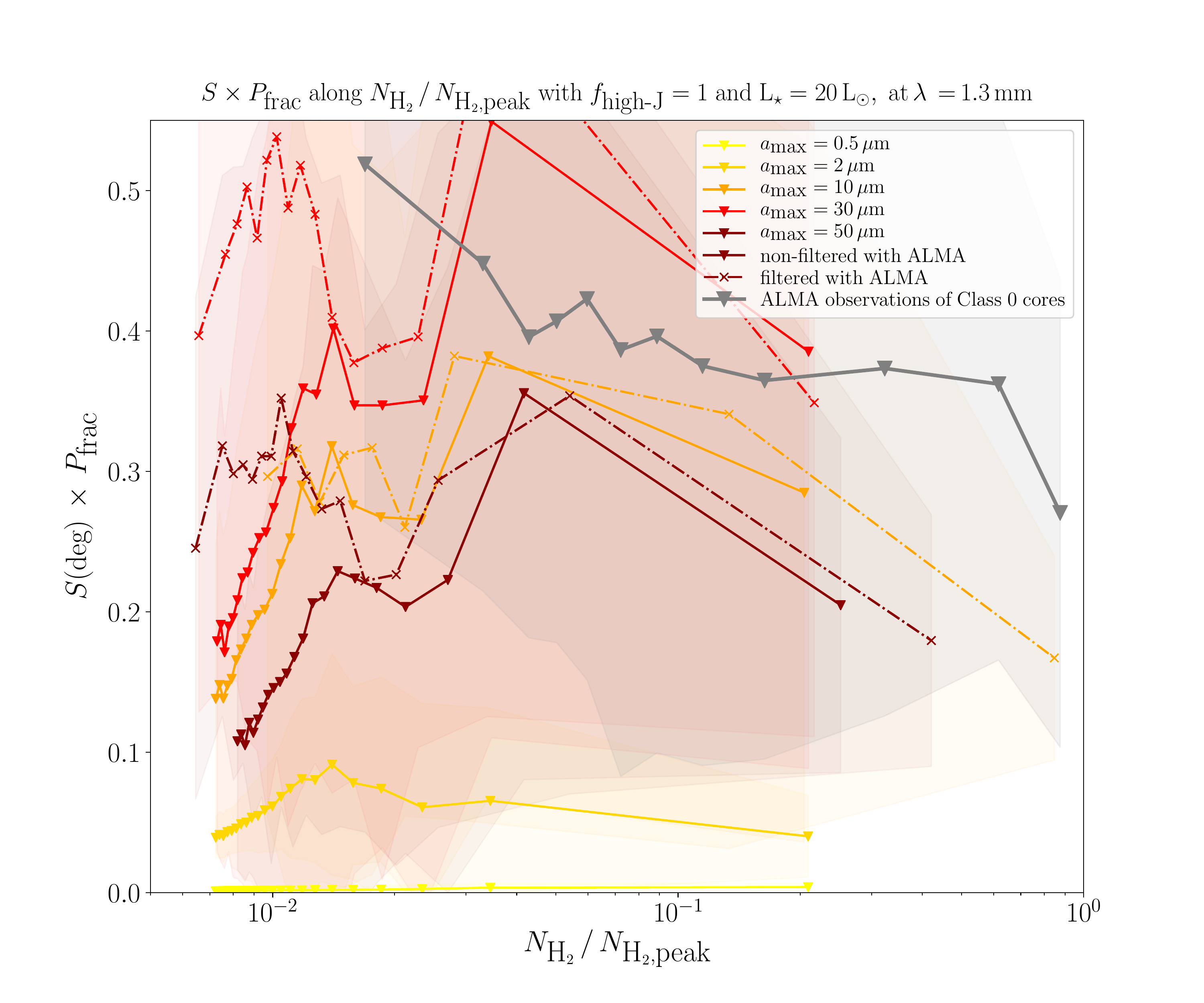}}
\subfigure{\includegraphics[scale=\scaleGE,clip,trim= 1.6cm 0.5cm 2cm 2cm]{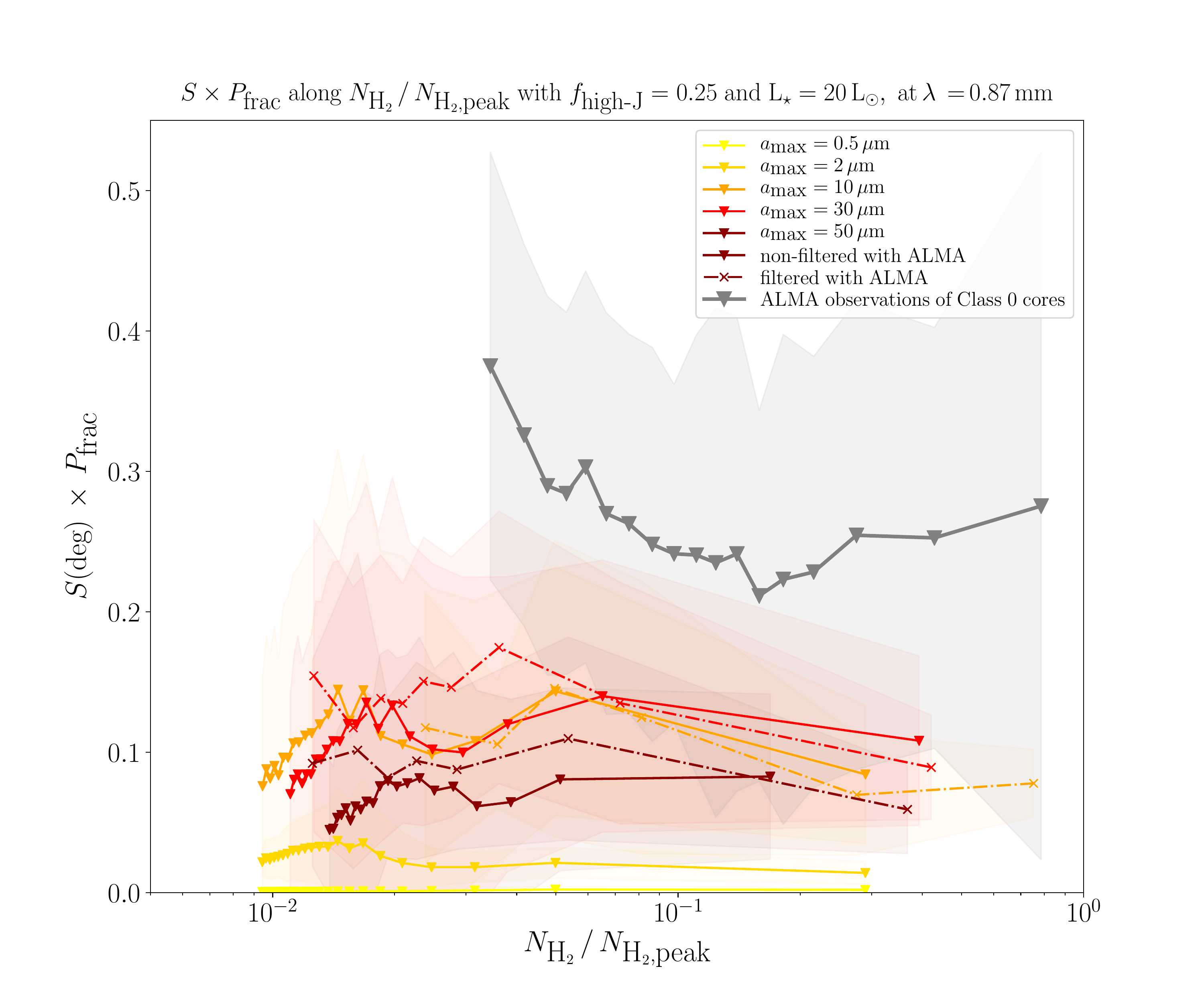}}
\subfigure{\includegraphics[scale=\scaleGE,clip,trim= 3.58cm 0.5cm 2cm 2cm]{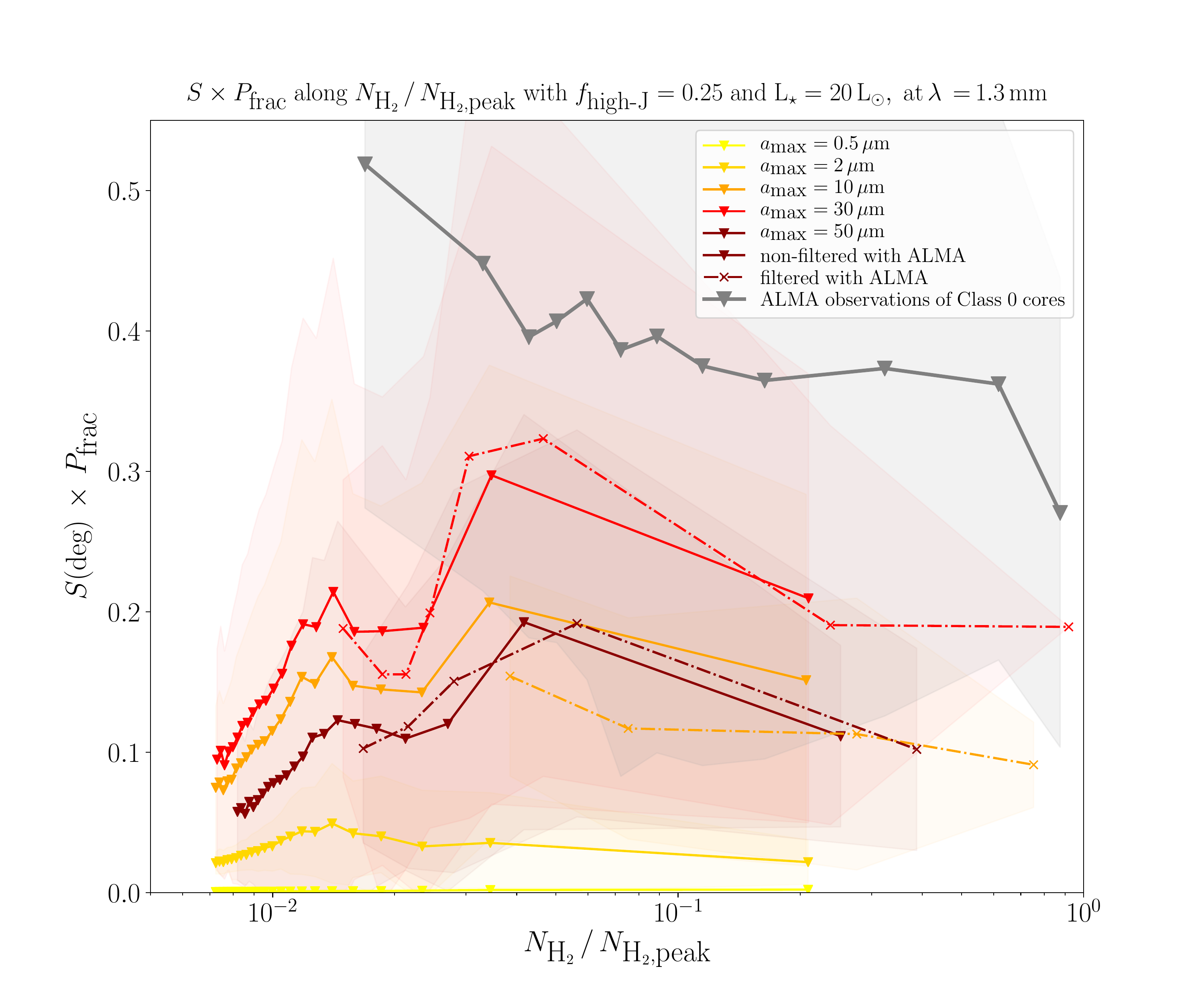}}
\caption{\footnotesize Comparisons on the evolution of $\StimesP$ as a function of normalized column density $N_{{\textrm{H}_2}}/N_{{\textrm{H}_2,\textrm{peak}}}$, between ALMA observations, and our models, before and after spatial filtering. The left (right) column corresponds to results of the radiative transfer performed at 0.87 mm (1.3mm). Same as Figure \ref{fig:RT_maps_GE1} for the sets I and IV (see Table \ref{t.RT_details}).}
\label{fig:RT_maps_GE1_app}
%\end{minipage}
\end{figure*}

Similar to the Figures presented in Section \ref{sec:align_eff}, Figure \ref{fig:RT_maps_GE1_app} presents the evolution of $\StimesP$ as a function of normalized column density $N_{{\textrm{H}_2}}/N_{{\textrm{H}_2,\textrm{peak}}}$, in ALMA observations and our models, before and after spatial filtering, for the sets I and IV of radiative transfer runs. In this Figure, we vary the a$_\textrm{max}$ values in the range 0.5, 2, 10, 30, 50 $\mu$m, with $L_\star$ = 20 L$\odot$. When $f_{\textrm{high-J}}\,=\,1$, the implementation of large grains with a$_\textrm{max} \,\geq\,10\,\mu$m allows to reproduce marginally the average level of the $\StimesP$ of ALMA observations. As noticed in Fig. \ref{fig:RT_maps_vamax_diffB}, we note also that the level of grain alignment efficiency traced by $\StimesP$ decreases when a$_\textrm{max}$\,=50\, $\mu$m, at 0.87 and 1.3 mm. When $f_{\textrm{high-J}}\,=\,0.25$, none of the a$_\textrm{max}$ values we implement produces high enough $\StimesP$ values to match ALMA observations.

\clearpage
\section{The effects of $f_{\textrm{high-J}}$ on the radiative transfer results}
\label{sec:app_RT_highJ}

\begin{figure*}[h!]
%\noindent\begin{minipage}{\textwidth}
\centering
% \vspace{-0.5cm}
\includegraphics[scale=0.63,clip,trim= 3.5cm 0cm 3.5cm 0cm]{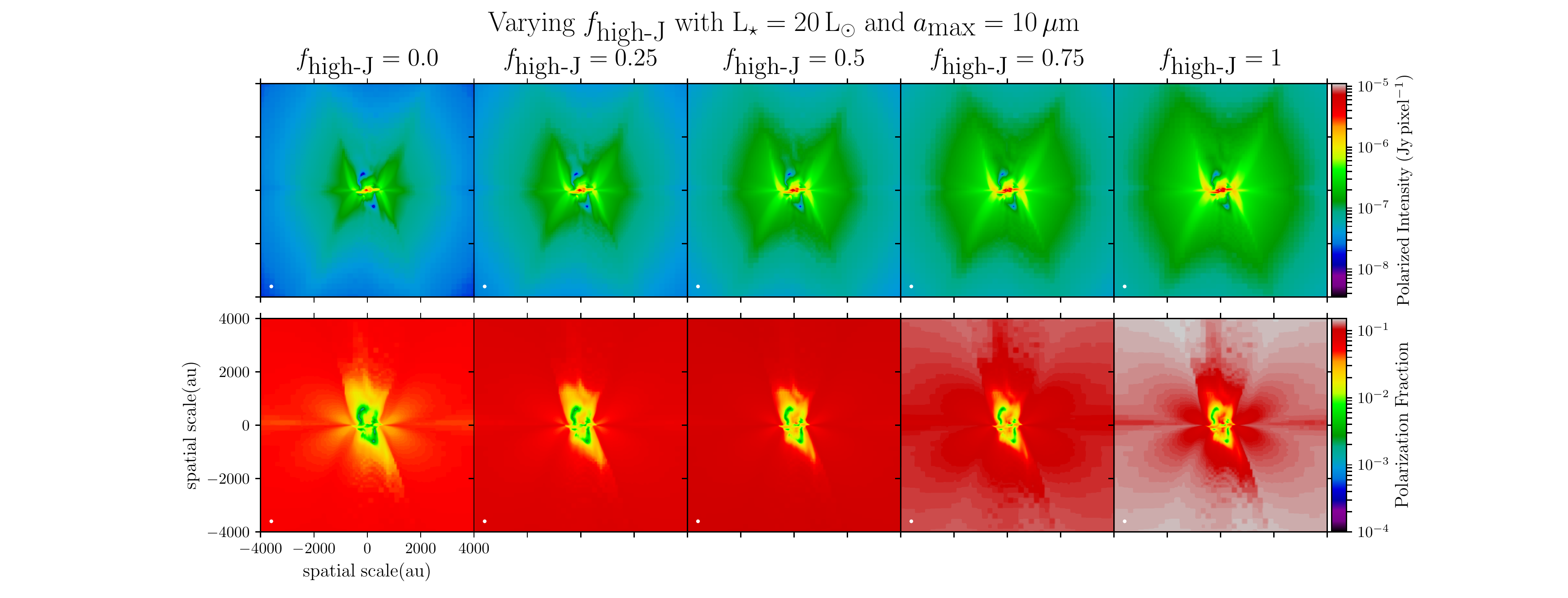}
\vspace{-0.2cm}
\captionof{figure}{\small Effects of the $f_{\rm{high}-J}$ parameter on the radiative transfer results at 0.87 mm, with the fixed parameters of $a_{\rm{max}}\,=\,10\,\mu$m, and $L_\star$\,=\,20\,L$_\odot$. This correspond to the set II of Table \ref{t.RT_details}. Each column is one radiative transfer run of POLARIS, with $f_{\rm{high}-J}\,=\,$0.0, 0.25, 0.5, 0.75, and 1.0. The top row is the polarized intensity $P$, the bottom row is the polarization fraction.}
\label{fig:RT_maps_vHJ}
%\end{minipage}
\end{figure*}

We refer to the recent study by \citet{ChauGiang2022} who thoroughly explored the polarized dust emission produced by super-paramagnetic grains. Yet we explore in this appendix the effects of changing the $f_{\rm{high}-J}$ parameter on the results of the radiative transfer calculations. In Figure \ref{fig:RT_maps_vHJ}, we vary the $f_{\rm{high}-J}$ parameter, from 0 to 1, with $a_{\rm{max}}\,=\,10\,\mu$m and $L_\star$\,=\,20\,L$_\odot$. 
The polarized intensity and polarization fraction increase with increasing fraction of grains at the high-$J$ attractor point, the fraction of grains (among those that are aligned) that can be considered perfectly aligned if their internal alignment is efficient \citep{Hoang2022,ChauGiang2022}.
The resulting polarization fraction is strongly affected by this parameter. Inside the central $\sim$\,2000\,au, $f_{\rm{high}-J}\,=\,0$ causes the polarization fraction to not be higher than 3\%, while $f_{\rm{high}-J}\,=\,1$ causes the polarization fraction to reach $\sim\,12\%$ in the outflow cavity walls.

\begin{figure*}[h!]
%\noindent\begin{minipage}{\textwidth}
\centering
 \vspace{0.4cm}
\includegraphics[scale=\scaleGE,clip,trim= 1.6cm 0.5cm 2cm 2cm]{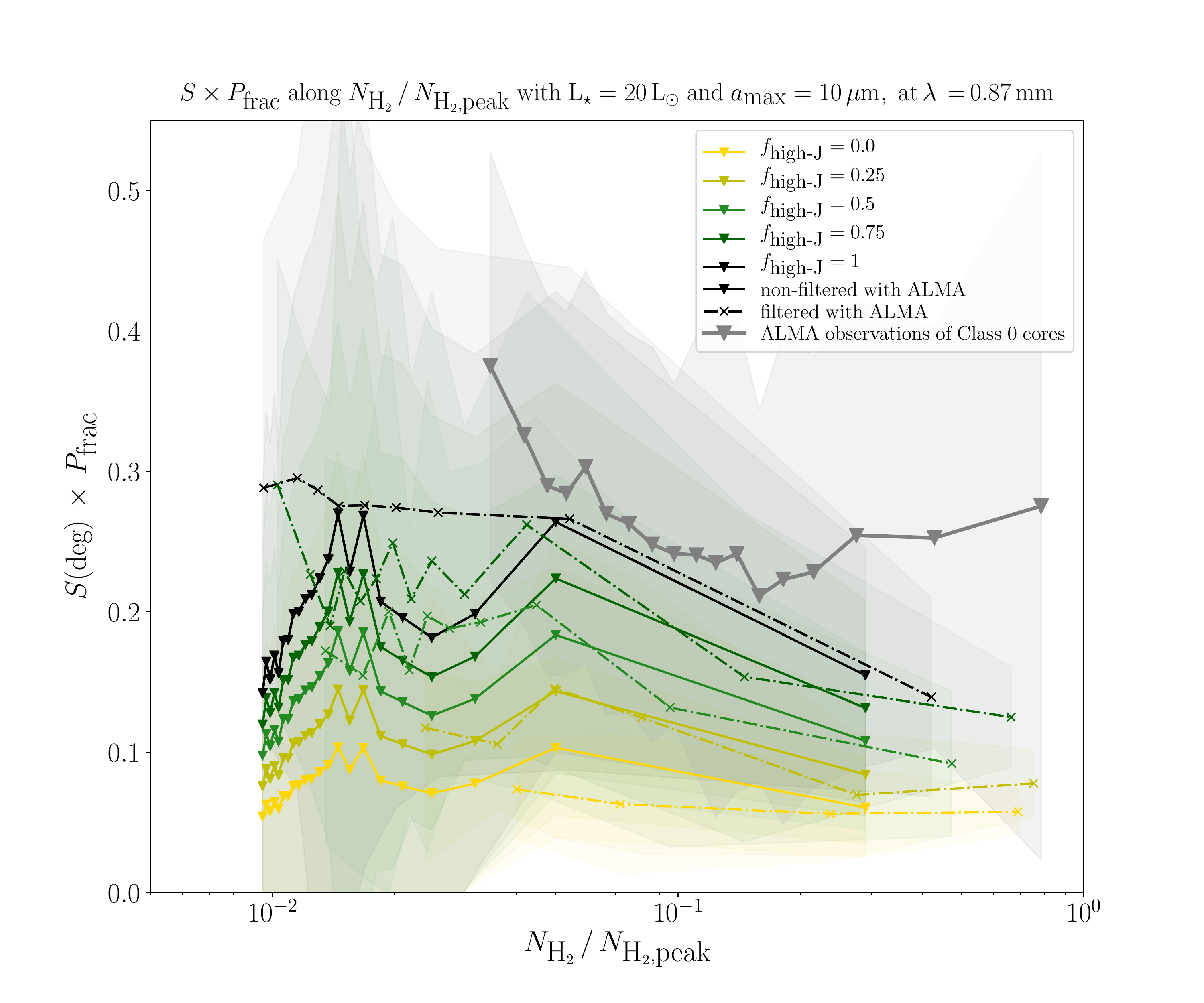}
\includegraphics[scale=\scaleGE,clip,trim= 3.58cm 0.5cm 2cm 2cm]{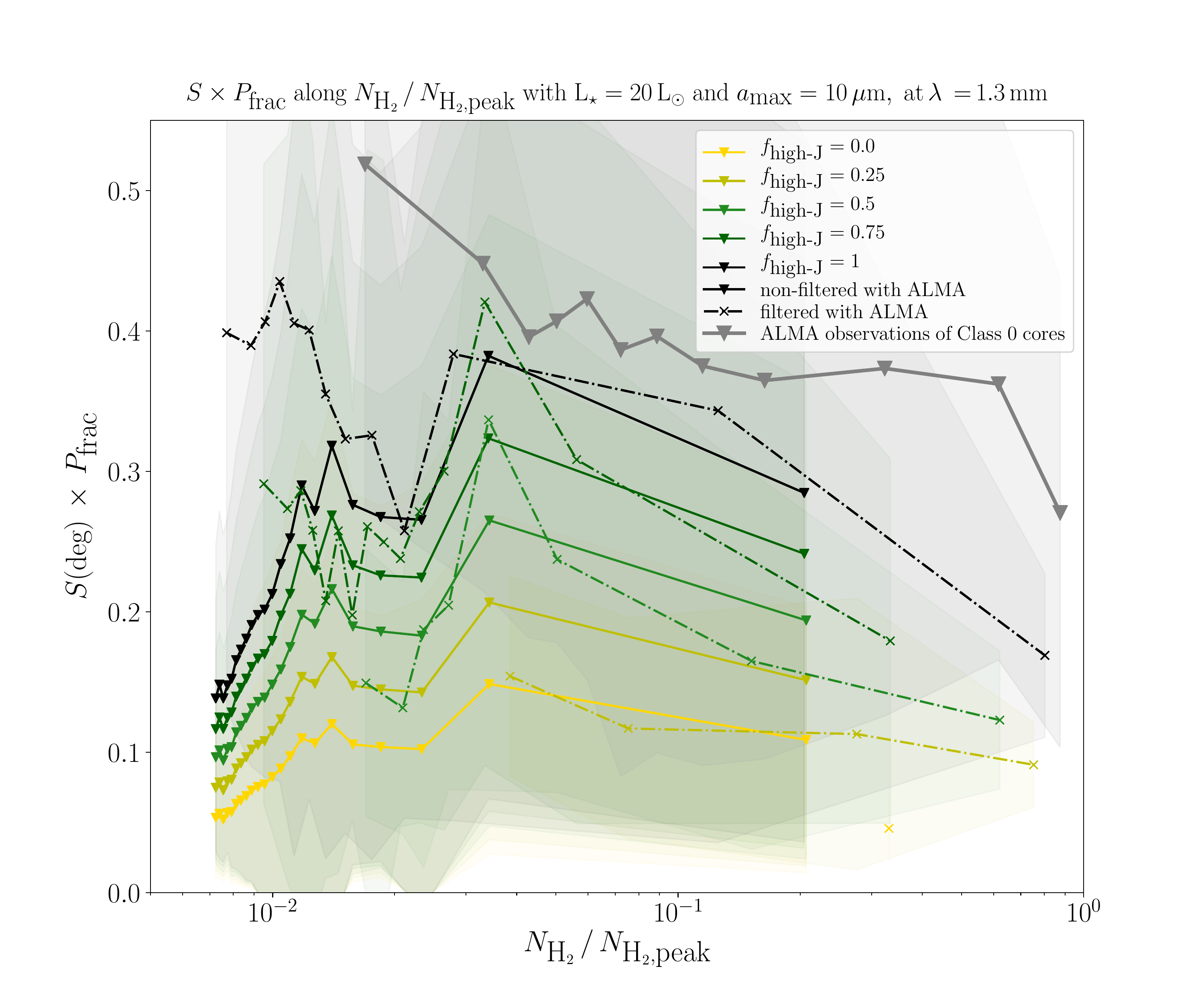}
\vspace{-0.2cm}
\captionof{figure}{\footnotesize Comparisons on the evolution of $\StimesP$ as a function of normalized column density $N_{{\textrm{H}_2}}/N_{{\textrm{H}_2,\textrm{peak}}}$, between ALMA observations, and our models, before and after spatial filtering. Same as Figure \ref{fig:RT_maps_GE1} for the set II (see Table \ref{t.RT_details}).}
\label{fig:RT_maps_GE2_app}
%\end{minipage}
\end{figure*}

Figure \ref{fig:RT_maps_GE2_app} presents the evolution of $\StimesP$ as a function of normalized column density $N_{{\textrm{H}_2}}/N_{{\textrm{H}_2,\textrm{peak}}}$, in ALMA observations and our models, before and after spatial filtering, for the set II of radiative transfer runs. In this Figure, we vary the $f_{\textrm{high-J}}$ values in the range 0.0, 0.25, 0.5, 0.75, 1.0., with $L_\star$ = 20 L$\odot$ and $a_{\rm{max}}\,=\,$10\,$\mu$m. Only values of $f_{\textrm{high-J}}$ close to 1 could produce high enough $\StimesP$ values in order to match ALMA observations. 

%\clearpage
\section{Correction for the optical thickness on the normalized column density of radiative transfer models.}
\label{sec:app_RT_alpha}

The distribution of normalized gas column density is more peaked in our model than in the distribution derived from ALMA observations in \citet{LeGouellec2020}. This difference can be explained by optically thick emission in the line of sights located toward the total intensity peak in ALMA maps, precluding us from accessing the true value of the column density peak. This effect results in the distributions of $\StimesP$ as a function of normalized column density $N_{{\textrm{H}_2}}/N_{{\textrm{H}_2,\textrm{peak}}}$ for our models being shifted to the left compared to the ALMA distribution.

\begin{figure*}[!bh]
\centering
\hspace{-0.64cm}
\subfigure{
\includegraphics[scale=0.3,clip,trim= 0cm 0cm 0cm 0cm]{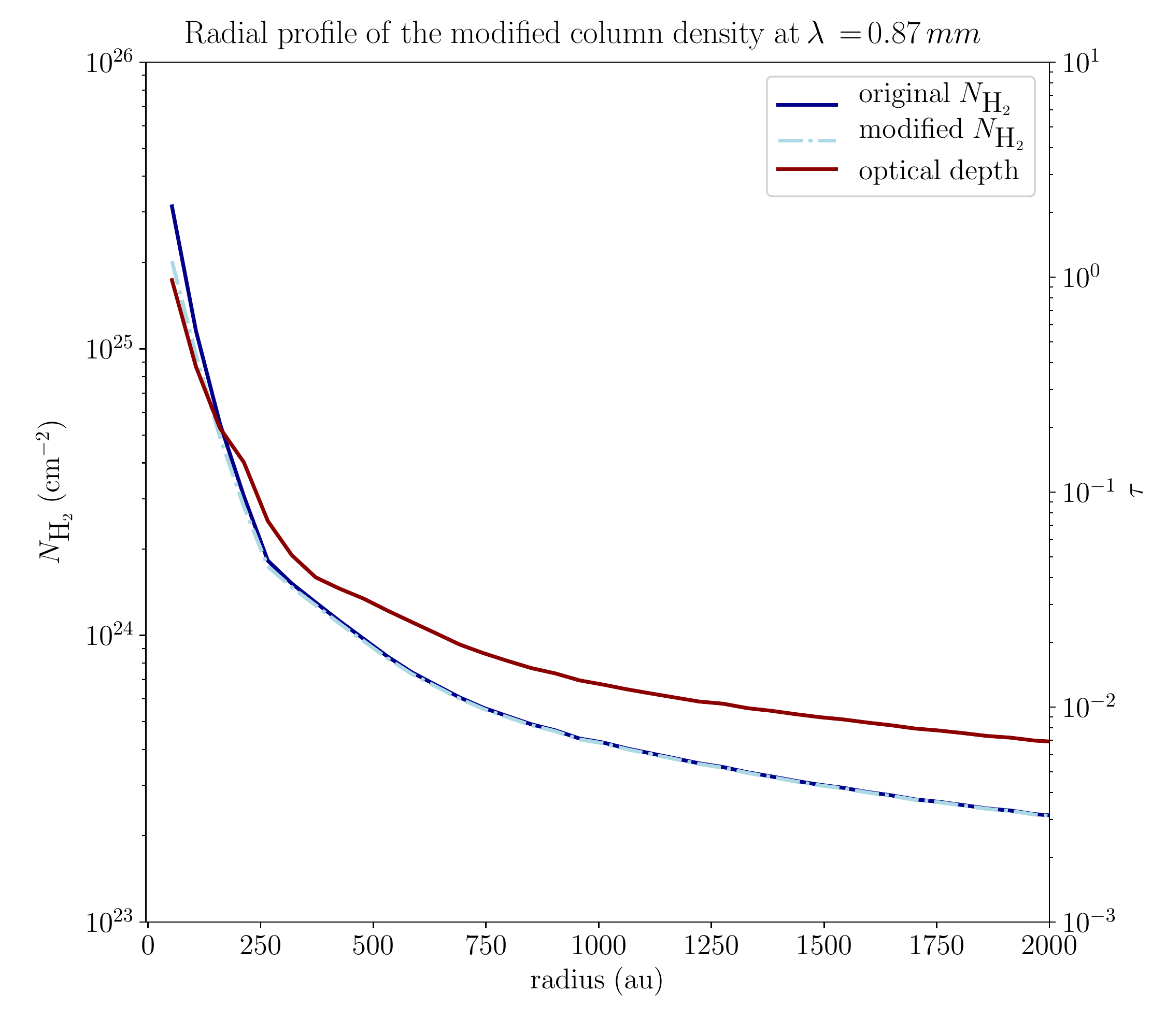}}
\subfigure{
\includegraphics[scale=0.3,clip,trim= 0cm 0cm 0cm 0cm]{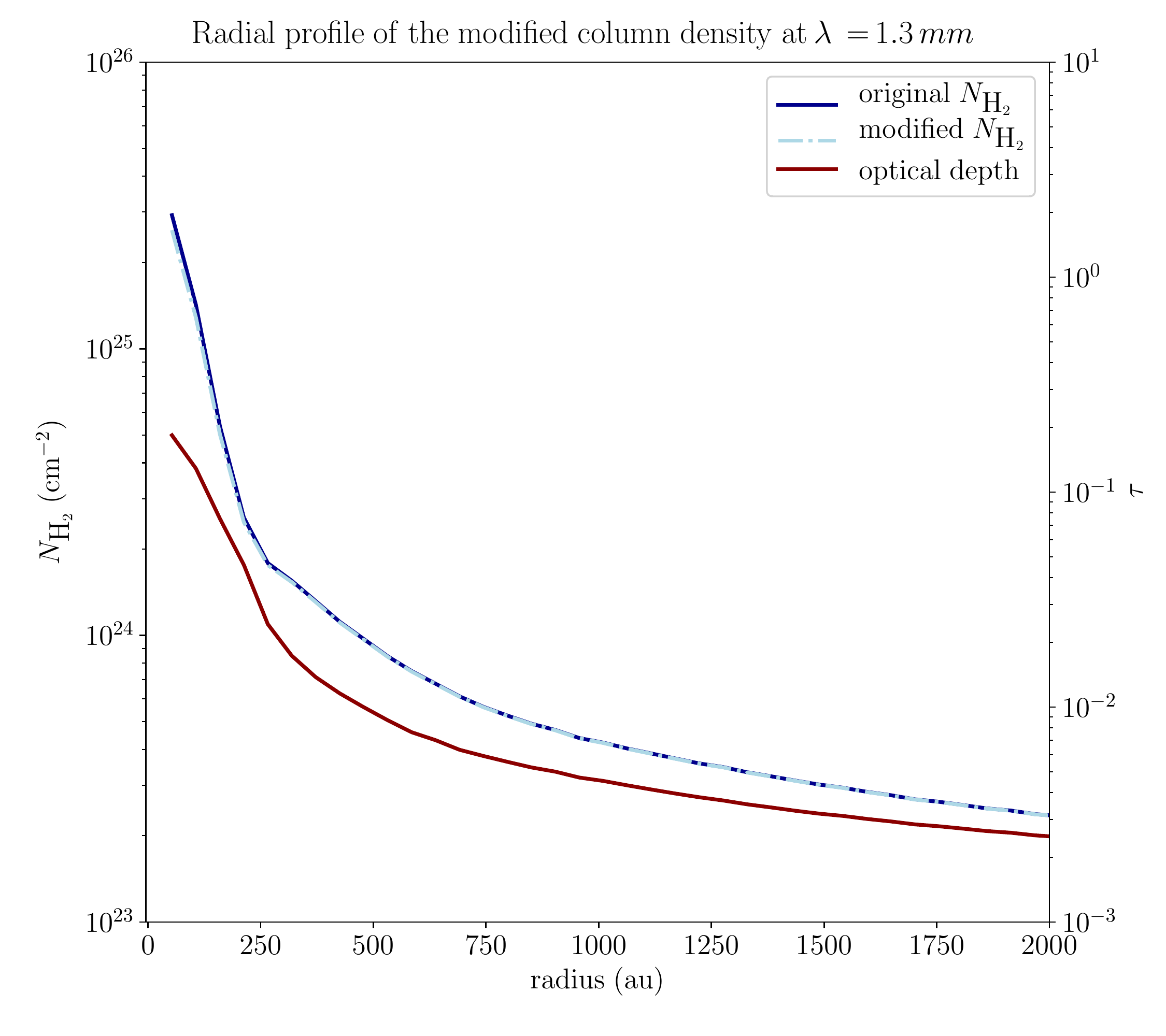}}
\caption[Alpha maps]{\small Correction for the optical thickness on the gas column density of our radiative transfer models, computed with $a_\textrm{max}$\,=\,10 $\mu$m, at $\lambda\,=\,0.87$ mm and $\lambda\,=\,1.3$ mm, respectively. The dot-dash light (solid dark) blue line shows the azimuthally averaged modified (original) gas column density as a function of radius. The dark red solid line shows the azimuthally averaged optical depth $\tau$ as a function of radius.
}
 \label{fig:alpha_maps_app}
\end{figure*}

We applied a tentative correction to the column density map in our model to make the comparisons with the ALMA observations more reliable. This correction factor is computed from the optical depth $\tau$ obtained in each radiative transfer calculation (see Fig. \ref{fig:polaris_canonical}). In each pixel, the gas column density is multiplied by $(1-e^{-\tau_{\lambda}})/\tau_{\lambda}$, to take into account the error made in assuming an optically thin emission computing $N_{{\textrm{H}_2}}/N_{{\textrm{H}_2,\textrm{peak}}}$ in ALMA observations. This correction is a first order approximation, and represents the error made assuming an optically thin emission while $\tau$ is no longer close to 0. Figure \ref{fig:alpha_maps_app} presents the radial distribution of $\tau_{\lambda}$, and the distribution of original and corrected $N_{{\textrm{H}_2}}$, at $\lambda\,=\,1.3$ and 0.87 mm.

\end{appendix}

\end{document}